%% file: U_om.tex
\documentclass[aps,prl,reprint,twocolumn,superscriptaddress,floatfix,amsmath,amssymb]{revtex4} 

\pdfoutput=1

\usepackage[latin1]{inputenc}
\usepackage{xcolor}
\usepackage{graphicx}%
\usepackage{dcolumn}%
\usepackage{bm}%
\usepackage{latexsym}
\usepackage{amsmath}
\usepackage{extarrows}
\usepackage{natbib}
\usepackage{braket}
\usepackage{url}
\usepackage{tikz}
\usepackage{hyperref}

\hypersetup{
	pdftoolbar=true,        
	pdfmenubar=true,        
	pdffitwindow=false,     
	pdfstartview={FitH},    
	pdftitle={My title},    
	pdfauthor={Author},     
	pdfsubject={Subject},   
	pdfcreator={Creator},   
	pdfproducer={Producer}, 
	pdfnewwindow=true,      
	colorlinks=true,       
	linkcolor=blue,          
	citecolor=magenta,        
	filecolor=magenta,      
	urlcolor=cyan           
}

\let\emptyset\varnothing

\begin{document}

	\title{Towards a minimal description of dynamical correlation in metals}
	
	\newcommand{\epfl}{Theory and Simulation of Materials (THEOS), and National Centre for Computational Design and Discovery of Novel Materials (MARVEL), \'Ecole Polytechnique F\'ed\'erale de Lausanne, Lausanne, Switzerland}
        \newcommand{\champi}{Coll\`ege Champittet, Pully, Switzerland}
	\newcommand{\etsf}{European Theoretical Spectroscopy Facility (ETSF)}
        \newcommand{\LMS}{Laboratory for Materials Simulations, Paul Scherrer Institute, Villigen, Switzerland}

	\author{Marco Vanzini}
	\email[]{marco.vanzini@champittet.ch}
	\affiliation{\epfl}
	\affiliation{\etsf}
        \affiliation{\champi}
	
	\author{Nicola Marzari}
        \email[]{nicola.marzari@epfl.ch}
	\affiliation{\epfl}
        \affiliation{\LMS}
	
	\date{\today}
	
	\begin{abstract}
		Dynamical correlations and non-local contributions beyond static mean-field theories are of fundamental importance for describing the electronic structure of correlated metals. Their effects are usually described with many-body approaches and non-local dynamical self-energies. We suggest here a class of simple model self-energies that are a generalization of the static DFT + Hubbard approach. This formulation, for simplicity called DFT$+U(\omega)+V$, 
		provides an intuitive physical picture, a lightweight implementation, and displays very good agreement with experimental data.
	\end{abstract}
	
	\maketitle

	Computational physics offers an invaluable tool to test our understanding of materials \cite{Marzari2021,PhysRevLett.108.068701,Jain2016}. Several calculated quantities can be compared to the outcome of experiments, yielding often quantitative, besides qualitative, agreement. In particular, first-principles approaches are particularly appealing as they do not rely on any adjustable parameter. The standard method of choice is density-functional theory (DFT) \cite{HK1964,KS1965,parr1994density}, a powerful and popular mean-field theory that often yields reliable ground-state properties. Even if available in principle, it is practically difficult to get information on electronic excited states \cite{PhysRevLett.49.1691,PhysRevLett.51.1888}. The latter can be measured in angle-resolved photoemission experiments (ARPES) \cite{hufner2003photoelectron,RevModPhys.75.473}, and are a primary source of information on the electronic structure of materials \cite{Onida2002}. 
	To obtain an accurate description of electronic spectra \cite{PhysRevB.58.15565}, different and advanced theories are needed, even for simple systems \cite{PhysRevLett.107.166401,zhou2018dispersing}. Two of these, GW \cite{Hedin1965,AryaGunn,Hyber,Schilfgaarde2006,10.3389/fchem.2019.00377} and DMFT, although often considered as alternative routes, \cite{Metzner,Georges,RevModPhys.68.13,RevModPhysKotliar}, find their common root in many-body perturbation theory \cite{fetter2003quantum,Strinati,luciabook}. Together with their refinements and extensions, they are the state-of-the-art tools in first-principles computations.
	A key class of compounds for which these methods should be applied are correlated materials \cite{Dagotto257,Tokura2003}, that can offer several promising technological applications \cite{app1,C1EE02717D}; usually, these are systems with partially filled $d$ or $f$ orbitals.
	DFT$+U$ \cite{Anisimov_1991,PhysRevB.48.16929,PhysRevB.57.1505} and its extensions \cite{Campo} aimed to include a static Hubbard interaction $U$ in a mean-field way. Its value, that can be tuned to get additional insight, can also be determined ab-initio \cite{PhysRevB.41.514,PhysRevB.57.4364,PhysRevB.43.7570,PhysRevB.39.9028,PhysRevB.70.195104,PhysRevB.71.035105,PhysRevB.98.085127,PhysRevB.103.045141}. In reality, the Hubbard terms mostly remove the self-interaction error \cite{PhysRevB.23.5048,doi:10.1063/1.4866996} from DFT \cite{PhysRevLett.97.103001} and often improve the description of electronic interactions in insulators \cite{PhysRevB.60.10763,PhysRevB.74.155108}. In metals, however, dynamical correlation effects are essential \cite{PhysRevB.67.153106}. 
	Two examples we will consider are the perovskites SrVO$_3$ and LaNiO$_3$: the first is a paradigmatic correlated metal, while the second is the only nickelate that doesn't undergo a metal-insulator transition at low temperature. 
	At the quasiparticle (QP) level, correlation is responsible for the reduction of the bandwidth close to the Fermi surface, with respect to, \textit{e.g.}, Kohn-Sham DFT states. The ratio between the true and the mean-field bandwidth is a measure of the effective mass $m^*$, which is of fundamental importance for physical understanding and technological applications. 
	This quantity is usually slightly overestimated in DMFT \cite{PhysRevLett.92.176403,PhysRevB.88.235110}, and considerably underestimated in GW
	\cite{GattiGuzzo, Tomczak_2012, Tomczak_review,PhysRevB.88.235110}. The latter does not capture the importance of localized physics \cite{PhysRevB.87.115110}, which is often recovered, at least conceptually, by the inclusion of vertex corrections. The former, instead, does not include non-diagonal contributions in the self energy, which can play an important role. 
 Both effects are taken into account by GW+DMFT  \cite{PhysRevLett.90.086402} that yields very good agreement with experiments \cite{PhysRevB.88.235110}. 
	%
	Beyond QPs, ARPES experiments reveal additional features in the form of \textit{satellites} \cite{PhysRevLett.69.1796,PhysRevB.52.13711}. These are purely many-body structures rooted in the quantum correlation between electrons. By definition, they are missing in static approaches like DFT or DFT$+U$, but are found in both DMFT and GW. Their interpretation 
	as Hubbard bands \cite{PhysRevLett.76.4781} or plasmonic excitations \cite{GattiGuzzo,PhysRevB.87.115110} is still under debate \cite{PhysRevMaterials.1.043803,PhysRevB.94.201106}. 
	In fact, the physics of both SrVO$_3$ and LaNiO$_3$, which is ruled by dynamical and non-local interactions and is very well reproduced by the (computationally expensive) GW+DMFT, is already contained in the simpler but still expensive GW, which qualitatively reproduces most of the experimental findings for these metals.
	 Therefore, it is tempting to simplify, and at the same time model and refine the GW approximation \cite{Fiorentini,Hyber,Gygi,PhysRevLett.74.2323} when applied to localized states \cite{PhysRevB.87.115110}. This approach is inspired by Refs. \cite{Anisimov_1997, PhysRevB.82.045108}, where it has been shown that a localized version of COHSEX, the static version of GW, can be identified with the simpler DFT$+U$. However, as mentioned above, dynamical screening effects (and, second, non-local interactions) cannot be discarded in correlated metals \cite{PhysRevB.79.195110}, and even DFT+DMFT calculations should include a dynamic $U(\omega)$, interestingly related to the GW interaction \cite{PhysRevB.70.195104}, to account at the same time for both band renormalization and satellites 	\cite{Huang_2012,PhysRevB.85.035115}. In this work, we propose a form for the self energy that keeps the full frequency-dependence of GW as well as the simplicity and transparency of DFT$+U$ or DFT$+U$$+V$ \cite{Campo,PhysRevB.98.085127}. Due to their similarities, we have called the resulting approach DFT+$U(\omega)(+V)$.
	
	\section{Theory} \label{page:theory}
	The GW approximation for the exchange-correlation part of the self energy $\Sigma_{\rm xc}$
	can be regarded as the first term in a perturbative expansion in $G$ of the Hedin equation for $\Sigma$ \cite{Hedin1965,luciabook}. Both the Green's function $G$ and the screened interaction $W$ are built on top of the eigen-solutions of a mean field Hamiltonian $\hat h_0$, like the DFT one. The latter yields, for each single-particle eigenstate $\psi_{s}(\boldsymbol{x})\equiv\braket{\boldsymbol{x}|s}$ \footnote{The label $s$ stands, \textit{e.g.}, for $(n,\boldsymbol{k})$.}, an eigenenergy $\varepsilon_s$, solution of the equation $\hat h_0\ket{s}=\varepsilon_s\ket{s}$, and, at $T=0$, an occupation number $n_s=\theta(\mu_0-\varepsilon_s)$, with $\mu_0$ the Fermi energy. Regarding $W$, a RPA approximation is usually considered \cite{luciabook}: the bare Coulomb interaction $v$ is screened by neutral excitations (labelled by $\lambda$) of energy $\omega_{\lambda}=E_{\lambda}-E_0$ (with $E_{\lambda}$ eigenenergy of the many-body Hamiltonian) and strength $W_{\lambda}^{\rm p}$, obtained as interband transitions between occupied and empty states \cite{luciagw,PhysRevLett.109.126408,0953-8984-26-17-173202,Hedin_1999}. With the spectral decompositions for the single particle Green's function $G$ and for the polarization part of the screened interaction $W_{\rm p}=W-v$,
	\begin{gather}
		G\left(\boldsymbol{x},\boldsymbol{x}',\omega\right)=
		\sum_{s}\frac{\psi_{s}(\boldsymbol{x})\psi^*_{s}(\boldsymbol{x}')}{\omega-\varepsilon_{s}+i\eta\operatorname{sign}(\varepsilon_{s}-\mu)}
		\label{eq:Gsp}
		\\
		W_{\rm p}\left(\boldsymbol{x},\boldsymbol{x}',\omega\right)=
		\sum_{\lambda\neq0}
		\frac{2\omega_{\lambda}W_{\lambda}^{\rm p}\left(\boldsymbol{x},\boldsymbol{x}'\right)}
		{\omega^2-\left(\omega_{\lambda}-i\eta\right)^2},
		\label{eq:WPsp}
	\end{gather}
	(with $\eta$ a small positive number), 
	the GW self energy $\Sigma^{\rm GW}_{\rm xc}$, which is the convolution between $G$ and $W=v+W_{\rm p}$
	\begin{equation*}
		\Sigma^{\rm GW}_{\rm xc}\left(\boldsymbol{x},\boldsymbol{x}',\omega\right)=
		i\int\frac{d\omega'}{2\pi}G\left(\boldsymbol{x},\boldsymbol{x}',\omega+\omega'\right)W\left(\boldsymbol{x},\boldsymbol{x}',\omega'\right)e^{i\omega'\eta},
	\end{equation*}
	can be expressed analytically as:
	\begin{multline*}
		\Sigma^{\rm GW}_{\rm xc}\left(\boldsymbol{x},\boldsymbol{x}',\omega\right)=
		\sum_{s} \psi_{s}(\boldsymbol{x})\psi^*_{s}(\boldsymbol{x}')
		\Bigl\{-n_{s} v\left(\boldsymbol{x},\boldsymbol{x}'\right)+\Bigr.
		\\
		+\Bigl.
		\sum_{\lambda\neq0}
		\Bigl[
		\frac{n_{s}}{\omega-\varepsilon_{s}+\omega_{\lambda}-i\eta}+\frac{1-n_{s}}{\omega-\varepsilon_{s}-\omega_{\lambda}+i\eta}
		\Bigr]
		W_{\lambda}^{\rm p}\left(\boldsymbol{x},\boldsymbol{x}'\right)
		\Bigr\}.
	\end{multline*}
	The first line is the static Fock exchange contribution, that acts only on the occupied states for which $n_s\neq0$, while the second line represents the correlation part of the self energy, in which occupied and empty states are treated symmetrically. 
	
	\paragraph{The GW self-energy matrix elements}
	The matrix elements of the GW self energy between two generic states $\ket{\alpha}$ and $\ket{\beta}$ can be expressed in terms of the matrix elements of the $\lambda$ component of the polarization part of the Coulomb  interaction $W_{\lambda,\alpha\beta}^{{\rm p},ss}$ and of the bare interaction $v_{\alpha\beta}^{ss}$, both generically defined as 
	$f_{ad}^{bc}=\bra{ab}\hat{f}\ket{cd}=\int d^3xd^3x'\phi_a^*(\boldsymbol{x})\phi_c(\boldsymbol{x})f\left(\boldsymbol{x},\boldsymbol{x}'\right)\phi_b^*(\boldsymbol{x}')\phi_{d}(\boldsymbol{x}')$ \cite{luciabook}:
	\begin{multline*}
		\bra{\alpha}\hat \Sigma^{\rm GW}_{\rm xc}\left(\omega\right)\ket{\beta}=
		\sum_s\Bigl\{-n_sv_{\alpha\beta}^{ss}+\Bigr.
		\\
		+\Bigl.
		\sum_{\lambda\neq0}\left[
		\frac{n_s}{\omega-\varepsilon_s+\omega_{\lambda}-i\eta}+\frac{1-n_s}{\omega-\varepsilon_s-\omega_{\lambda}+i\eta}
		\right]
		W_{\lambda,\alpha\beta}^{{\rm p},ss}
		\Bigr\}.
	\end{multline*}
	Gathering together the $n_s$ terms reconstructs the matrix element of the fully screened interaction $W$ 
	plus a correction:
	\begin{multline}
		\bra{\alpha}\hat \Sigma^{\rm GW}_{\rm xc}\left(\omega\right)\ket{\beta}=
		\sum_s\Bigl\{-n_sW_{\alpha\beta}^{ss}\left(\omega-\varepsilon_s\right)
		+\Bigr.
		\\
		+\Bigl.
		\sum_{\lambda\neq0}\frac{W_{\lambda,\alpha\beta}^{{\rm p},ss}}{\omega-\varepsilon_s-\omega_{\lambda}+i\eta}
		\Bigr\}.
		\label{eq:SM1}
	\end{multline}

\paragraph{COHSEX:}
The COHSEX approximation \cite{Hedin1965,Hyber,PhysRevB.34.4405,Hedin_1999}, from which DFT$+U$ can be derived \cite{Anisimov_1997, PhysRevB.82.045108}, is a static approximation to GW, and it can be obtained from the latter when only $\omega=\varepsilon_s$ is retained in the frequency argument of $W$. Its matrix element reads:
\begin{equation*}
	\bra{\alpha}\hat \Sigma_{\rm xc}^{\rm COHSEX}\ket{\beta}=
	\sum_s\Bigl\{-n_sW_{\alpha\beta}^{ss}\left(0\right)
	+
	\sum_{\lambda\neq0}\frac{W_{\lambda,\alpha\beta}^{{\rm p},ss}}{-\omega_{\lambda}+i\eta}
	\Bigr\}.
\end{equation*}
The latter term involves a sum over all possible transitions. However, via the spectral representation of $W_{\rm p}$, Eq. \eqref{eq:WPsp}, it can be resummed to yield the simple quantity $\frac{1}{2}W_{\alpha\beta}^{{\rm p},ss}(0)$, so that the expression above assumes the expected scissor-like \cite{Fiorentini} form, that pushes up and down empty and occupied states respectively:	
\begin{equation}
	\bra{\alpha}\hat \Sigma_{\rm xc}^{\rm COHSEX}\ket{\beta}=
	\sum_s\Bigl\{\Bigl(\frac{1}{2}-n_s\Bigr)W_{\alpha\beta}^{ss}\left(0\right)
	-\frac{1}{2}
	v_{\alpha\beta}^{ss}
	\Bigr\},
	\label{eq:SM2}
\end{equation}	
particularly suited for the link to DFT$+U$ (see below). 

\paragraph{GW as a dynamical COHSEX:}
Motivated by the form of Eq. \eqref{eq:SM2}, it is useful to recast the GW matrix element in Eq. \eqref{eq:SM1}
as:
	\begin{multline}
	\bra{\alpha}\hat \Sigma_{\rm xc}^{\rm GW}\left(\omega\right)\ket{\beta}=
	\sum_s\Bigl\{\left(\frac{1}{2}-n_s\right)W_{\alpha\beta}^{ss}\left(\omega-\varepsilon_s\right)\Bigr.-\frac{1}{2}
	v_{\alpha\beta}^{ss}+
	\\
	\Bigl.+
	\sum_{\lambda\neq0}\frac{\left(\omega-\varepsilon_s\right)W_{\lambda,\alpha\beta}^{{\rm p},ss}}{\left(\omega-\varepsilon_s\right)^2-\left(\omega_{\lambda}-i\eta\right)^2}
	\Bigr\}.
	\label{eq:GWme}
	\end{multline}
	The terms in the first line constitute a natural dynamic generalization of COHSEX, Eq. \eqref{eq:SM2}, to which the whole expression reduces once the approximation $\omega=\varepsilon_s$ is considered. The last term, in fact, is a dynamical correction that goes to zero in that limit, but it is not obvious how to express it as a closed expression in the matrix elements of $\hat W$ and $\hat v$ only. However, in the spirit of having a simple final expression, by multiplying and dividing it by $\omega_{\lambda}$ and replacing the $\omega_{\lambda}$ at the denominator by a constant $\omega_0$, whose value we will discuss in the following,
	we can make use of the exact spectral decomposition of Eq. \eqref{eq:WPsp} to get:
	\begin{multline}
	\bra{\alpha}\hat \Sigma_{\rm xc}\left(\omega\right)\ket{\beta}=
	\sum_s\Bigl\{\Bigl[\frac{1}{2}\Bigl(1+\frac{\omega-\varepsilon_s}{\omega_{0}}\Bigr)-n_s\Bigr]\times 
	\Bigr.\\
	\Bigl.
	\times W_{\alpha\beta}^{ss}\left(\omega-\varepsilon_s\right)
	-\frac{1}{2}\Bigl(1+\frac{\omega-\varepsilon_s}{\omega_{0}}\Bigr)v_{\alpha\beta}^{ss}
	\Bigr\}.
	\label{eq:almost}
	\end{multline}
	The introduction of $\omega_0$, which plays an analogous role to the one introduced in \cite{PhysRevLett.109.126408}, is, together with the localization process described below, the main approximation of this work; however, through this, we achieve the goal of expressing the GW self energy matrix element in a form that is the straightforward dynamical generalization of Eq. \eqref{eq:SM2}.

	\paragraph{Localization in the correlated subspace:}
	We haven't specified so far the states that should undergo the action of $\Sigma$. In this work, in the same spirit of DMFT, we restrict ourselves to correlated metals, where it is possible to identify a certain manifold of \textit{correlated} states. Those are often non-dispersing low-energy bands, not fully occupied nor fully empty, with a strong $d$ or $f$ character, that are not quantitatively well-described within DFT. We call that subspace $\mathcal{C}$, and we describe it by a set of localized orbitals $\{\ket{I,m}\}$, where $I$ identifies the atom and $m$ the orbital quantum numbers \cite{Campo}, $\mathcal{C}=\cup_{I}\mathcal{C}_I$. For the purpose of the present derivation, it is not important if these are atomic orbitals, Wannier functions or other. However, if these were Wannier functions \cite{Marzari_review}, they could be built from a larger set of bands, $\mathcal{W}\supseteq\mathcal{C}$, to improve their localization. For instance, if $\mathcal{C}$ is a certain $d$ manifold, Wannier functions can be built from $\mathcal{C}$ itself (what is called the $d$ \textit{model}) or a larger manifold that contains all bands with which the $d$ states are mostly entangled (often $p$ bands, hence the name $dp$ model) \cite{Amadon_2008}. 
	
	Following Refs. \cite{Anisimov_1997, PhysRevB.82.045108}, each state $\ket{s}$ can be decomposed into a purely itinerant component $\ket{s^{\rm it}}$ and a localized contribution $\sum_{I,m\in\mathcal{C}}\ket{I,m}\braket{I,m|s}$. We assume that any matrix element of any Coulomb interaction involving an itinerant and a localized contribution be zero. Thus, the matrix element of the self energy between the states $\ket{I,m}$ and $\ket{J,m'}\in\mathcal{C}$ reads:
	\begin{multline}
		\bra{I,m}\hat \Sigma_{\rm xc}\left(\omega\right)\ket{J,m'}=
		\sum_{s,Km_1,Lm_2\in\mathcal{C}}\braket{K,m_1|s}\braket{s|L,m_2}\times
		\\
		\times\Bigl\{\Bigl[\frac{1}{2}\Bigl(1+\frac{\omega-\varepsilon_s}{\omega_{0}}\Bigr)-n_s\Bigr]W_{Im,Jm'}^{Lm_2,Km_1}\left(\omega-\varepsilon_s\right)+\Bigl.
		\\
		\Bigr.-\frac{1}{2}\Bigl(1+\frac{\omega-\varepsilon_s}{\omega_{0}}\Bigr)
		v_{Im,Jm'}^{Lm_2,Km_1}
		\Bigr\}.
		\label{eq:ohmygosh}
	\end{multline}
	
	Of this general formula we will consider the simplest situation in which only the diagonal elements of the interactions are retained \cite{Campo}, $W_{Im,Jm'}^{Lm_2,Km_1}(\omega)\approx\delta_{KI}\delta_{m_1m}\delta_{LJ}\delta_{m_2m'}W_{Im,Jm'}^{Jm',Im}(\omega)$, and similarly for $v$. 
	
	\begin{figure}[t!]
		\includegraphics[width=1.0\columnwidth]{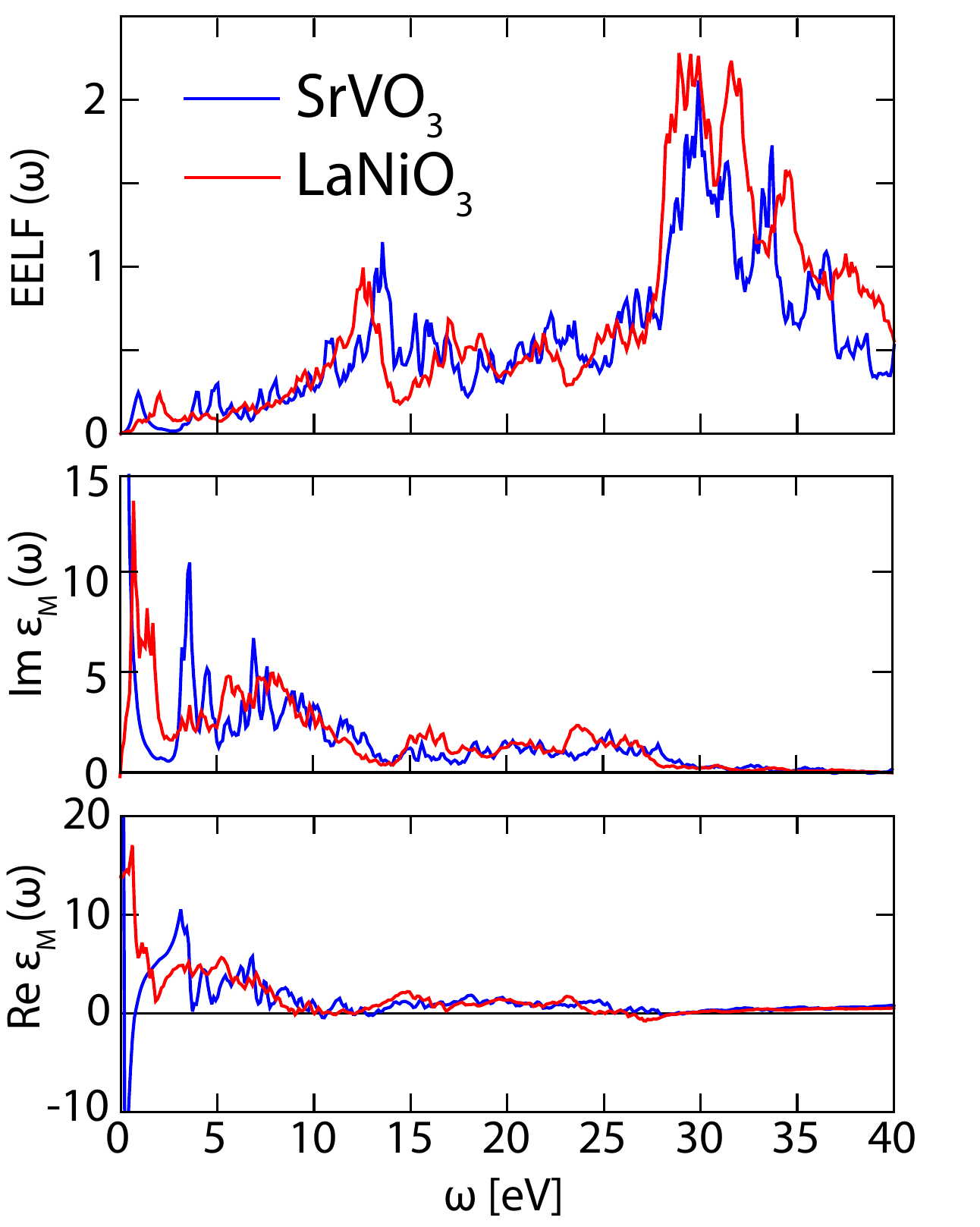}
		\caption{(Color online) Electron energy loss function $-{\rm Im}\epsilon_{\rm M}^{-1}(\omega)$ (EELF), and imaginary and real parts of the macroscopic dielectric function $\epsilon_{\rm M}(\omega)$, in the RPA approximation, as a function of frequency, for both SrVO$_3$ (blue) and LaNiO$_3$ (red). Electron-hole transitions as well as plasmons are responsible for the structure of $-{\rm Im}\epsilon_{\rm M}^{-1}(\omega)$ \cite{Ehrenreich}.}
		\label{fig:eps}
	\end{figure}
	
\begin{figure}[t!]
		\includegraphics[width=1.0\columnwidth]{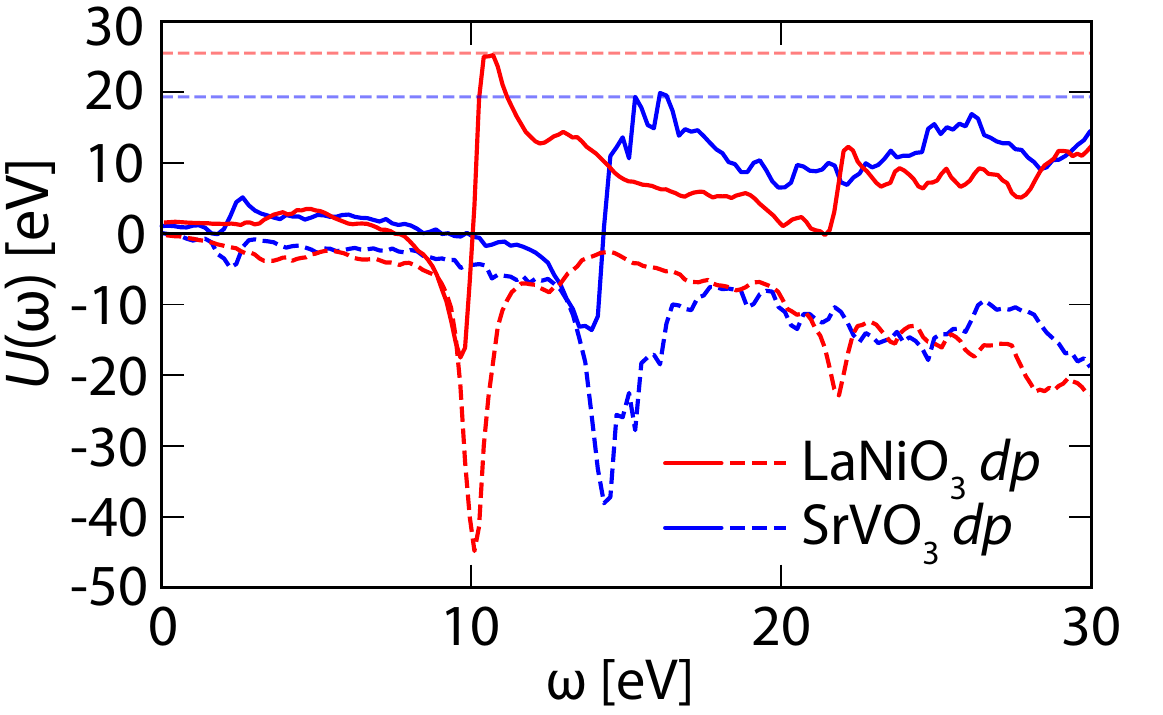}
		\caption{(Color online) $U(\omega)$ in SrVO$_3$ (blue), and in LaNiO$_3$ (red); the real part is continuous and the imaginary part is dashed. The Wannier functions have been built from the $d$ and $p$ states. The two horizontal lines identify the static value of the bare Coulomb interaction $U_{\infty}$, 19.16 and 25.51 eV respectively.}
		\label{fig:Uom}
	\end{figure}

\begin{figure}[t!]
		\includegraphics[width=1.0\columnwidth]{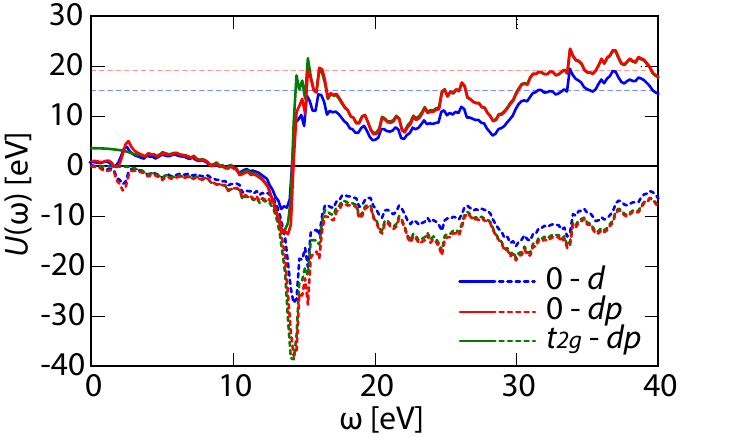}
		\caption{(Color online) Real (continuous) and imaginary (dashed) parts of $U(\omega)$ for SrVO$_3$, built from different sets of Wannier functions: the $d$ states only (blue) and the $d$ and $p$ states (red). The two horizontal lines identify the static value of the bare Coulomb interaction $U_{\infty}$ in the $d$ and $dp$ model, 15.10 eV and 19.16 eV, respectively. For comparison, we also show the cRPA result for Wannier functions in the $dp$ model with the $t_{2g}-t_{2g}$ transitions removed (green lines).}
		\label{fig:SrVO3_Uom}
	\end{figure}

	\paragraph{On-site contributions:}
	To further simplify \cite{Amadon_2014}, we define $U^I(\omega)$ as the average of those matrix elements on the site $I$ of the correlated atom $I$ (\textit{e.g.}, vanadium or nickel):
	\begin{equation}
		U^I\left(\omega\right):=\frac{1}{N_I^2}\sum_{mm'\in\mathcal{C}_I}W_{Im,Im'}^{Im',Im}\left(\omega\right), 
		\label{eq:defU}
	\end{equation}
	and analogously for $U^I_{\infty}$, defined as the average of the matrix elements of the bare Coulomb interaction $v$, with $N_I$ the total number of $m$ states for the atom $I$, see Fig. \ref{fig:Uom}. 
	This definition is very similar to the one usually adopted in cRPA \cite{PhysRevB.70.195104}, the difference stemming from the fact that here we do not discard any excitation channel, but the full RPA $\epsilon^{-1}(\omega)$ is retained (see Fig. \ref{fig:eps} and Fig. \ref{fig:SrVO3_Uom}). 
	With the introduction of $U^I(\omega)$ in place of the tensor $W_{Im,Im'}^{Im',Im}(\omega)$, we greatly simplify the site and orbital dependency of the on-site $J=I$ self energy. In fact, introducing the matrix elements of the mean-field density matrix and Hamiltonian,
	\begin{gather*}
		\sum_{s}n_s\ket{s}\bra{s}=\hat{\gamma}_{0};\qquad \sum_{s}\varepsilon_s\ket{s}\bra{s}=\hat{h}_{0},
	\end{gather*}
	in the localized basis:
	\begin{equation}
	\begin{aligned}
		n^{IJ}_{mm'}=\bra{I,m}\hat{\gamma}_{0}\ket{J,m'}=\sum_{s}n_s\braket{I,m|s}\braket{s|J,m'}
		\\
		\varepsilon_{mm'}^{IJ}=\bra{I,m}\hat{h}_{0}\ket{J,m'}=\sum_{s}\varepsilon_s\braket{I,m|s}\braket{s|J,m'},
		\label{eq:nemmp}
	\end{aligned}
	\end{equation}
	the self energy Eq. \eqref{eq:ohmygosh} becomes (when $J=I$):
	\begin{multline}
		\bra{I,m}\hat \Sigma_{\rm xc}\left(\omega\right)\ket{I,m'}=
		\sum_{m''}\\
		\Biggl\{
		\frac{1}{2}\Bigl(\delta_{mm''}+\frac{\omega\delta_{mm''}-\varepsilon^{II}_{mm''}}{\omega_0}\Bigr)
		-n^{II}_{mm''}\Biggr\}\times
		\\
		\times\bra{I,m''}U^I(\omega-\hat{h}_{0})\ket{I,m'}+\\
		-\frac{1}{2}\Bigl(\delta_{mm'}+\frac{\omega\delta_{mm'}-\varepsilon_{mm'}^{II}}{\omega_0}\Bigr)U^I_{\infty},
	\end{multline}
	where $U^I(\omega-\hat{h}_{0})$ should be interpreted as a power expansion series, and the validity of the identity ${\bf 1}_{\mathcal{C}_I}=\sum_{m''}\ket{I,m''}\bra{I,m''}$ depends on the disentanglement of the $\mathcal{C}_I$ submanifold from the rest of the bands.
	In this situation, to a good approximation the off-diagonal elements of both matrices $n^{II}_{mm'}$ and $\varepsilon^{II}_{mm'}$ in $\mathcal{C}_I$ can be considered negligible with respect to the diagonal ones \cite{Tomczak_2012,PhysRevB.86.195136}, as we have verified for both SrVO$_3$ and LaNiO$_3$. As a consequence, the self-energy matrix $\bra{I,m}\hat \Sigma_{\rm xc}\left(\omega\right)\ket{I,m'}\equiv\Sigma^I_{{\rm xc}\,{mm'}}\left(\omega\right)$ becomes diagonal and reads:
	\begin{multline}
		\Sigma^I_{{\rm xc}\;m}\left(\omega\right)=
		\left[
		\frac{1}{2}\Bigl(1+\frac{\omega-\varepsilon^I_{m}}{\omega_0}\Bigr)
		-n^I_{m}\right]U^I(\omega-\varepsilon^I_{m})+\\
		-\frac{1}{2}\Bigl(1+\frac{\omega-\varepsilon^I_{m}}{\omega_0}\Bigr)U^I_{\infty},
		\label{eq:fullsigma}
	\end{multline}
	with the shorthands $n^I_{m}$ and $\varepsilon^I_{m}$ for $n^{II}_{mm}$ and $\varepsilon_{mm}^{II}$ respectively. Finally, the self-energy operator assumes the form of a projector operator onto the correlated subspace $\mathcal{C}$ only:
	\begin{equation}
		\hat\Sigma_{\rm xc}\left(\omega\right)=\sum_{I,m}
		\Sigma_{{\rm xc},m}^I\left(\omega\right)\ket{I,m}\bra{I,m}.
		\label{eq:Sigma_proj}
	\end{equation}

	\paragraph{Inclusion of inter-site terms:}
	To complete 
	our minimal description of correlated metals, off-site effects beyond the on-site $U^I(\omega)$ shall be taken into account, combining $\bra{I,m}\hat{\Sigma}(\omega)\ket{I,m}$ with the matrix element of the self energy between different but neighboring atoms $I$ and $J$, $\bra{I,m}\hat{\Sigma}\ket{J,m'}$, with $J\neq I$. In this case, the matrices $n^{IJ}_{mm'}$ and $\varepsilon_{mm'}^{IJ}$ connect states of different atoms (\textit{e.g.}, the transition-metal $d$ states and the ligand $p$ states), hence their physics lays intrinsically in the cross terms (in general, at fixed $I$ and $J$, these are not even square matrices in $m$ and $m'$). Although a dynamical approach is possible, the simplifications we have implemented above cannot hold anymore, and the resulting expression would not be transparent. On the other hand, taking the \textit{static} COHSEX approximation $\omega=\varepsilon_s$ in Eq. \eqref{eq:ohmygosh} and keeping only the diagonal elements of the interactions results in:
	\begin{multline*}
		\bra{I,m}\hat \Sigma_{\rm xc}\ket{J,m'}=
		\\
		=\Bigl[\frac{1}{2}\delta_{IJ}\delta_{mm'}-n_{mm'}^{IJ}\Bigr]W_{Im,Jm'}^{Jm',Im}\left(0\right)-\frac{1}{2}\delta_{IJ}\delta_{mm'}
		v_{Im,Jm'}^{Jm',Im}.
	\end{multline*}
	The on-site $J=I$ terms, that would yield the static DFT$+U$, have been replaced by the frequency-dependent self energy of Eq. \eqref{eq:fullsigma}. The $J\neq I$ terms, instead, contribute as:
	\begin{equation*}
		\bra{I,m}\hat \Sigma_{\rm xc}\ket{J,m'}\stackrel{I\neq J}{=}
		-n_{mm'}^{IJ}
		W_{Im,Jm'}^{Jm',Im}\left(0\right).
	\end{equation*}
	Analogously to $U^I(\omega)$, we average the matrix $W_{Im,Jm'}^{Jm',Im}(0)$, defining the off-site Hubbard interactions:
	\begin{equation}  V^{IJ}:=\frac{1}{N_IN_J}\sum_{mm'}W_{Im,Jm'}^{Jm',Im}\left(\omega=0\right),
		\label{eq:V}
\end{equation}
 which represents the strength of the fully screened Coulomb interaction between atoms $I$ and $J$ at zero frequency. Finally, this contribution to the self-energy correction reads:
	\begin{equation}
		\hat v_V:=-\sum_{I\neq J}V^{IJ}\sum_{mm'}n_{mm'}^{IJ}\ket{I,m}\bra{J,m'}
		\label{eq:VV}
	\end{equation}
	in agreement with the off-site generalization of DFT$+U$ proposed in \cite{Campo}, but with a natural definition of $V$ as a screened many-body quantity.
	With this additional term, that gathers the most important static but inter-site contributions discarded in the derivation of DFT+$U(\omega)$, we aim to cover the same non-local physics \cite{RevModPhys.90.025003,PhysRevB.85.045109} included via a screened exchange term in SEx-DDMFT \cite{PhysRevLett.113.266403,van_Roekeghem_2014} or the full GW+DMFT \cite{Tomczak_review}. Here, however, both the static inter-site as well as the dynamical on-site terms have the same GW origin (and both do contain non-local physics). Furthermore, we have exactly disentangled inter- from on-site contributions, and the fact that the former is static and the latter dynamic descends here from a simplicity argument: while frequency-dependence naturally arises in the on-site part, it is less straightforward in the inter-site contribution.
	
	\paragraph{The Hartree term:}
	To treat the whole electron-electron interactions on the same footing, also the mean-field Hartree contribution should be expressed in terms of the localized orbitals \cite{Anisimov_1997, PhysRevB.82.045108}. In this representation, the density reads $n_{\rm loc}(\boldsymbol{x}):=\sum_{Km_1,Lm_2}\psi_{K,m_1}(\boldsymbol{x})n^{KL}_{m_1m_2}\psi^*_{L,m_2}(\boldsymbol{x})$, and the matrix element of the Hartree operator is:
	\begin{multline*}
		\bra{I,m}\hat v_{\rm H}^{\rm loc}\ket{J,m'}
		=\int d^3x\, \psi^*_{Im}(\boldsymbol{x})v_{\rm H}^{\rm loc}(\boldsymbol{x})\psi_{Jm'}(\boldsymbol{x})=
		\\
		=\int d^3xd^3x'\, \psi^*_{Im}(\boldsymbol{x})n_{\rm loc}(\boldsymbol{x}')v(\boldsymbol{x},\boldsymbol{x}')\psi_{Jm'}(\boldsymbol{x})=
		\\
		=\sum_{Km_1,Lm_2}n^{KL}_{m_1m_2}v_{Im,Km_1}^{Lm_2,Jm'},
	\end{multline*}
	which becomes the simpler $\bra{I,m}\hat v^{\rm H}_{\rm loc}\ket{J,m'}
	=\delta_{IJ}\delta_{mm'}\bigl\{n^IU^I_{\infty}+\sum_{K\neq I}n^KV^{IK}_{\infty}\bigr\}$ within the same approximations as above, with $n^I=\sum_{m\in\mathcal{C}_I}n^{II}_{mm}$ the total occupancy of the correlated manifold $\mathcal{C}_I$. As expected, the Hartree term describes an electrostatic physics ($\delta_{IJ}$), depending on both the total charge on the same site $I$ ($n^I$) and on different sites $K\neq I$ ($n^K$), that interact through the on-site ($U^I$) and off-site ($V^{IK}$) bare ($\infty$) interactions.
	
	\paragraph{Self-interaction correction:}
	The diagonal matrix element of the full self energy  $\hat{\Sigma}^I=\hat v_{\rm H}^{{\rm loc},I}+\hat \Sigma_{\rm xc}^I$ (Hartree \textit{and} exchange-correlation) acting on states in $\mathcal{C}_I$ of the atom $I$ can thus be written as:
	\begin{multline}
		\Sigma_m\left(\omega\right)=(n-n_m)U_{\infty}+\sum_{K\neq I}V^{IK}_{\infty}n^K+
		\\
		+\left[
		\frac{1}{2}\Bigl(1+\frac{\omega-\varepsilon_{m}}{\omega_0}\Bigr)
		-n_{m}\right]\Bigl(U(\omega-\varepsilon_{m})-U_{\infty}\Bigr),
		\label{eq:PSigma}
	\end{multline}
	where the superscript $I$ is implied everywhere. 
	In the infinite-frequency limit $\omega\to\infty$, when particles do not have time to polarize, the previous expression becomes static and real, and it reduces to $(n-n_m)U_{\infty}$ (neglecting the $V$-part, whose interpretation is as above and is clearly self-interaction free). Its orbital dependence removes the self-interaction error of a naive Hartree-like contribution $nU_{\infty}$: when the state $\ket{m}$ is empty, this term describes the action of $n$ particles interacting through the bare Coulomb interaction $U_{\infty}$. Conversely, if $\ket{m}$ is occupied, only $n-1$ particles do enter the self energy \cite{NicolaGW100}.
	Thus, as expected in the infinite-frequency limit \cite{Farid}, the full self energy tends to a self-interaction-free, orbital-dependent Hartree-Fock-like term:
	\begin{equation}
		\Sigma_m\left(\omega\right)\stackrel{\omega\to\infty}{\longrightarrow}v^{\rm Hx}_m\equiv(n-n_m)U_{\infty}+\sum_{K\neq I}V^{IK}_{\infty}n^K
		\label{eq:supmatstat}
	\end{equation} 
	This observation allows us to identify the second line of Eq. \eqref{eq:PSigma}, that is complex valued, frequency dependent, and non-zero for finite frequencies only, with the purely correlation part of the self energy. 
	
	\paragraph{Double counting:} 
	The self energy in Eq. \eqref{eq:PSigma} should be used perturbatively as a correction to the $\mathcal{C}$ part of the spectrum of $\hat h_0$.
	When considering the state $\ket{m}$, all electron-electron interaction effects are accounted for by $\Sigma_m\left(\omega\right)$, Eq. \eqref{eq:PSigma}, and Eq. \eqref{eq:VV} for the off-site contributions. However, these are also included in $\hat{h}_{0}$, although only at the mean-field level. The double-counting correction $\hat v_{\rm DC}$, that is static, local and orbital-independent, aims exactly at removing the latter. Discussing the different expressions for $\hat v_{\rm DC}$ proposed in the literature \cite{Anisimov_1991,KAROLAK201011,Ryee,PhysRevLett.115.196403,PhysRevB.67.153106} is beyond the scope of this paper; we mention here that for metals the around mean-field (AMF) form $v^{\rm AMF}_{\rm DC}=U(n-\bar n)$ \cite{Anisimov_1991}, with $\bar n=n/N$, should be preferred \cite{Deng2012} over the most popular fully localized limit (FLL) $v^{\rm FLL}_{\rm DC}=U(n-\frac{1}{2})$ \cite{AMF}. In a dynamical approach like the present one, one can naturally identify at least two static values for $U$: the fully-screened $U(0)$ and the bare $U_{\infty}$. The latter is the interaction strength between non-interacting electrons, and it is the one to be chosen when we remove mean-field terms. An intriguing way to define it is taking it as the orbital-average of the purely static (\textit{i.e.}, what survives in the limit $\omega\to\infty$), purely local (\textit{i.e.}, on-site) part of $\hat \Sigma^I+\hat{v}_V$, namely $\bigl<(n-n_m)U_{\infty}+\sum_{K\neq I}V^{IK}_{\infty}n^K\bigr>_m$. Therefore: 
	\begin{equation*}
		v_{\rm DC} \equiv v_{0}^{\rm Hxc}= U_{\infty}(n-\bar n)+\sum_{K\neq I}V^{IK}_{\infty}n^K
	\end{equation*}
	The first term is nothing but the around mean-field (AMF) form of the double-counting term \cite{Anisimov_1991}, best suited for metals, but with $U_{\infty}$, the bare interaction, in place of $U$, at variance with methods that employ a single value for $U$. The $V$ contribution is the same as the one that stems from the double-counting energy $E_{\rm DC}^V=\frac{1}{2}\sum_{I\neq J}V^{IJ}n^In^J$ proposed in \cite{Campo}, but with $V\to V_{\infty}$. The $U$ term is very similar to the one in Eq. \eqref{eq:supmatstat}. Their difference $U_{\infty}(\bar n-n_m)$ can be viewed as the purely correlation contribution $v_m^{\rm c}$ in the mean-field Hamiltonian, whose orbital-dependency exactly balances the one in $v^{\rm Hx}_m$ to yield an orbital-independent $v_{0}^{\rm Hxc}$. In other words, the previous formulation can be viewed as an additional source of dynamical correlations - namely the second line of Eq. \eqref{eq:PSigma} - on top of the static $v_m^{\rm c}=U_{\infty}(\bar n-n_m)$, already accounted for in $\hat h_{0}$. Finally, the on-site self-energy correction $\Delta\Sigma_m\left(\omega\right):=\Sigma_m\left(\omega\right)-v_{\rm DC}$ can be written as:
	\begin{multline}
		\Delta\Sigma_m\left(\omega\right)=(\bar n-n_m)U_{\infty}+
		\\
		+\left[
		\frac{1}{2}\Bigl(1+\frac{\omega-\varepsilon_{m}}{\omega_0}\Bigr)
		-n_{m}\right]\Bigl(U(\omega-\varepsilon_{m})-U_{\infty}\Bigr),
		\label{eq:PSigmaCorr}
	\end{multline}
	and it contains purely exchange and correlation contributions. This is the main result of this work.
	
	\paragraph{Physical content:}
	The advantage of having explicit and compact expressions is, beyond simplicity, understanding. In the static limit of $\omega=\varepsilon_s$ GW reduces to COHSEX, which leads to DFT$+U$ within the above procedure \cite{PhysRevB.82.045108}. In the same limit $\Delta\Sigma_m\left(\omega\right)$, Eq. \eqref{eq:PSigmaCorr}, goes to $\bigl(\frac{1}{2}-n_m\bigr)U(0)-\bigl(\frac{1}{2}-\bar n\bigr)U_{\infty}$, which has the same orbital dependence of DFT$+U$ in the AMF formulation, $v_{m}^{\rm AMF}=\bigl(\frac{1}{2}-n_m\bigr)U-\bigl(\frac{1}{2}-\bar n\bigr)U$ \cite{luciabook}; there, a single value of $U$ averages  the two $U(0)$ and $U_{\infty}$. In its simplest formulation, the AMF DFT$+U$ functional can also be written as $v_{m}^{\rm AMF}=\bigl(\bar n-n_m\bigr)U$, namely the first term in Eq. \eqref{eq:PSigmaCorr}, with $U\to U_{\infty}$. Eq. \eqref{eq:PSigmaCorr} can thus be interpreted as the unscreened DFT$+U$ result, that removes the self-interaction error from $\hat h_0$, modified by correlation for finite values of $\omega$.	
	 In the limit of frequencies close to $\varepsilon_{m}$ a perturbative expansion of the QP equation $E^{\rm QP}_m=\varepsilon_m+\Delta\Sigma_m(E^{\rm QP}_m)$ can be performed,
	 resulting in $E_m^{\rm QP}=\varepsilon_m+Z\left[\left(\frac{1}{2}-n_m\right)U(0)-\left(\frac{1}{2}-\bar n\right)U_{\infty}\right]$, with the renormalization factor $Z=(1-\partial\Delta\Sigma/\partial\omega|_{\varepsilon_m})^{-1}\approx(1+(U_{\infty}-U(0))/2\omega_0)^{-1}$. As expected, frequency dependence counteracts the non-locality of DFT$+U$, by reducing the value of $U(0)$ and $U_{\infty}$: the more efficient the screening, namely the difference between $U_{\infty}$ and $U(0)$, the larger the reduction of $U$, with $\omega_0$ the typical energy over which screening takes place. 	
	In addition, the strong renormalization of the correlated low-energy bands is due to the linear dependence of the real part of the self energy in the vicinity of the Fermi level. Such a structure stems from the pre-factor $-\frac{\omega-\varepsilon_m}{2\omega_0}$, always negative as $\Re U(\omega)<U_{\infty}$ in the low-energy regime. This allows us to identify $\omega_0$ -- that sets the bandwidth reduction in a $U(\omega)$ approach --  with the subplasmon energy needed to reproduce the outcome of a GW result when using a plasmon-pole model in GW itself \cite{GattiGuzzo}. 
	With such an identification, we recover a parameter-free theory again.
	
	The simplicity of Eq. \eqref{eq:PSigmaCorr} and the resulting clarity in the physics should not hide the simplifications needed to obtain it, namely the introduction of $\omega_0$ in passing from Eq. \eqref{eq:GWme} to Eq. \eqref{eq:almost} and the drop of any coupling between $\mathcal{C}$ and all other states in the self energy. The most visible practical consequence is the lack of a strict time-ordered structure in the resulting self energy. In fact, for the exact G$_0$W$_0$ self energy, 
	${\rm Im}\bra{r}\hat\Sigma_{G_0W_0}(\omega)\ket{r'}=
	\sum_s
	\left[\theta(\omega-\varepsilon_s)-n_s\right]
	{\rm Im}W_{rr'}^{ss}(\omega-\varepsilon_s)$. Summing over all states $s$ results in the expected Fermi liquid behaviour ${\rm Im}\Sigma(\omega)\sim-{\rm sign}(\omega-\mu)(\omega-\mu)^2$ in the vicinity of $\mu$ \cite{PhysRev.118.1417}.
	With Eq. \eqref{eq:PSigmaCorr}, however, 
	${\rm Im}\Delta\Sigma_m\left(\omega\right)=
	\bigl[\frac{1}{2}\bigl(1+\frac{\omega-\varepsilon_{m}}{\omega_0}\bigr)-n_{m}\bigr]
	{\rm Im}U(\omega-\varepsilon_{m})$, a compact function of $\omega$. In a way, the sum of $\theta$ functions centered on $\varepsilon_s$ has reduced to the single line $\frac{1}{2}\bigl(1+\frac{\omega-\varepsilon_{m}}{\omega_0}\bigr)$, which still returns $1/2$ for $\omega=\varepsilon_m$ and, together with ${\rm Im}U(\omega)\sim-|\omega|$, yields a quadratic behaviour of the self energy, positive for $\omega<\varepsilon_m-2\omega_0(\frac{1}{2}-n_m)$ and negative for $\omega>\varepsilon_m$. This is a proper Fermi liquid time-ordered self energy (with a renormalized $\mu$), in a perfectly half-filled metal with $n_m=1/2$. However, such a solution, besides being artificial, wouldn't reproduce the asymmetrical treatment of occupied and empty states that we see, \textit{e.g.}, in the position of the two satellites of SrVO$_3$. It is the scissor-like action that Eq. \eqref{eq:PSigmaCorr} inherits from DFT$+U$ that restores this asymmetry \cite{Tomczak_review}, at the price of a non-time-ordered self-energy in the interval $[\varepsilon_m-2\omega_0(\frac{1}{2}-n_m),\varepsilon_m]$. This issue, practically outdone by a large enough value of the regularizing parameter $\eta$ in Eq. \eqref{eq:Gsp} and \eqref{eq:WPsp}, reflects the contrasting effects of metallicity (the frequency-dependence) and localization (the scissor action of  DFT$+U$) already present in GW, but enhanced and clarified by the simple self energy of Eq. \eqref{eq:PSigmaCorr}. 
	
	\begin{figure}[t!]
		\includegraphics[width=1.0\columnwidth]{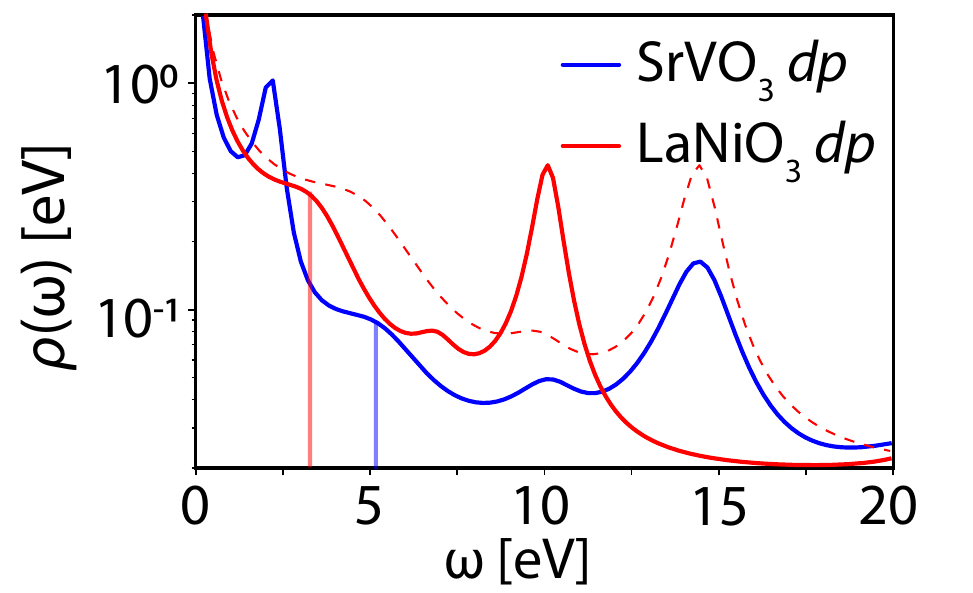}
		\caption{(Color online) Density of screening modes  $\rho(\omega)={\rm Im} U(\omega)/\omega^2$ in SrVO$_3$ (blue), and in LaNiO$_3$ (red). The vertical lines highlight the relevant screening mode $\omega_0$ we use in this work for each material, $\omega_0\approx 5$ eV and $\omega_0\approx 3$ eV respectively.}
		\label{fig:Uom_dens_modes}
	\end{figure}

\section{Application to metallic perovskites}
To test the approach presented here, we study the correlated bands of two paradigmatic perovskites,  SrVO$_3$ and LaNiO$_3$, that have already been extensively studied both experimentally and theoretically. 

\begin{figure}[t!]
	\includegraphics[width=1.0\columnwidth]{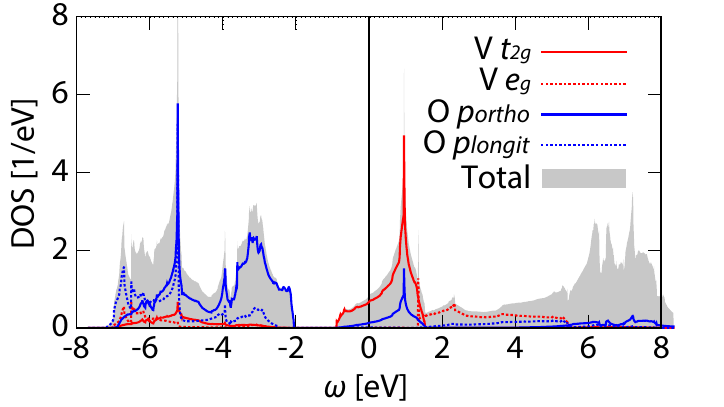}
	\caption{(Color online) PBE density of states of SrVO$_3$, in gray. The $m$-resolved projected DOS are shown for V, $t_{2g}$ and $e_{g}$ states, and O, orthogonal and longitudinal (with respect to the V $e_{g}$ states) orbitals, see text. The empty states at high energy are mainly Sr $4d$ states.}
	\label{fig:SrVO3_PBE_DOS}
\end{figure}
\begin{figure}[t!]
	\includegraphics[width=1.0\columnwidth]{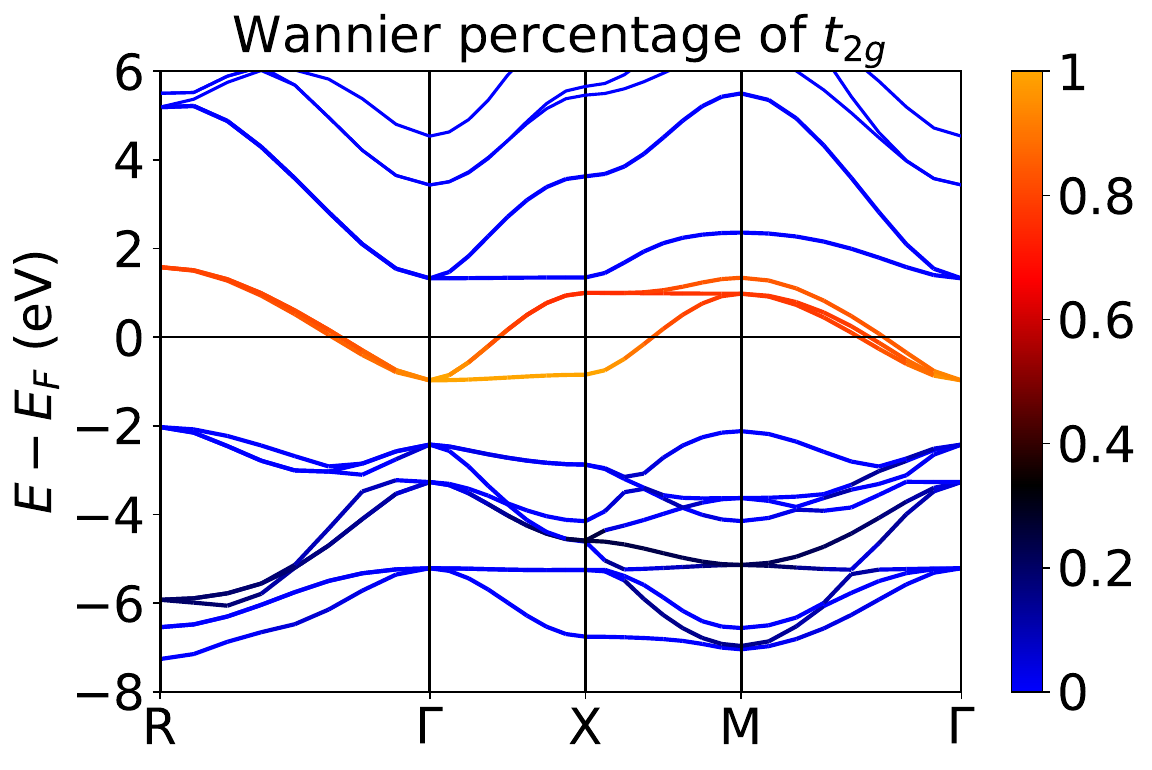}
	\caption{(Color online) PBE band structure of SrVO$_3$, highlighting the percentage of $t_{2g}$ character.}
	\label{fig:SrVO3_PBE_BS}
\end{figure}

\begin{figure}[t!]
	\includegraphics[width=1.0\columnwidth]{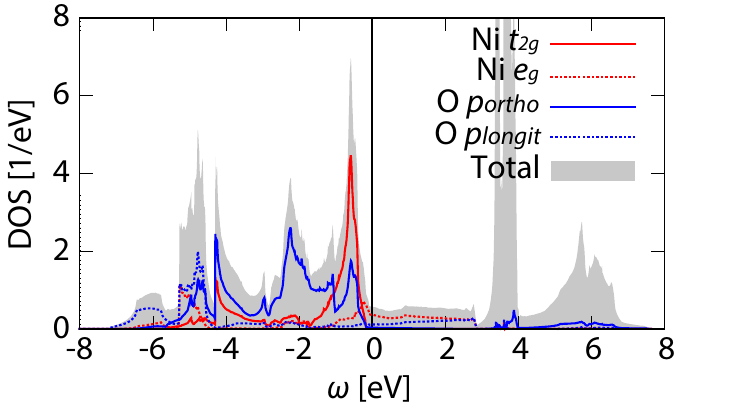}
	\caption{(Color online) PBE density of states of cubic LaNiO$_3$, in gray. The $m$-resolved projected DOS are shown for Ni and O. The empty states at high energy are mainly La $4f$ (the peak at $\sim$ 3.5 eV) and $5d$ states.}
	\label{fig:LaNiO3_cub_PBE_DOS}
\end{figure}
\begin{figure}[t!]
	\includegraphics[width=1.0\columnwidth]{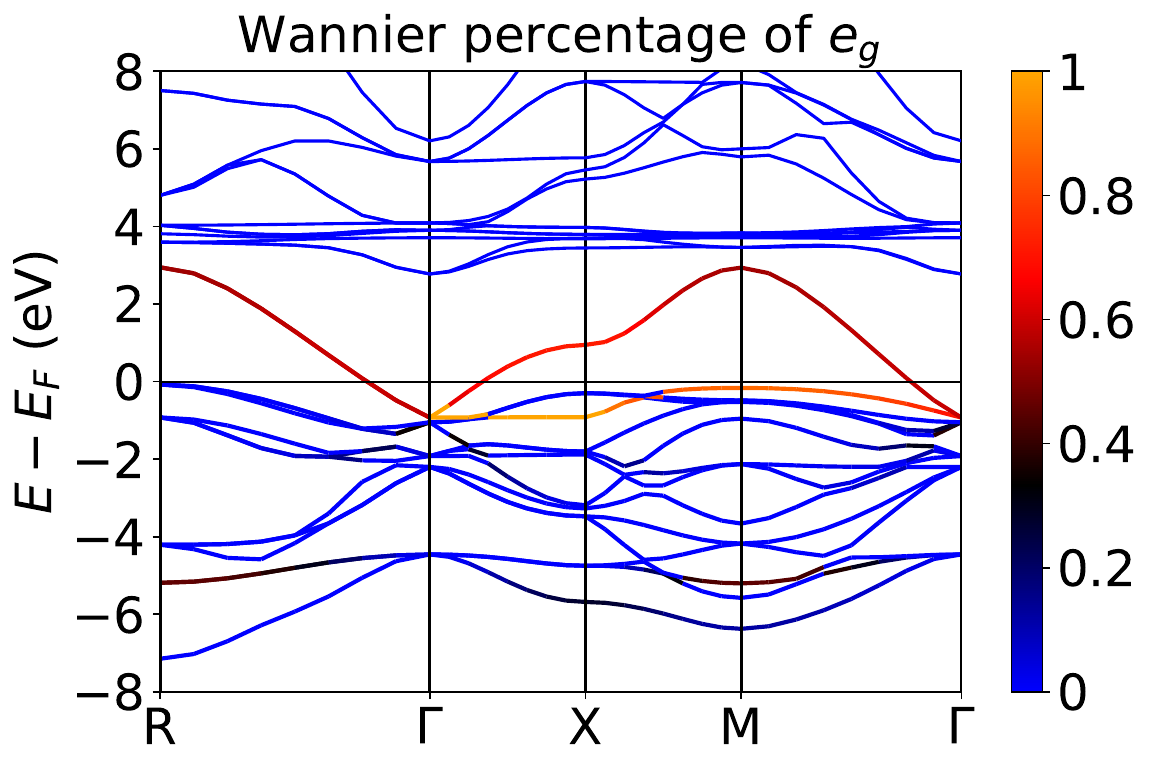}
	\caption{(Color online) PBE band structure of cubic LaNiO$_3$, highlighting the percentage of $e_g$ character.}
	\label{fig:LaNiO3_cub_PBE_BS}
\end{figure}

\subsection{Experimental findings}
SrVO$_3$ crystallizes in a cubic structure, with a paramagnetic metallic ground state. ARPES results from SrVO$_3$ show two main features around the Fermi energy: a dispersing band, whose width is $\sim$ 0.45 $\div$ 0.7 eV \cite{PhysRevB.80.235104,PhysRevLett.95.146404}, and an incoherent feature at $-2$ $\div$ $-1.5$ eV, with a weak dispersion of 0.1 eV but a dispersing intensity that finds its maximum at the $\Gamma$ point \cite{PhysRevB.80.235104, Tomczak_2012}. The importance of this satellite - interpreted as a lower Hubbard band - over the quasiparticle has been questioned in \cite{PhysRevB.94.241110}, where it has been shown that its intrinsic weight \cite{zhou2018dispersing,PhysRevB.92.245109} could be much less strong than previously thought, because of possible oxygen vacancies in the samples \cite{PhysRevB.94.201106} and surface contributions that add up in the incoherent part of the spectrum \cite{PhysRevLett.80.2885,PhysRevLett.93.156402,PhysRevLett.95.146404}. Also what was interpreted as an upper Hubbard band (see below), between 2.7 and 3.5 eV \cite{PhysRevB.52.13711}, lies in the region of the $e_g$ states, and would not be thus visible in IPES experiments \cite{Tomczak_2012,Tomczak_review}.

LaNiO$_3$, at variance with other rare-earth nickelates that exhibit metal-insulator transitions when cooled down to low temperatures, always displays a paramagnetic metallic ground state \cite{PhysRevB.45.8209}. Its crystal structure is rhombohedral $\rm R\bar 3 c$, but as a first approximation ($t = 0.97$ \cite{Gou2011}, $\beta=90.41^{\circ}$ \cite{Eguchi2009}) we will consider the high-temperature undistorted cubic structure $\rm Pm\bar3m$ of lattice parameter $a=7.2887a_0$, for simplicity but also to be as close as possible to the perfectly cubic compound SrVO$_3$.
Bulk LaNiO$_3$ was studied by PES in \cite{Horiba2007} and ARPES in \cite{Eguchi2009}, showing a momentum dependent mass renormalization at the Fermi energy. From thermal properties, it is expected $m^*/m\sim 10$, while from PES as well as from mean-field calculations in the cubic structure, it is around 3 \cite{Eguchi2009}. This value is very anisotropic and strongly depends on the path in the Brillouin zone. In fact, for the path shown in \cite{PhysRevB.92.245109} and using the cubic structure as a starting point, $m^*/m\sim 7$ \cite{KingNature} ($m^*/m=3.1\pm 0.5$ if the rhombohedral structure is employed).

\subsection{Theoretical state-of-the-art approaches}
For both systems, a non-magnetic Kohn-Sham solution within the PBE functional \cite{PhysRevLett.77.3865} yields a metallic ground state \cite{GUAN20102011,ALVES20192952,Sarma1995}, in agreement with experiments. 
In both perovskites the low-energy valence band is mainly constituted by oxygen $2p$ states hybridized with the 3$d$ states of vanadium and nickel respectively, see Fig. \ref{fig:SrVO3_PBE_DOS} and Fig. \ref{fig:LaNiO3_cub_PBE_DOS}. The crystal field splits both $p$ and $d$ states, gathering them into different groups depending on the reciprocal orientation of the orbitals: two $d$ orbitals ($e_g$) have lobes pointing towards the oxygens (one $p$ orbital per atom, which we call \textit{longitudinal}), and are thus more hybridized and dispersing. The other three $d$ ($t_{2g}$) and six $p$ (\textit{orthogonal}) orbitals are less overlapping, more localized and at lower energy.
Hybridization between the transition metal $3d$ states and oxygen ligands plays a fundamental role at the Fermi level. While the $p$ bands get completely filled, the nominal occupancy of the $d$ states is different for the two systems: $t_{2g}^1e_g^0$ in SrVO$_3$ and $t_{2g}^6e_g^1$ in LaNiO$_3$. As correlation mainly affects partially filled orbitals, it is larger for the $t_{2g}$ bands in SrVO$_3$ \cite{PhysRevB.53.7158,PhysRevResearch.2.013191} and for the $e_{g}$ bands in LaNiO$_3$ \cite{Horiba2007}. In fact, at the DFT level, these manifolds are not quite in agreement with the experimental results: the $t_{2g}$ bandwidth for SrVO$_3$ is 2.5 $\div$ 2.6 eV at the DFT level, twice as larger as measured; analogously for the $e_g$ band of LaNiO$_3$, which is too dispersing.

\begin{figure}[t!]
	\includegraphics[width=1.0\columnwidth]{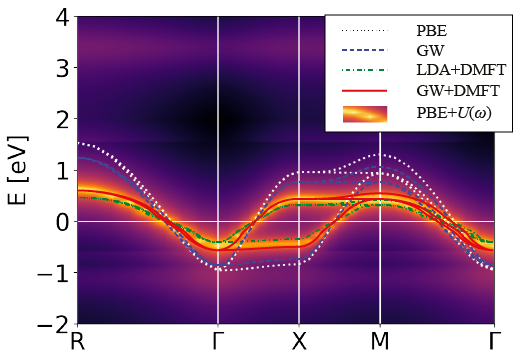}
	\caption{(Color online) The $t_{2g}$ manifold of SrVO$_3$ in DFT-PBE (white dotted) and PBE$+U(\omega)$ (color map). For comparison, we have also included results from \cite{PhysRevB.88.235110}, namely GW, DFT+DMFT and GW+DMFT.}
	\label{fig:end}
\end{figure}

For both systems, the low-energy physics requires a dynamical treatment of correlation beyond and alternative to hybrids or DFT$+U$ \cite{Janson2018,Gou2011,PhysRevB.86.235117}. SrVO$_3$, being the prototype of correlated metals, has been extensively studied by DFT+DMFT \cite{PhysRevLett.93.156402,PhysRevLett.92.176403,Karolak_2011,PhysRevB.73.155112,Amadon_2008,PhysRevB.94.241110} and GW+DMFT \cite{Tomczak_2012,Tomczak_review,PhysRevB.88.235110,PhysRevResearch.2.013191}. 
While a GW calculation doesn't reduce the band width enough (by a factor of 0.7-0.8 instead of 0.5, $m^*/m\sim1.3$, bandwidth of around 2 eV) \cite{GattiGuzzo, Tomczak_2012, Tomczak_review,PhysRevB.88.235110}, LDA+DMFT does the opposite, to 0.9 eV ($m^*/m\sim2.2$ \cite{PhysRevLett.92.176403}). The joint frequency-dependence and non-locality of GW+DMFT yields an acceptable value of 1.2 eV (0.5 eV in the occupied part) \cite{PhysRevB.88.235110}, as does a localized version of GW, 1.3 eV \cite{PhysRevB.87.115110}. Both GW and GW+DMFT place a high-energy electron-gas plasmon peak at $\sim$ 16 $\div$ 17 eV \cite{Tomczak_2012}, seen in the experimental EELS spectra of the related compound SrTiO$_3$ \cite{PhysRevB.62.7964}. Around the main $t_{2g}$ bands, GW shows a $t_{2g}$ excitation between 2 and 4 eV \cite{Tomczak_2012,GattiGuzzo,PhysRevB.87.115110}, though hidden by $e_g$ bands, and a lower satellite between $\sim -2$ and $-3$ eV \cite{Tomczak_review,GattiGuzzo,PhysRevB.87.115110}.
GW+DMFT displays a lower Hubbard band (LHB) at $-1.6$ eV \cite{Tomczak_2012}, at higher energy than LDA+DMFT with a static $U$, that places the Hubbard bands at $-1.8$ eV and $+3$ eV \cite{PhysRevLett.92.176403}. Using the value of $U=3.5$ eV from cRPA, the LHB is barely a shoulder within LDA+DMFT.
Finally, GW+EDMFT is shown to be not that different from $G_0W_0$ \cite{PhysRevB.94.201106}, pushing towards the interpretation of the upper and lower Hubbard bands as plasmonic features. The intensity modulation of the LHB shown in experiments is confirmed.

\begin{figure}[t!]
	\includegraphics[width=1.0\columnwidth]{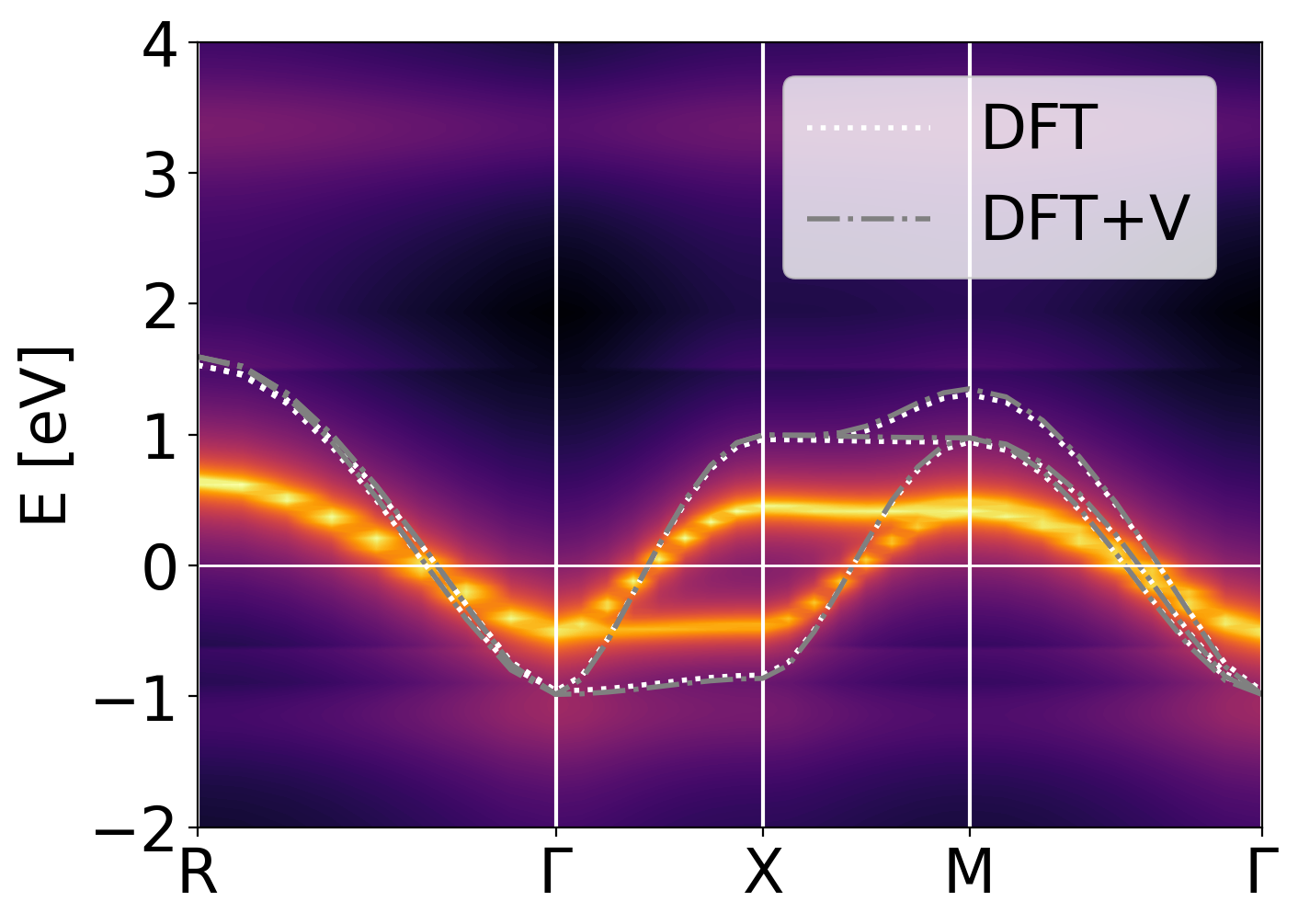}
	\caption{(Color online) The $t_{2g}$ manifold of SrVO$_3$ in PBE (white dotted), PBE$+V$ (gray dashed-dotted) and PBE$+V+U(\omega)$ (color map).}
	\label{fig:endV}
\end{figure}

In LaNiO$_3$, self-consistent GW yields $m^*/m\sim 1.3$ \cite{PhysRevB.90.035127}, in a good direction but definitely not large enough. On the contrary, DMFT \cite{Deng2012,PhysRevB.92.245109} reproduces well the mass enhancement, $m^*/m\sim 3$, and displays kinks in the $e_g$ bands at $\sim -0.2$ eV. These kinks stem from the onset of non-Fermi liquid behaviour in the self energy, which is no more linear (parabolic) in its real (imaginary) part. 

\subsection{DFT$+U(\omega)$ workflow}
In both perovskites we can obtain both a set of Wannier functions and the screened interaction from the PBE mean field solution $\{\varepsilon_{n\boldsymbol{k}},\psi_{n\boldsymbol{k}}(\boldsymbol{x})\}$, as explained above. We choose to build a set of maximally localized Wannier functions $\{\ket{I,m}\}$ \cite{Marzari_review,PhysRevB.77.085122} from the whole $p$ and $d$ states, due to their large hybridization, even if the self-energy correction is applied only to the partially filled submanifold of the $d$ bands only, the $t_{2g}$ states in SrVO$_3$ and the $e_{g}$ in LaNiO$_3$. 

\begin{figure}[t!]
	\includegraphics[width=1.0\columnwidth]{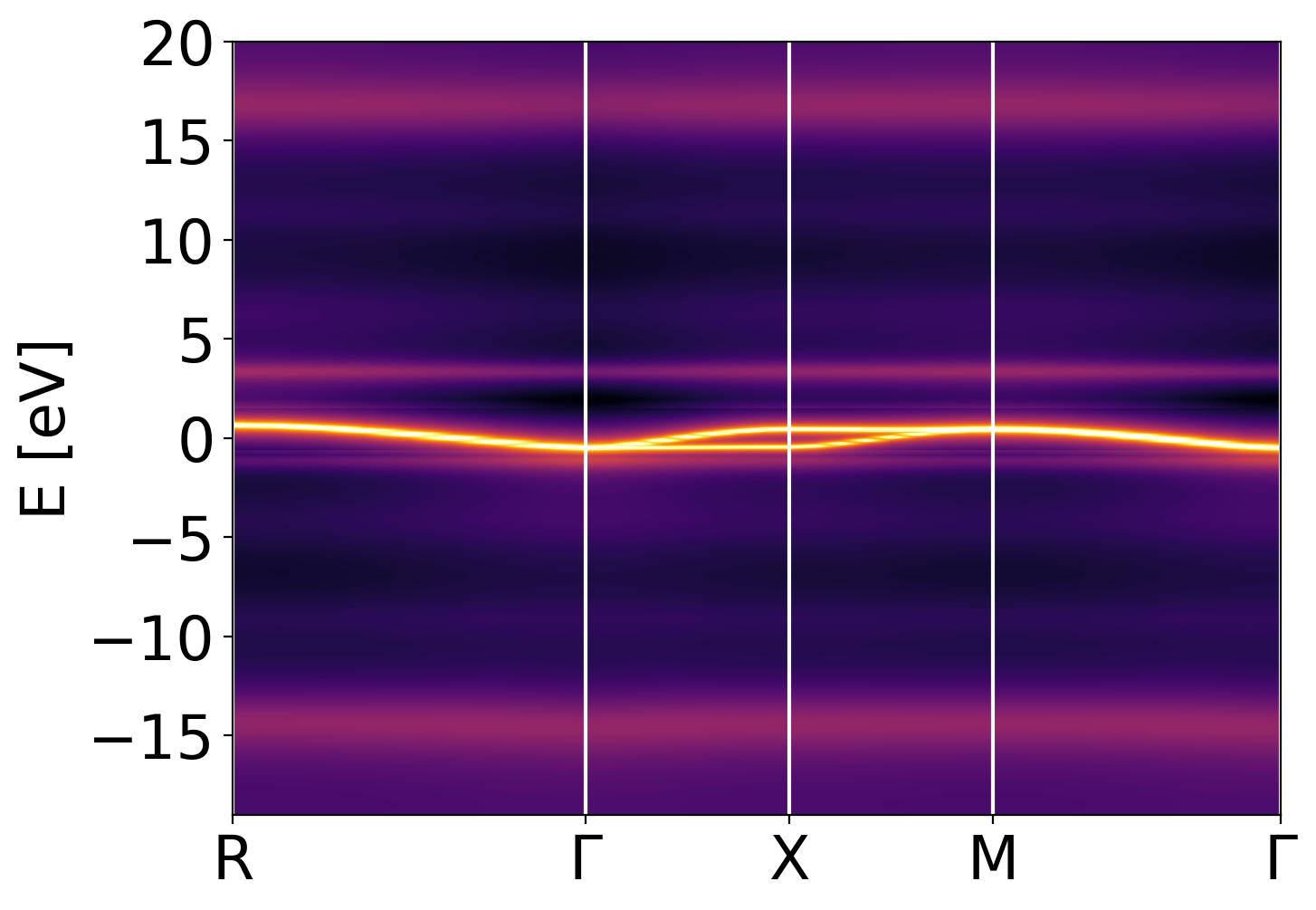}
	\caption{(Color online) The $t_{2g}$ manifold of SrVO$_3$ in PBE$+V+U(\omega)$ (color map), on a wider energy range.}
	\label{fig:endV2}
\end{figure}

The occupation and Hamiltonian matrices are built following Eq. \eqref{eq:nemmp}, and the RPA $U(\omega)$ from Eq. \eqref{eq:defU}, centered on V and Ni respectively. The very localized $4f$ states of La do not contribute much to screening, thus $U(\omega)$ is similar for the two systems, see Fig. \ref{fig:Uom}. As explained above, the parameter $\omega_0$ is a neutral excitation energy that sets the characteristic energy of screening. In SrVO$_3$, we consider the value $\omega_0=5$ eV, as this is the plasmon-pole energy \cite{Hyber} able to reproduce the full GW result \cite{GattiGuzzo}. It is also the most relevant excitation at low energies \cite{Tomczak_review} that can be inferred from the density of screening modes, $\rho(\omega)={\rm Im} U(\omega)/\omega^2$ \cite{PhysRevB.85.035115}, (the other one, at 2 eV, is responsible for the subplasmon satellites rather than band renormalization, see below); see also Fig. \ref{fig:Uom_dens_modes}. Analogously, we have performed plasmon-pole GW calculations for LaNiO$_3$ and compared with a full frequency calculation \cite{PhysRevB.90.035127}, obtaining $\omega_0=3$ eV. Again, this is the relevant screening mode at low energy, see Fig. \ref{fig:Uom_dens_modes}. It's interesting the two $\omega_0$s correspond to each other, once we stretch the LaNiO$_3$ frequency axis in such a way that the two main plasmon peaks superimpose (dashed red line in Fig. \ref{fig:Uom_dens_modes}). We stress that taking for $\omega_0$ the plasmon-pole energy of a GW calculation is a well-defined procedure that, although possibly time-consuming, keeps this approach parameter-free and fully ab-initio.

With these ingredients at hand, we can build the self-energy correction from Eq. \eqref{eq:PSigmaCorr}, use it perturbatively as a one-shot \cite{PhysRevB.94.201106} to correct the Kohn-Sham states:
\begin{equation*}
	E_{n\boldsymbol{k}} (\omega)= \varepsilon_{n\boldsymbol{k}}+\sum_{I,m}\left|\braket{I,m|n\boldsymbol{k}}\right|^2\Delta\Sigma_m(\omega),
\end{equation*}
and evaluate the spectral function: 
\begin{equation*}
	A_{n\boldsymbol{k}} (\omega)=-\frac{1}{\pi}\operatorname{sign}(\omega-\mu)\operatorname{Im}\frac{1}{\omega-E_{n\boldsymbol{k}} (\omega)},
\end{equation*}
with the new chemical potential $\mu$ set by conserving the Fermi wavevector, as suggested in \cite{Tomczak_review}. In fact, as in G$_0$W$_0$, the chemical potential is not conserved by the renormalization brought by the self energy. Another method to set the Fermi energy is by counting the electrons from the integrated DOS; for these systems, the two methods differ by $\sim0.1$ eV (see also \cite{PhysRevB.88.235110}).

\begingroup 
\squeezetable 
\begin{table}
		\begin{tabular}{ccccc}
			 & bw [eV] & $Z$ & LS  [eV] & US  [eV] \\ \hline
		   exp. \cite{PhysRevLett.93.156402} &  & &  $-1.6$ & \\
		   exp. \cite{PhysRevB.80.235104} & 0.44 & $\sim0.5$ & $-1.5$ & \\
		   exp. \cite{PhysRevLett.95.146404} & $\sim0.7$ & 0.55 & $-1.5$ \\ \hline
		   DFT-PBE & 0.958 & 1 & $\emptyset$ & $\emptyset$ \\
		   DFT+GW \cite{GattiGuzzo,Tomczak_review} & 0.8 & 0.77 & $-2$ & $2.2$ \\
		   DFT+DMFT \cite{PhysRevLett.93.156402} &&0.47&$-2$&$2.5$\\
		   DFT+GW+DMFT \cite{Tomczak_review} & 0.5 & 0.5 & $-1.6$ & 2\\
		   DFT+GW+DMFT \cite{PhysRevB.88.235110} & 0.6 & 0.6 & $-1.5$ &  2.5 \\
		   DFT+GW+EDMFT \cite{PhysRevB.94.201106} & & &$-1.7$& 2.8\\
		   \hline
		   DFT+$U(\omega)$ & 0.48 & 0.50& -1.10 & 3.42 \\
		   DFT+$V_{\rm LR}+U(\omega)$ & 0.60 & 0.62 & -1.34 & 3.18 \\
		   DFT+$V+U(\omega)$ & 0.49 & 0.51 & -1.15 & 3.37 \\
		\end{tabular}
\caption{Bandwidth of the occupied $t_{2g}$ manifold (bw), renormalization factor $Z$, lower (LS) and upper (US) satellites in SrVO$_3$. Experimental results, state-of-the-art theories and three different flavours of this work are compared.}
\label{Tab:1}
\end{table} 
\endgroup
	
	\subsection{SrVO$_3$}
	We have applied the protocol described above to the $t_{2g}$ manifold of SrVO$_3$. The main action of the $U(\omega)$ self energy on the PBE solution is to considerably shrink the coherent part of the manifold, from a DFT value of $2.49$ to $1.11$ eV in the overall $t_{2g}$ bandwidth. In particular, the bottom of the $t_{2g}$ manifold lies now at $-0.48$ eV (from a DFT value of $-0.96$ eV), in very good agreement with both experiments and GW+DMFT calculations, see Table \ref{Tab:1} and Fig. \ref{fig:end}. 
	This result clearly improves what can be obtained at the GW level, where the renormalization brought is only of 0.5 eV in the whole $t_{2g}$ manifold \cite{GattiGuzzo}. However, this self energy should not be considered (only) as an approximated GW, in the same way in which LDA+U is not (just) an approximated COHSEX. The additional ingredient is the localization procedure, that \textit{adds}, rather than removes, physics. In fact, we have added this piece of physics directly in the expression of GW, yielding a self energy that contains both effects. Along the same lines, Ref. \cite{PhysRevB.87.115110} has artificially selected the local part only of the GW self-energy, that yields indeed the experimental renormalization. Finally, from this point of view one would interpret also the DMFT correction to the GW solution as the addition of localized vertex diagrams that restore a greater locality in the physics of GW. 
	
	The  effective mass resulting from DFT+$U(\omega)$ is $m^*/m=2.00$ in the occupied part of the spectrum, corresponding to a renormalization factor $Z_{U(\omega)}=0.50$, in agreement with the value obtained from the derivative of the self energy and DFT+DMFT results \cite{PhysRevB.88.235110}. 
	A renormalization factor smaller than one depends on the loss of electronic charge, typical of non-conservative self energies like, \textit{e.g.}, G$_0$W$_0$, and the emergence of satellites. In fact, from Fig. \ref{fig:endV2} we can observe that part of the spectral weight is transferred to high energy structures. Among others, two, at $-14.6$ and $+16.7$ eV, slightly dispersing, stem from the large plasmon in $U(\omega)$ at 14 eV, in agreement with usual GW calculations \cite{Tomczak_review,Tomczak_2012} and EELS experiments on the isostructural material SrTiO$_3$ \cite{PhysRevB.62.7964}. 
	Closer to the Fermi energy, a non-dispersing lower satellite appears at $-1.09$ eV, at slightly too high energy with respect to experiments and DMFT results. The misplacement of this feature is most likely inherited from GW which, although usually very good in describing QPs \cite{Schilfgaarde2006}, has a tendency to miss the exact positions of satellites \cite{luciabook}. However, the dispersing intensity of the satellite, which is largest at $\Gamma$, is in agreement with experimental results and previous findings \cite{PhysRevB.82.085119,PhysRevB.80.235104, Tomczak_2012}. 	
	The intensity of this peak is extremely small with respect to DMFT calculations. However,  refined photoemission experiments have revealed how that feature could have been previously overestimated \cite{PhysRevLett.93.156402,PhysRevB.94.241110,PhysRevB.94.201106,PhysRevB.80.235104}. Finally, in the empty parts of the spectrum, there is a noticeable satellite at 3.41 eV, which will however be covered by $e_g$ bands \cite{Tomczak_review}. 
	
	It is easy to show that these two satellites stem from the lowest energy peak of $U(\omega)$, at $\sim 2$ eV, the one removed in cRPA. Moreover, although the position of the two satellites may be overestimated in energy, their relative distance, 4.5 eV, is the same as the one obtained in GW+EDMFT \cite{PhysRevB.94.201106} or the simpler DMFT \cite{PhysRevLett.93.156402}, which is again roughly the same that one obtains in GW \cite{GattiGuzzo}. These two observations push towards the interpretation of these satellites as sub-plasmons due to intraband excitations rather than Hubbard bands \cite{PhysRevB.94.201106}. Finally, note that DFT+$U(\omega)$ yields an asymmetric position, with respect to the quasiparticle band, for the two couples of satellites, the low- (at $-1.1$ and $3.4$ eV) and the high-energy ones (at $-14.6$ and $+16.7$ eV). This asymmetry, confirmed by other theories, is missed by GW.
	
	 \paragraph{Inclusion of Hubbard $V$:}
  An intersite Hubbard $V$ parameter can be calculated \textit{ab-initio} from linear-response theory $V=V_{\rm LR}$ \cite{Campo,PhysRevB.98.085127,PhysRevB.103.045141}. It results in a negligible value when considering nearest vanadium atoms. While the effects of $V_{\rm LR}^{\rm V-Sr}=0.11$ eV and $V_{\rm LR}^{\rm O-Sr}=0.37$ eV are tiny on the band structure, a relevant modification goes with the inclusion of $V_{\rm LR}^{\rm V-O}=1.69$ eV \footnote{The value of $V$ we have obtained has the same order of magnitude of other correlated metals \cite{Kulik,Campo,PhysRevB.85.045109}.}\footnote{For evaluating $V$, we have employed ortho-normalized atomic orbitals.}. This is the major expected inter-site effect, as vanadium is surrounded by six oxygen atoms, as pointed out in \cite{PhysRevB.79.033104}. 
	 The correlated $t_{2g}$ manifold widens from $2.49$ to $3.19$ eV, an effect mirrored by the non-local part of $GW$ \cite{Tomczak_review}.
	 The overall effect is a larger dispersion of the bands due to an enhancement of the V-O bonding. 
	 When $\hat{h}_{\rm DFT+V}$ is used as the starting mean-field Hamiltonian \cite{PhysRevB.85.045109}, the resulting renormalization is too weak, as the bottom of the $t_{2g}$ bands goes to $-0.6$ eV. We suppose that this drawback is due to using an energy-related parameter into a spectrum-related approach. More deeply, all parameters introduced are effective quantities for which a certain prescription for use is due. Mixed approaches might be powerful, but not consistent. In fact, the definition of $V$ in Eq. \eqref{eq:V} is through the RPA $\hat W(\omega=0)$ and, at the level of the more widespread $U$, it is well known that $U_{\rm RPA}<U_{\rm cRPA}<U_{\rm LR}$. In the case of SrVO$_3$, in fact, $U_{\rm RPA}=U(\omega=0)=1.04$ eV and $U_{\rm LR}=7.65$ eV (in passing, we note that all these values are not intrinsic nor universal, but depend on the localized orbital manifolds chosen \cite{PhysRevB.98.085127}; these large differences arise also from the fact that $V_{\rm LR}$ aims to correct self-interaction in the energy functional \cite{PhysRevLett.97.103001}, while $U_{\rm RPA}/U_{\rm cRPA}$ address spectral properties). To evaluate the $V$ of Eq. \eqref{eq:V}, we hence propose to use the simplified formula $V=\frac{U}{U_{\rm LR}}V_{\rm LR}$, where the linear response quantities are evaluated through the procedure of \cite{Campo,PhysRevB.98.085127,PhysRevB.103.045141}, $U$ is the average of the on-site matrix elements of $\hat W(\omega=0)$ and $V$ is the unknown corresponding off-site average. The underlying assumption is of course that $U$ and $V$ are proportionally related to each other in different theories. With the values written above, we get $V=0.23$ eV, a much smaller value of $V_{\rm LR}$; the change in the spectrum is minimum as seen in Fig. \ref{fig:endV}, with a tiny renormalization factor $Z_V=1.03$: inter-site effects are important but not fundamental for the $t_{2g}$ manifold, which explains the early successes of base DMFT in reproducing these features. 
  The bottom of the quasiparticle band now lies at $-0.49$ eV, and the full $t_{2g}$ bandwidth is 1.107 eV. This implies an effective mass $m^*/m=1.96$ corresponding to a renormalization factor $Z=0.51$ in the occupied part of the manifold, still in perfect agreement with state-of-the-art calculations \cite{PhysRevB.88.235110,Tomczak_review} and experimental findings \cite{PhysRevLett.93.156402,PhysRevLett.95.146404}. Note that this is also equal to the product $Z_V\times Z_{U(\omega)}=1.03\times0.50$ \cite{PhysRevB.87.115110}, which again highlights the separability, to a good approximation, of dynamical and non-local effects. On the other hand, it should be noted that the latter are already present in the $U(\omega)$ only self-energy that, at zero frequency, reduces to DFT+U and hence already contains much of the non-locality of COHSEX \cite{Tomczak_review}.	Therefore, it would be more correct to talk about dynamical on-site and static off-site interactions.
	 Finally, note that the satellites gain some spectral weight and their positions go to $-1.15$ and $3.36$ eV, in a good direction towards agreement with other theories.	 
	 
	 \begin{figure}[t!]
	 	\includegraphics[width=1.0\columnwidth]{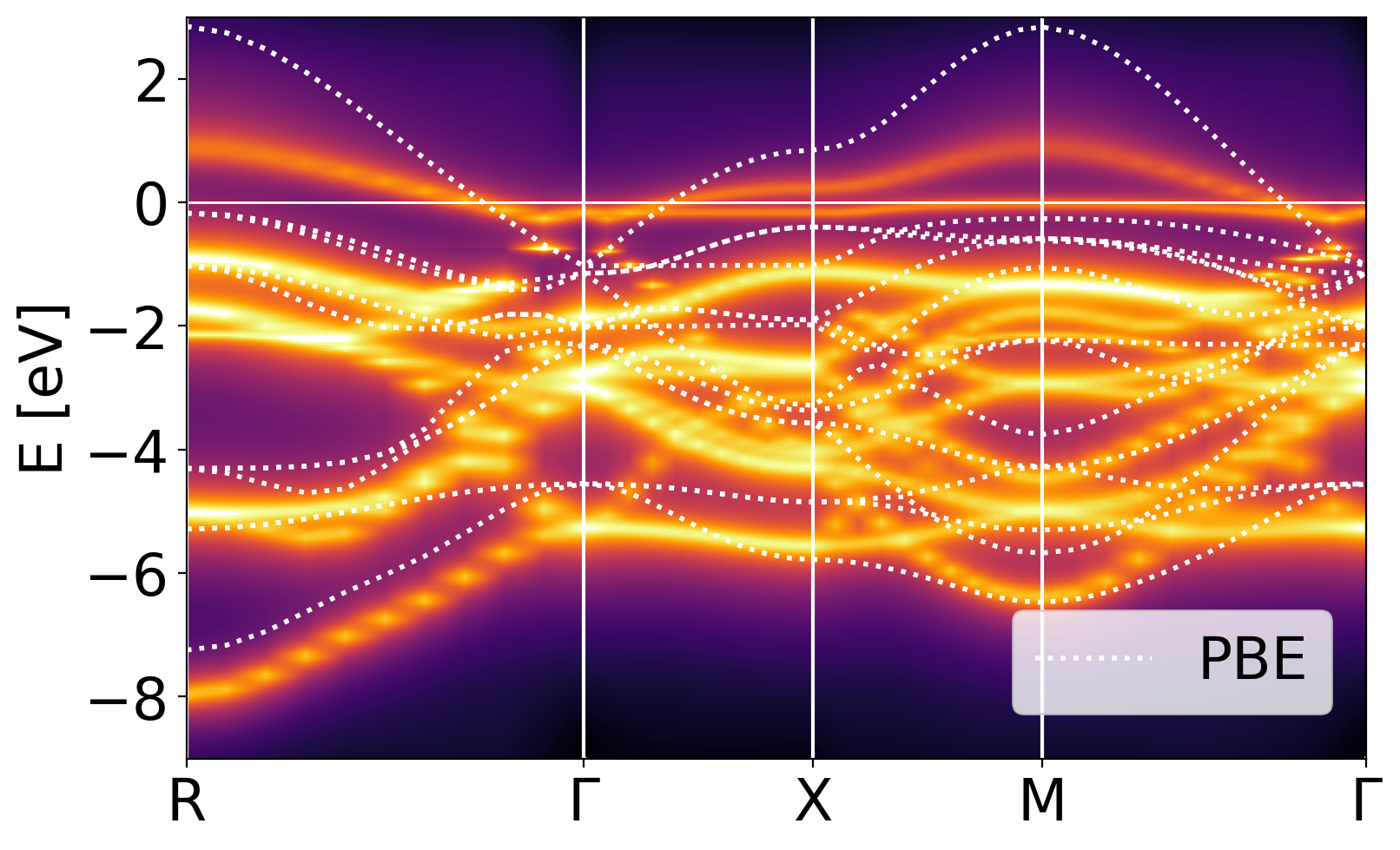}
	 	\includegraphics[width=1.0\columnwidth]{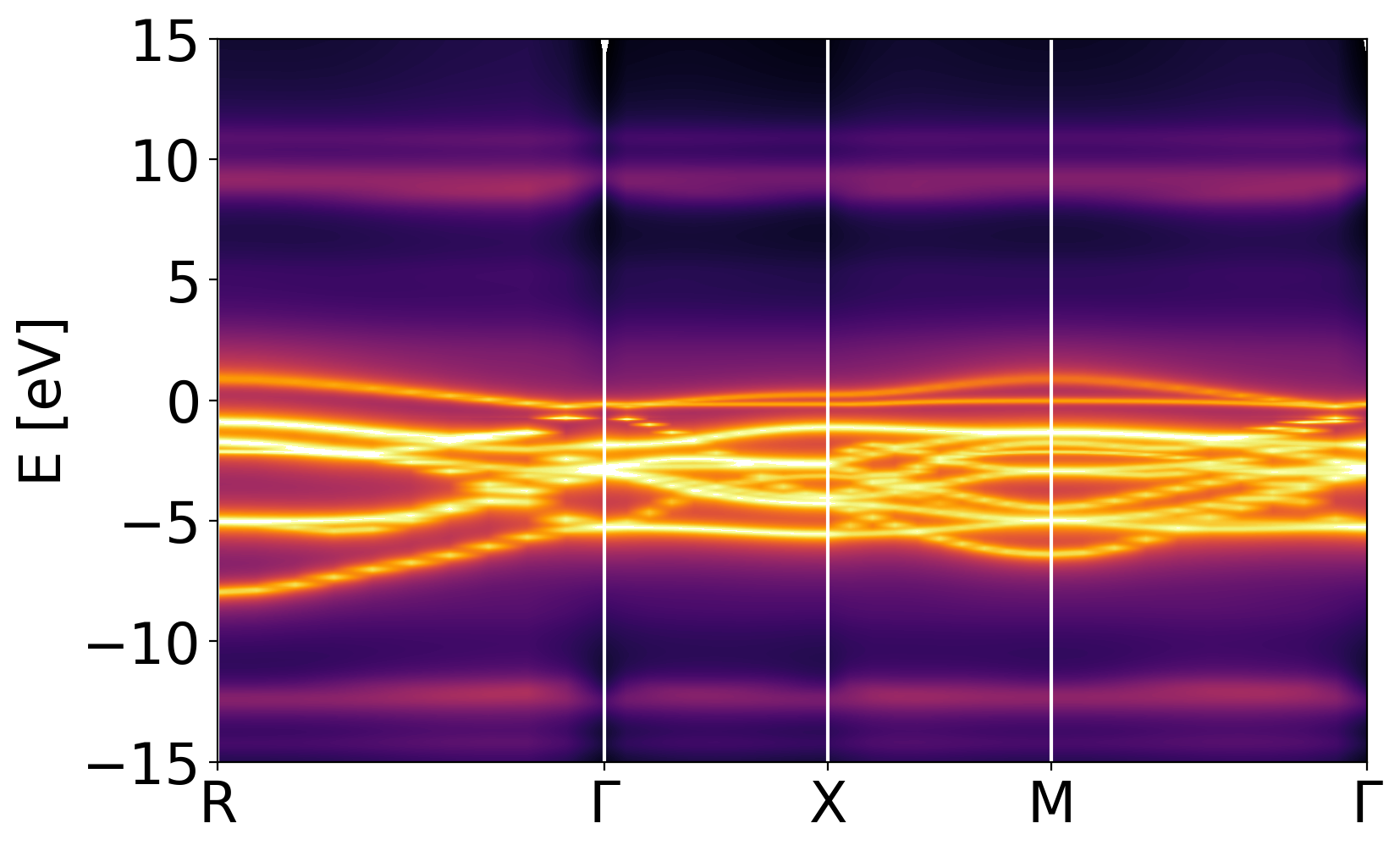}
	 	\caption{(Color online) Top panel, $p$ and $d$ bands of LaNiO$_3$ in DFT$+U(\omega)$ (color map), and DFT (white dotted). Bottom panel, same, on a wider energy range.}
	 	\label{fig:endLA}
	 \end{figure}

\subsection{LaNiO$_3$}
By contrast to SrVO$_3$ the $e_g$ bands, usually considered as the correlated manifold in LaNiO$_3$, are not well separated from the rest of the valence bands (see Fig. \ref{fig:LaNiO3_cub_PBE_BS}). The effects of the self energy, Eq. \eqref{eq:PSigmaCorr}, will therefore spread all over the valence manifold, according to the importance of the $e_g$ character of the different Kohn-Sham states. Therefore, also high-energy QPs are affected (see Fig. \ref{fig:endLA}).

We will focus on the low-energy region, where a precise knowledge of the effective mass is most needed and detailed ARPES measurements are available. As in the case of SrVO$_3$ and as discussed in the theory section, the main effect of the self energy of Eq. \eqref{eq:PSigmaCorr} is to increase the effective mass by weakening the dispersing character of the correlated bands or, equivalently, enhancing their localization. This is shown in particular for the two experimental paths we consider: the first \cite{PhysRevB.92.245109} is in the $k_y$ direction, with $k_x=\pi/2a$ and $k_z=0.7\pi/a$, with $a$ the pseudo-cubic lattice vectors obtained when considering a rhombohedral structure; the second path \cite{Eguchi2009} is in the $\Gamma\rm X$ direction. 

\begin{figure}[t!]
	\includegraphics[width=1.0\columnwidth]{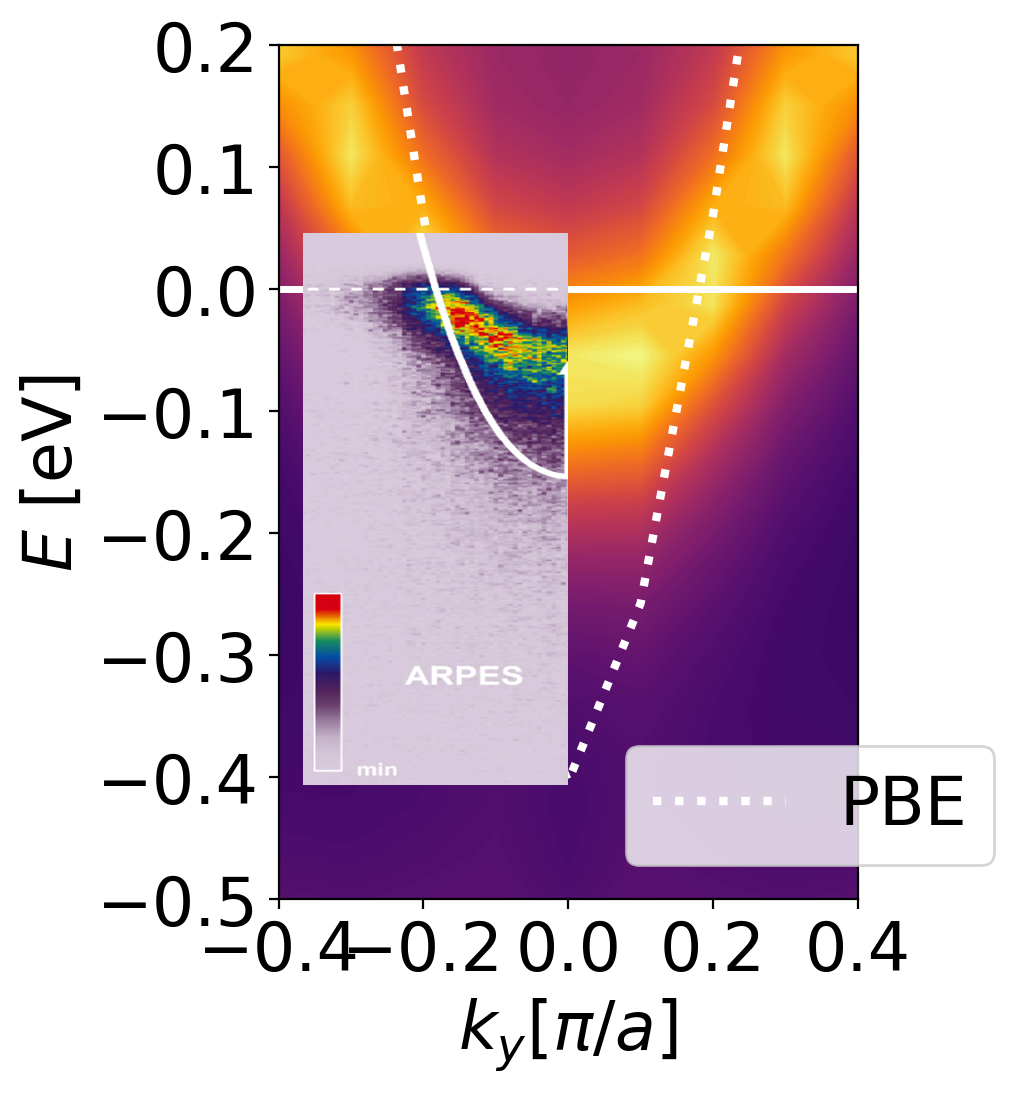}
	\caption{(Color online) The band structure of LaNiO$_3$ around the Fermi level in DFT-PBE (white dotted) and DFT$+U(\omega)$ (color map). In the inset, the experimental results from Ref. \cite{PhysRevB.92.245109} and, in white, the DFT result in the rhombohedral structure.}
	\label{fig:LaParab}
\end{figure}

\begin{figure}[t!]
	\includegraphics[width=1.0\columnwidth]{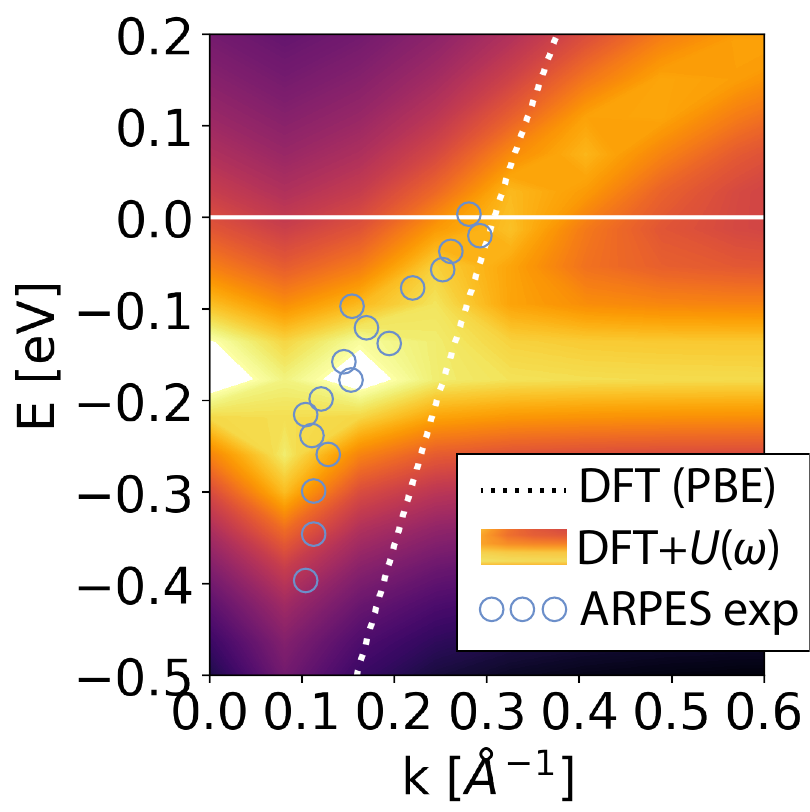}
	\caption{(Color online) The band structure of LaNiO$_3$ along the direction $\Gamma \rm X$ in DFT-PBE (white dotted) and DFT$+U(\omega)$ (color map). The cyan circles are the ARPES experiments from \cite{Eguchi2009} as reproduced in \cite{Deng2012}.}
	\label{fig:LaGX}
\end{figure}

For the latter case, the value of the renormalization can be derived from the different slopes of the DFT and DFT$+U(\omega)$ bands that cross the Fermi level along $\Gamma\rm X$. We obtain an effective mass $m^*/m=3.5$, corresponding to a renormalization $Z=0.3$. For the other path, we take for the renormalization  the ratio between the distance of the extrema of the parabolas from the Fermi level, and we get $Z=0.134$ and $m^*/m=7.5$. This $k$-dependent renormalization is confirmed by other approaches and, more importantly, by ARPES experiments \cite{Eguchi2009}, as can be seen from the inset of Fig. \ref{fig:LaParab} and the circles in Fig. \ref{fig:LaGX}. In particular, the bottom of the parabolic band has been measured to be 50 meV away from the Fermi level, and we get 54 meV with the present approach. Also the kink behavior of the experimental band in Fig. \ref{fig:LaGX} at $-2$ eV is somehow captured by this approach. 
The $Z=0.3$ renormalization around the $\Gamma$ point is confirmed by ARPES \cite{Eguchi2009} and DMFT results (on the same cubic structure we use \cite{Deng2012}), as well as the kink feature.
Analogously to GW results \cite{PhysRevB.90.035127}, the two upper bands of mostly $e_g$ character are now decoupled from the lower bands. However, due to the localization procedure, the renormalization is much stronger, and goes from an overall GW reduction of 1.2 eV \cite{PhysRevB.90.035127} to a DFT$+U(\omega)$ reduction of 2.7 eV. Also the downshift of the rest of the valence manifold is reproduced, from about half an eV in GW to 0.75 eV here.

In Fig. \ref{fig:endLA} we can note the lower intensity of the $e_g$ bands with respect to the others. In fact, a dynamical renormalization of these bands goes together with the transfer of electronic charge to incoherent features. However, by contrast to the previous perovskite, here there are no important satellites at low energy. That can be understood from an analysis of the self energy and, in particular, the different behaviors of $U(\omega)$ for the two systems (see Fig. \ref{fig:Uom}, and also the dielectric functions of Fig. \ref{fig:eps}). The high-energy features are instead similar for the two systems, that share the same average electronic density: for SrVO$_3$ the Wigner-Seitz radius is $r_s = 1.31$ and the plasma frequency is $\omega_{\rm P}= 31.56$ eV, while for LaNiO$_3$ $r_s = 1.25$ and $\omega_{\rm P}= 33.61$ eV; these translate in the same main loss peak at around 30 eV (Fig. \ref{fig:eps}). In the low energy regime, instead, the two systems do differ: in fact, in $\operatorname{Im}U(\omega)$, the distinct peak at $\sim 2$ eV of SrVO$_3$ doesn't have a clear counterpart in LaNiO$_3$. One could surmise that its true counterpart could be the shoulder at 3 eV; however, the latter should be rather matched with the shoulder at 5 eV for SrVO$_3$. This can be seen in different ways: first, these shoulders are responsible for the renormalization of the correlated bands; they are the plasmon-pole-model energies to be used in a GW calculation; they correspond to each other in the density of the screening modes (Fig. \ref{fig:Uom_dens_modes}), once the energy scales are stretched in such a way to have the two main plasmons in the same position. More deeply, they have the same physical origins as inter-band transitions, while the 2 eV peak of SrVO$_3$ can be considered as an intra-band Drude term (see, \textit{e.g.}, the divergence of the imaginary part of the macroscopic dielectric function in Fig. \ref{fig:eps}). The latter further screens electron-hole excitations, resulting in a smaller value of $U(\omega=0)=1.04$ eV, to be compared to $U(\omega=0)=1.59$ eV in LaNiO$_3$.
Therefore, the lack of a strong, intra-band excitation in LaNiO$_3$ seems to be the reason of the absence of a visible low energy plasmon.

On the high-energy side, instead, non-dispersing plasmons do show up, stemming from the main excitation in $U(\omega)$ at $\sim 10$ eV. As expected, due to the different structure of $U(\omega)$ in the two perovskites, plasmons are here closer to their quasiparticles, at $\sim -12$ and $+10$ eV respectively (see Fig. \ref{fig:endLA}, bottom panel).

\subsubsection{Inclusion of Hubbard V}
We can include intersite interactions via Eq. \eqref{eq:VV} also for LaNiO$_3$. As for the value of $V$ between nickel and oxygen, a linear-response calculation yields $V_{\rm LR}=1.19$ eV, together with $U_{\rm LR}=10.77$ eV. Including the intersite $V$ term results in a slightly weaker renormalization of the $e_g$ bands; in particular, the bottom of the parabola would now be at 83 meV. As in the case of SrVO$_3$, $V_{\rm LR}$ is not the one prescribed by Eq. \eqref{eq:V}; to get the latter in a simple way, we employ the proportionality relation $V=\frac{U}{U_{\rm LR}}V_{\rm LR}$, with $U=U(\omega=0)=1.5920$ eV, to obtain $V=0.176$ eV. With such a value inserted in $\hat h_{{\rm DFT}+V}$ as a starting point Hamiltonian, we obtain 64 meV for the vertex of the parabolic band, and no significant modification along $\Gamma\rm X$.

An exact quantitative agreement with experiments for LaNiO$_3$ is beyond the scope of this paper, as it would require to take into account at least the rhombohedral structure of the crystal. The already very good reproduction of experimental features is notable, as well as the momentum dependent renormalization of the $e_g$ bands, $m^*/m=3.5$ around $\Gamma$ and $m^*/m=6.3$ around the parabolic band at Fermi. In addition, the good match with the DMFT results of Ref. \cite{Deng2012} along $\Gamma \rm X$ highlights again the power of such streamlined approach.

\section{Conclusions}
Electronic correlations are known to play an important role in solids with partially filled $d$ or $f$ orbitals, where the atomic, localized physics competes with the dispersive nature of the solid-state bands. These effects are usually considered to be captured by strong vertex corrections to GW, leading to highly sophisticated approaches like GW+DMFT. However, for the cases studied, the same localization features can be included at the GW level itself, using localized basis functions and suppressing cross contributions with plane-wave-like terms. Moreover, in order to get a simple and transparent framework, we have proposed a self-energy expression, Eq. \eqref{eq:PSigmaCorr}, that contains both the plasmonic physics of GW and the scissor action of COHSEX, and can be thought as a dynamical generalization of DFT+U. The application of this self-energy to the correlated manifold of two metallic perovskites shows the power of the present approach, yielding results in extremely good agreement both with state-of-the-art theories and experiments. In particular, it can predict the renormalization of the band at the Fermi level without adjustable parameters; perhaps equally important, the simplicity of the self-energy allows a transparent understanding of the processes involved in these systems, as well as a lightweight implementation and fairly negligible computational costs, best suited for materials discovery \cite{Marzari2021}, material characterization and technological applications. 

	
\section{Acknowledgements}
The authors would like to thank David O' Regan, Tommaso Chiarotti and Mario Caserta for fruitful discussions.

	\appendix
	
	\section{Appendix}
	
		\begin{figure}[t!]
		\includegraphics[width=1.0\columnwidth]{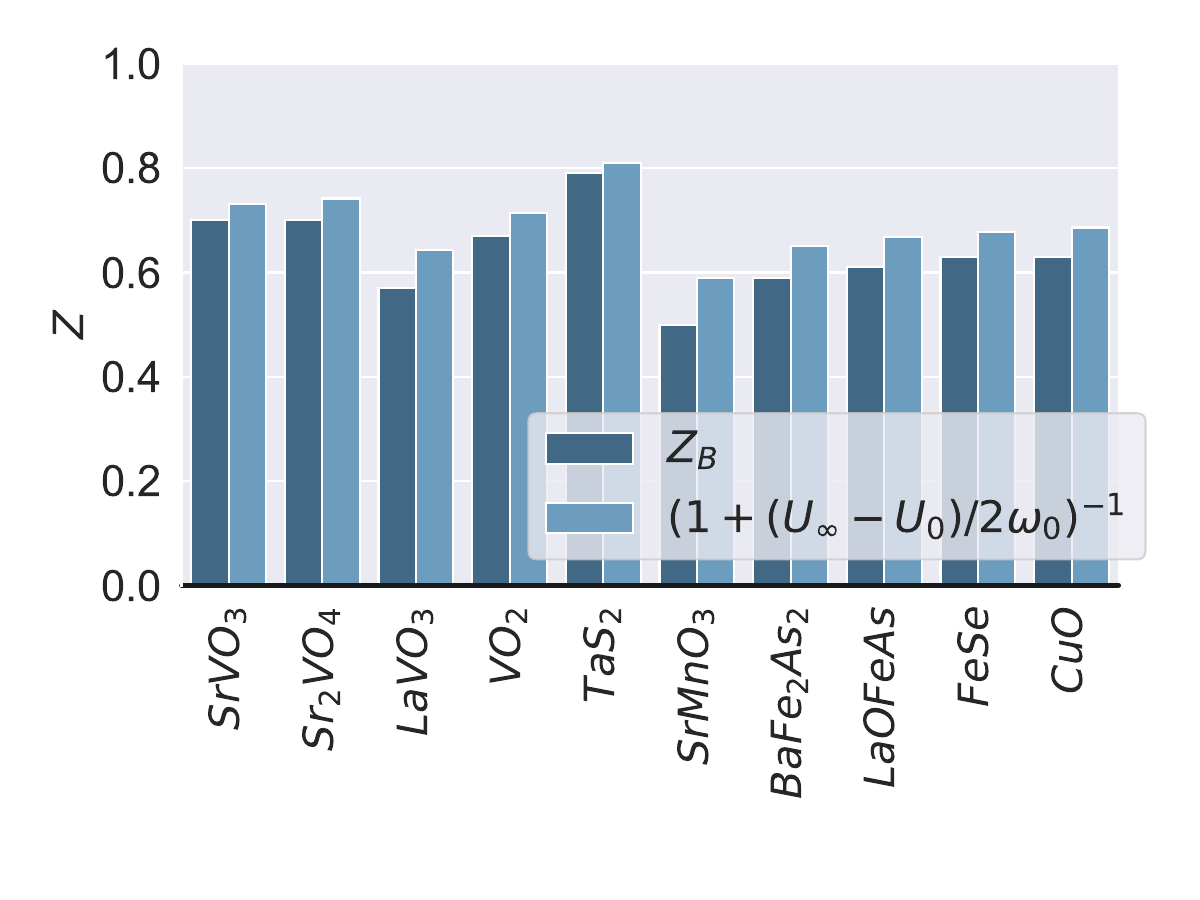}
		\caption{(Color online) The $Z_B$ factor obtained in \cite{PhysRevLett.109.126408} compared with the one obtained via our suggested expression $Z\approx(1+(U_{\infty}-U(0))/2\omega_0)^{-1}$, with $U_{\infty}$, $U_0$ and $\omega_0$ taken from \cite{PhysRevLett.109.126408}.}
		\label{fig:materials}
	\end{figure}
	
	\begin{figure}[t!]
		\includegraphics[width=1.0\columnwidth]{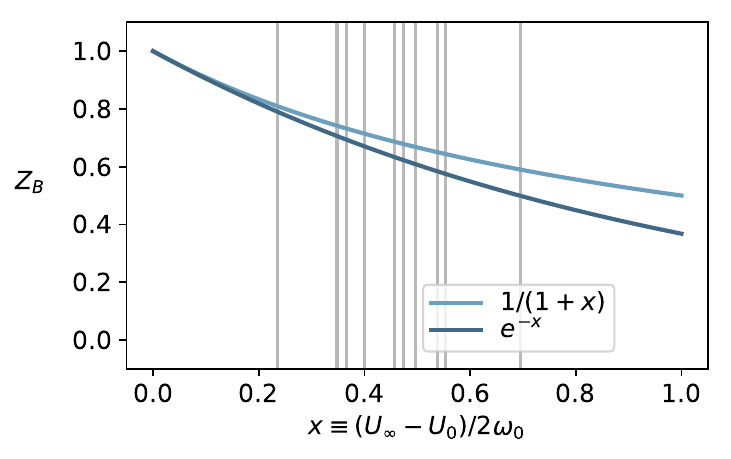}
		\caption{(Color online) The two formulas for $Z_B$, in dark blue the one from Ref. \cite{PhysRevLett.109.126408} and in light blue  $(1+(U_{\infty}-U(0))/2\omega_0)^{-1}$, as a function of  $x:=(U_{\infty}-U(0))/2\omega_0$, for the different materials considered in Ref. \cite{PhysRevLett.109.126408} (gray vertical lines).}
		\label{fig:materials_2}
	\end{figure}

	\subsection{Renormalization factor}
	In the examples above we have shown how the main feature brought by the $U(\omega)$ self-energy of Eq. \eqref{eq:PSigmaCorr} is a renormalization of the bands around the Fermi level. As stated above, this is $Z=(1-\partial\Delta\Sigma/\partial\omega|_{\varepsilon_m})^{-1}\approx(1+(U_{\infty}-U(0))/2\omega_0)^{-1}$. It is interesting to note that this same relation approximately holds also for the bandwidth renormalization introduced in Ref. \cite{PhysRevLett.109.126408}. There, $\omega_0$ plays an analogous but different role, as it is an average excitation energy which sets the important screening to renormalize, at the one-particle level, the low-energy bands:
	\begin{equation}	\omega_0=\frac{\int_0^{+\infty}d\omega\omega{\rm Im}U_{\rm cRPA}(\omega)/\omega^2}{\int_0^{+\infty}d\omega{\rm Im}U_{\rm cRPA}(\omega)/\omega^2}
    \label{eq:omomom0}
	\end{equation}
	As a result, the renormalization $Z_B={\rm exp}\frac{1}{\pi}\int_0^{+\infty}d\omega\,{\rm Im}U_{\rm cRPA}(\omega)/\omega^2$ is weaker than the full $Z$, and an additional DMFT calculation is responsible for further shrinking the band. In our approach, instead, a single renormalization with the full RPA $U(\omega)$ and a different $\omega_0$ account for the whole reduction of the band. However, the physics is similar (renormalization due to coupling with bosons) and thus it is not surprising that our renormalization formula $Z\approx(1+(U_{\infty}-U(0))/2\omega_0)^{-1}$ holds also in that case. This is shown in the comparison of Fig. \ref{fig:materials} for the different materials studied in Ref. \cite{PhysRevLett.109.126408}. In fact, the expression proposed in Ref. \cite{PhysRevLett.109.126408}, Eq. \eqref{eq:omomom0}, reduces to $Z=e^{-x}$, with $x:=(U_{\infty}-U(0))/2\omega_0$, while ours is $Z=1/(1+x)$, and the two are asymptotically equal for small values of $x$, as shown in Fig. \ref{fig:materials_2}.

	\subsection{Computational details}
	For the calculation of the PBE ground state and the Wannier functions $\{\ket{m}\}$ we have used the open-source code {\sc Quantum ESPRESSO}, version 6.4.1 \cite{QE-2009,QE-2017}, interfaced to {\sc Wannier90}, version 3.0.0 \cite{Pizzi_2020}, with ultrasoft PBE pseudopotentials \cite{PSP} from the Materials Cloud SSSP library \footnote{https://www.materialscloud.org/discover/sssp/table/precision}.
	For $U(\omega)$, we have employed the open-source code {\sc abinit}, version 8.10.1 \cite{Gonze2020}, where cRPA calculations are already implemented  \cite{Amadon_2014,Torrent2008} using projected local orbitals Wannier functions \cite{Amadon_2008}. We have used PAW atomic data \cite{JOLLET20141246} with PBE potentials  from \footnote{https://www.abinit.org/downloads/PAW2.}.
	
	\paragraph{SrVO$_3$:}
	Strontium vanadate crystallizes in an undistorted perovskite structure \cite{Itoh}, with a simple cubic unit cell $Pm\bar3m$ of experimental lattice constant $a=3.842\AA$ \cite{CHAMBERLAND1971243}. For the calculation of $U(\omega)$, we consider a shifted Monkhorst-Pack $6\times6\times6$ grid \cite{PhysRevB.13.5188} with a Fermi-Dirac smearing of 0.1 eV, 75 bands, a cut-off of 12 Ha for the wavefunctions, and 200 frequencies. We use a cut-off energy of 6 Ha for the dielectric tensor $\epsilon(\omega)$ and 20 Ha for $U(\omega)$.
	
	\paragraph{LaNiO$_3$:}
	At low temperature, the crystal structure of the paramagnetic lanthanum nickelate is $R\bar3c$, with a slight distortion with respect to a perfect cube \cite{PhysRevB.46.4414,MASYS2015153}, with lattice constant $a=5.433\AA$, and pseudocubic length of $3.842\AA$.	
	To better compare with SrVO$_3$, we consider the high-temperature undistorted cubic structure $\rm Pm\bar3m$ \cite{HAMADA19931157} with lattice parameter $a=3.857\AA$.
	For evaluating $U(\omega)$, we have taken a shifted Monkhorst-Pack $8\times8\times8$ grid with Fermi-Dirac smearing of 0.01 eV, 70 bands, a cut-off of 29 Ha for the wavefunctions, and 200 frequencies. The cut-off energy for the dielectric tensor $\epsilon(\omega)$ is 5 Ha and 9 Ha for $U(\omega)$.

	\bibliographystyle{apsrev4-1}
	
 \input{U_om.bbl}
\end{document}

%% file: U_om.bbl
%

%% file: U_om.bbl
\begin{thebibliography}{130}%
\makeatletter
\providecommand \@ifxundefined [1]{%
 \@ifx{#1\undefined}
}%
\providecommand \@ifnum [1]{%
 \ifnum #1\expandafter \@firstoftwo
 \else \expandafter \@secondoftwo
 \fi
}%
\providecommand \@ifx [1]{%
 \ifx #1\expandafter \@firstoftwo
 \else \expandafter \@secondoftwo
 \fi
}%
\providecommand \natexlab [1]{#1}%
\providecommand \enquote  [1]{``#1''}%
\providecommand \bibnamefont  [1]{#1}%
\providecommand \bibfnamefont [1]{#1}%
\providecommand \citenamefont [1]{#1}%
\providecommand \href@noop [0]{\@secondoftwo}%
\providecommand \href [0]{\begingroup \@sanitize@url \@href}%
\providecommand \@href[1]{\@@startlink{#1}\@@href}%
\providecommand \@@href[1]{\endgroup#1\@@endlink}%
\providecommand \@sanitize@url [0]{\catcode `\\12\catcode `\$12\catcode `\&12\catcode `\#12\catcode `\^12\catcode `\_12\catcode `\%12\relax}%
\providecommand \@@startlink[1]{}%
\providecommand \@@endlink[0]{}%
\providecommand \url  [0]{\begingroup\@sanitize@url \@url }%
\providecommand \@url [1]{\endgroup\@href {#1}{\urlprefix }}%
\providecommand \urlprefix  [0]{URL }%
\providecommand \Eprint [0]{\href }%
\providecommand \doibase [0]{http://dx.doi.org/}%
\providecommand \selectlanguage [0]{\@gobble}%
\providecommand \bibinfo  [0]{\@secondoftwo}%
\providecommand \bibfield  [0]{\@secondoftwo}%
\providecommand \translation [1]{[#1]}%
\providecommand \BibitemOpen [0]{}%
\providecommand \bibitemStop [0]{}%
\providecommand \bibitemNoStop [0]{.\EOS\space}%
\providecommand \EOS [0]{\spacefactor3000\relax}%
\providecommand \BibitemShut  [1]{\csname bibitem#1\endcsname}%
\let\auto@bib@innerbib\@empty
\bibitem [{\citenamefont {Marzari}\ \emph {et~al.}(2021)\citenamefont {Marzari}, \citenamefont {Ferretti},\ and\ \citenamefont {Wolverton}}]{Marzari2021}%
  \BibitemOpen
  \bibfield  {author} {\bibinfo {author} {\bibfnamefont {N.}~\bibnamefont {Marzari}}, \bibinfo {author} {\bibfnamefont {A.}~\bibnamefont {Ferretti}}, \ and\ \bibinfo {author} {\bibfnamefont {C.}~\bibnamefont {Wolverton}},\ }\href {\doibase 10.1038/s41563-021-01013-3} {\bibfield  {journal} {\bibinfo  {journal} {Nature Materials}\ }\textbf {\bibinfo {volume} {20}},\ \bibinfo {pages} {736} (\bibinfo {year} {2021})}\BibitemShut {NoStop}%
\bibitem [{\citenamefont {Yu}\ and\ \citenamefont {Zunger}(2012)}]{PhysRevLett.108.068701}%
  \BibitemOpen
  \bibfield  {author} {\bibinfo {author} {\bibfnamefont {L.}~\bibnamefont {Yu}}\ and\ \bibinfo {author} {\bibfnamefont {A.}~\bibnamefont {Zunger}},\ }\href {\doibase 10.1103/PhysRevLett.108.068701} {\bibfield  {journal} {\bibinfo  {journal} {Phys. Rev. Lett.}\ }\textbf {\bibinfo {volume} {108}},\ \bibinfo {pages} {068701} (\bibinfo {year} {2012})}\BibitemShut {NoStop}%
\bibitem [{\citenamefont {Jain}\ \emph {et~al.}(2016)\citenamefont {Jain}, \citenamefont {Shin},\ and\ \citenamefont {Persson}}]{Jain2016}%
  \BibitemOpen
  \bibfield  {author} {\bibinfo {author} {\bibfnamefont {A.}~\bibnamefont {Jain}}, \bibinfo {author} {\bibfnamefont {Y.}~\bibnamefont {Shin}}, \ and\ \bibinfo {author} {\bibfnamefont {K.~A.}\ \bibnamefont {Persson}},\ }\href {\doibase 10.1038/natrevmats.2015.4} {\bibfield  {journal} {\bibinfo  {journal} {Nature Reviews Materials}\ }\textbf {\bibinfo {volume} {1}},\ \bibinfo {pages} {15004} (\bibinfo {year} {2016})}\BibitemShut {NoStop}%
\bibitem [{\citenamefont {Hohenberg}\ and\ \citenamefont {Kohn}(1964)}]{HK1964}%
  \BibitemOpen
  \bibfield  {author} {\bibinfo {author} {\bibfnamefont {P.}~\bibnamefont {Hohenberg}}\ and\ \bibinfo {author} {\bibfnamefont {W.}~\bibnamefont {Kohn}},\ }\href {\doibase 10.1103/PhysRev.136.B864} {\bibfield  {journal} {\bibinfo  {journal} {Phys. Rev.}\ }\textbf {\bibinfo {volume} {136}},\ \bibinfo {pages} {B864} (\bibinfo {year} {1964})}\BibitemShut {NoStop}%
\bibitem [{\citenamefont {Kohn}\ and\ \citenamefont {Sham}(1965)}]{KS1965}%
  \BibitemOpen
  \bibfield  {author} {\bibinfo {author} {\bibfnamefont {W.}~\bibnamefont {Kohn}}\ and\ \bibinfo {author} {\bibfnamefont {L.~J.}\ \bibnamefont {Sham}},\ }\href {\doibase 10.1103/PhysRev.140.A1133} {\bibfield  {journal} {\bibinfo  {journal} {Phys. Rev.}\ }\textbf {\bibinfo {volume} {140}},\ \bibinfo {pages} {A1133} (\bibinfo {year} {1965})}\BibitemShut {NoStop}%
\bibitem [{\citenamefont {Parr}\ and\ \citenamefont {Yang}(1994)}]{parr1994density}%
  \BibitemOpen
  \bibfield  {author} {\bibinfo {author} {\bibfnamefont {R.}~\bibnamefont {Parr}}\ and\ \bibinfo {author} {\bibfnamefont {W.}~\bibnamefont {Yang}},\ }\href {https://books.google.ch/books?id=mGOpScSIwU4C} {\emph {\bibinfo {title} {Density-Functional Theory of Atoms and Molecules}}},\ International Series of Monographs on Chemistry\ (\bibinfo  {publisher} {Oxford University Press},\ \bibinfo {year} {1994})\BibitemShut {NoStop}%
\bibitem [{\citenamefont {Perdew}\ \emph {et~al.}(1982)\citenamefont {Perdew}, \citenamefont {Parr}, \citenamefont {Levy},\ and\ \citenamefont {Balduz}}]{PhysRevLett.49.1691}%
  \BibitemOpen
  \bibfield  {author} {\bibinfo {author} {\bibfnamefont {J.~P.}\ \bibnamefont {Perdew}}, \bibinfo {author} {\bibfnamefont {R.~G.}\ \bibnamefont {Parr}}, \bibinfo {author} {\bibfnamefont {M.}~\bibnamefont {Levy}}, \ and\ \bibinfo {author} {\bibfnamefont {J.~L.}\ \bibnamefont {Balduz}},\ }\href {\doibase 10.1103/PhysRevLett.49.1691} {\bibfield  {journal} {\bibinfo  {journal} {Phys. Rev. Lett.}\ }\textbf {\bibinfo {volume} {49}},\ \bibinfo {pages} {1691} (\bibinfo {year} {1982})}\BibitemShut {NoStop}%
\bibitem [{\citenamefont {Sham}\ and\ \citenamefont {Schl\"uter}(1983)}]{PhysRevLett.51.1888}%
  \BibitemOpen
  \bibfield  {author} {\bibinfo {author} {\bibfnamefont {L.~J.}\ \bibnamefont {Sham}}\ and\ \bibinfo {author} {\bibfnamefont {M.}~\bibnamefont {Schl\"uter}},\ }\href {\doibase 10.1103/PhysRevLett.51.1888} {\bibfield  {journal} {\bibinfo  {journal} {Phys. Rev. Lett.}\ }\textbf {\bibinfo {volume} {51}},\ \bibinfo {pages} {1888} (\bibinfo {year} {1983})}\BibitemShut {NoStop}%
\bibitem [{\citenamefont {H{\"u}fner}\ and\ \citenamefont {Huber}(2003)}]{hufner2003photoelectron}%
  \BibitemOpen
  \bibfield  {author} {\bibinfo {author} {\bibfnamefont {S.}~\bibnamefont {H{\"u}fner}}\ and\ \bibinfo {author} {\bibfnamefont {T.}~\bibnamefont {Huber}},\ }\href {https://books.google.ch/books?id=WfOw6jP9-oIC} {\emph {\bibinfo {title} {Photoelectron Spectroscopy: Principles and Applications}}},\ Advanced Texts in Physics\ (\bibinfo  {publisher} {Springer},\ \bibinfo {year} {2003})\BibitemShut {NoStop}%
\bibitem [{\citenamefont {Damascelli}\ \emph {et~al.}(2003)\citenamefont {Damascelli}, \citenamefont {Hussain},\ and\ \citenamefont {Shen}}]{RevModPhys.75.473}%
  \BibitemOpen
  \bibfield  {author} {\bibinfo {author} {\bibfnamefont {A.}~\bibnamefont {Damascelli}}, \bibinfo {author} {\bibfnamefont {Z.}~\bibnamefont {Hussain}}, \ and\ \bibinfo {author} {\bibfnamefont {Z.-X.}\ \bibnamefont {Shen}},\ }\href {\doibase 10.1103/RevModPhys.75.473} {\bibfield  {journal} {\bibinfo  {journal} {Rev. Mod. Phys.}\ }\textbf {\bibinfo {volume} {75}},\ \bibinfo {pages} {473} (\bibinfo {year} {2003})}\BibitemShut {NoStop}%
\bibitem [{\citenamefont {Onida}\ \emph {et~al.}(2002)\citenamefont {Onida}, \citenamefont {Reining},\ and\ \citenamefont {Rubio}}]{Onida2002}%
  \BibitemOpen
  \bibfield  {author} {\bibinfo {author} {\bibfnamefont {G.}~\bibnamefont {Onida}}, \bibinfo {author} {\bibfnamefont {L.}~\bibnamefont {Reining}}, \ and\ \bibinfo {author} {\bibfnamefont {A.}~\bibnamefont {Rubio}},\ }\href {\doibase 10.1103/RevModPhys.74.601} {\bibfield  {journal} {\bibinfo  {journal} {Rev. Mod. Phys.}\ }\textbf {\bibinfo {volume} {74}},\ \bibinfo {pages} {601} (\bibinfo {year} {2002})}\BibitemShut {NoStop}%
\bibitem [{\citenamefont {Hedin}\ \emph {et~al.}(1998)\citenamefont {Hedin}, \citenamefont {Michiels},\ and\ \citenamefont {Inglesfield}}]{PhysRevB.58.15565}%
  \BibitemOpen
  \bibfield  {author} {\bibinfo {author} {\bibfnamefont {L.}~\bibnamefont {Hedin}}, \bibinfo {author} {\bibfnamefont {J.}~\bibnamefont {Michiels}}, \ and\ \bibinfo {author} {\bibfnamefont {J.}~\bibnamefont {Inglesfield}},\ }\href {\doibase 10.1103/PhysRevB.58.15565} {\bibfield  {journal} {\bibinfo  {journal} {Phys. Rev. B}\ }\textbf {\bibinfo {volume} {58}},\ \bibinfo {pages} {15565} (\bibinfo {year} {1998})}\BibitemShut {NoStop}%
\bibitem [{\citenamefont {Guzzo}\ \emph {et~al.}(2011)\citenamefont {Guzzo}, \citenamefont {Lani}, \citenamefont {Sottile}, \citenamefont {Romaniello}, \citenamefont {Gatti}, \citenamefont {Kas}, \citenamefont {Rehr}, \citenamefont {Silly}, \citenamefont {Sirotti},\ and\ \citenamefont {Reining}}]{PhysRevLett.107.166401}%
  \BibitemOpen
  \bibfield  {author} {\bibinfo {author} {\bibfnamefont {M.}~\bibnamefont {Guzzo}}, \bibinfo {author} {\bibfnamefont {G.}~\bibnamefont {Lani}}, \bibinfo {author} {\bibfnamefont {F.}~\bibnamefont {Sottile}}, \bibinfo {author} {\bibfnamefont {P.}~\bibnamefont {Romaniello}}, \bibinfo {author} {\bibfnamefont {M.}~\bibnamefont {Gatti}}, \bibinfo {author} {\bibfnamefont {J.~J.}\ \bibnamefont {Kas}}, \bibinfo {author} {\bibfnamefont {J.~J.}\ \bibnamefont {Rehr}}, \bibinfo {author} {\bibfnamefont {M.~G.}\ \bibnamefont {Silly}}, \bibinfo {author} {\bibfnamefont {F.}~\bibnamefont {Sirotti}}, \ and\ \bibinfo {author} {\bibfnamefont {L.}~\bibnamefont {Reining}},\ }\href {\doibase 10.1103/PhysRevLett.107.166401} {\bibfield  {journal} {\bibinfo  {journal} {Phys. Rev. Lett.}\ }\textbf {\bibinfo {volume} {107}},\ \bibinfo {pages} {166401} (\bibinfo {year} {2011})}\BibitemShut {NoStop}%
\bibitem [{\citenamefont {Zhou}\ \emph {et~al.}(2018)\citenamefont {Zhou}, \citenamefont {Reining}, \citenamefont {Nicolaou}, \citenamefont {Bendounan}, \citenamefont {Ruotsalainen}, \citenamefont {Vanzini}, \citenamefont {Kas}, \citenamefont {Rehr}, \citenamefont {Muntwiler}, \citenamefont {Strocov}, \citenamefont {Sirotti},\ and\ \citenamefont {Gatti}}]{zhou2018dispersing}%
  \BibitemOpen
  \bibfield  {author} {\bibinfo {author} {\bibfnamefont {J.~S.}\ \bibnamefont {Zhou}}, \bibinfo {author} {\bibfnamefont {L.}~\bibnamefont {Reining}}, \bibinfo {author} {\bibfnamefont {A.}~\bibnamefont {Nicolaou}}, \bibinfo {author} {\bibfnamefont {A.}~\bibnamefont {Bendounan}}, \bibinfo {author} {\bibfnamefont {K.}~\bibnamefont {Ruotsalainen}}, \bibinfo {author} {\bibfnamefont {M.}~\bibnamefont {Vanzini}}, \bibinfo {author} {\bibfnamefont {J.~J.}\ \bibnamefont {Kas}}, \bibinfo {author} {\bibfnamefont {J.~J.}\ \bibnamefont {Rehr}}, \bibinfo {author} {\bibfnamefont {M.}~\bibnamefont {Muntwiler}}, \bibinfo {author} {\bibfnamefont {V.~N.}\ \bibnamefont {Strocov}}, \bibinfo {author} {\bibfnamefont {F.}~\bibnamefont {Sirotti}}, \ and\ \bibinfo {author} {\bibfnamefont {M.}~\bibnamefont {Gatti}},\ }\href {http://adsabs.harvard.edu/abs/2017arXiv170802450V} {\bibfield  {journal} {\bibinfo  {journal} {ArXiv e-prints}\ } (\bibinfo {year} {2018})},\ \Eprint {http://arxiv.org/abs/1811.12217} {arXiv:1811.12217
  [cond-mat.str-el]} \BibitemShut {NoStop}%
\bibitem [{\citenamefont {Hedin}(1965)}]{Hedin1965}%
  \BibitemOpen
  \bibfield  {author} {\bibinfo {author} {\bibfnamefont {L.}~\bibnamefont {Hedin}},\ }\href {\doibase 10.1103/PhysRev.139.A796} {\bibfield  {journal} {\bibinfo  {journal} {Phys. Rev.}\ }\textbf {\bibinfo {volume} {139}},\ \bibinfo {pages} {A796} (\bibinfo {year} {1965})}\BibitemShut {NoStop}%
\bibitem [{\citenamefont {Aryasetiawan}\ and\ \citenamefont {Gunnarsson}(1998)}]{AryaGunn}%
  \BibitemOpen
  \bibfield  {author} {\bibinfo {author} {\bibfnamefont {F.}~\bibnamefont {Aryasetiawan}}\ and\ \bibinfo {author} {\bibfnamefont {O.}~\bibnamefont {Gunnarsson}},\ }\href@noop {} {\bibfield  {journal} {\bibinfo  {journal} {Rep. Prog. Phys.}\ }\textbf {\bibinfo {volume} {61}},\ \bibinfo {pages} {237} (\bibinfo {year} {1998})}\BibitemShut {NoStop}%
\bibitem [{\citenamefont {Hybertsen}\ and\ \citenamefont {Louie}(1986)}]{Hyber}%
  \BibitemOpen
  \bibfield  {author} {\bibinfo {author} {\bibfnamefont {M.~S.}\ \bibnamefont {Hybertsen}}\ and\ \bibinfo {author} {\bibfnamefont {S.~G.}\ \bibnamefont {Louie}},\ }\href {\doibase 10.1103/PhysRevB.34.5390} {\bibfield  {journal} {\bibinfo  {journal} {Phys. Rev. B}\ }\textbf {\bibinfo {volume} {34}},\ \bibinfo {pages} {5390} (\bibinfo {year} {1986})}\BibitemShut {NoStop}%
\bibitem [{\citenamefont {van Schilfgaarde}\ \emph {et~al.}(2006)\citenamefont {van Schilfgaarde}, \citenamefont {Kotani},\ and\ \citenamefont {Faleev}}]{Schilfgaarde2006}%
  \BibitemOpen
  \bibfield  {author} {\bibinfo {author} {\bibfnamefont {M.}~\bibnamefont {van Schilfgaarde}}, \bibinfo {author} {\bibfnamefont {T.}~\bibnamefont {Kotani}}, \ and\ \bibinfo {author} {\bibfnamefont {S.}~\bibnamefont {Faleev}},\ }\href@noop {} {\bibfield  {journal} {\bibinfo  {journal} {Phys. Rev. Lett.}\ }\textbf {\bibinfo {volume} {96}},\ \bibinfo {pages} {226402} (\bibinfo {year} {2006})}\BibitemShut {NoStop}%
\bibitem [{\citenamefont {Golze}\ \emph {et~al.}(2019)\citenamefont {Golze}, \citenamefont {Dvorak},\ and\ \citenamefont {Rinke}}]{10.3389/fchem.2019.00377}%
  \BibitemOpen
  \bibfield  {author} {\bibinfo {author} {\bibfnamefont {D.}~\bibnamefont {Golze}}, \bibinfo {author} {\bibfnamefont {M.}~\bibnamefont {Dvorak}}, \ and\ \bibinfo {author} {\bibfnamefont {P.}~\bibnamefont {Rinke}},\ }\href {\doibase 10.3389/fchem.2019.00377} {\bibfield  {journal} {\bibinfo  {journal} {Frontiers in Chemistry}\ }\textbf {\bibinfo {volume} {7}},\ \bibinfo {pages} {377} (\bibinfo {year} {2019})}\BibitemShut {NoStop}%
\bibitem [{\citenamefont {Metzner}\ and\ \citenamefont {Vollhardt}(1989)}]{Metzner}%
  \BibitemOpen
  \bibfield  {author} {\bibinfo {author} {\bibfnamefont {W.}~\bibnamefont {Metzner}}\ and\ \bibinfo {author} {\bibfnamefont {D.}~\bibnamefont {Vollhardt}},\ }\href {\doibase 10.1103/PhysRevLett.62.324} {\bibfield  {journal} {\bibinfo  {journal} {Phys. Rev. Lett.}\ }\textbf {\bibinfo {volume} {62}},\ \bibinfo {pages} {324} (\bibinfo {year} {1989})}\BibitemShut {NoStop}%
\bibitem [{\citenamefont {Georges}\ and\ \citenamefont {Kotliar}(1992)}]{Georges}%
  \BibitemOpen
  \bibfield  {author} {\bibinfo {author} {\bibfnamefont {A.}~\bibnamefont {Georges}}\ and\ \bibinfo {author} {\bibfnamefont {G.}~\bibnamefont {Kotliar}},\ }\href {\doibase 10.1103/PhysRevB.45.6479} {\bibfield  {journal} {\bibinfo  {journal} {Phys. Rev. B}\ }\textbf {\bibinfo {volume} {45}},\ \bibinfo {pages} {6479} (\bibinfo {year} {1992})}\BibitemShut {NoStop}%
\bibitem [{\citenamefont {Georges}\ \emph {et~al.}(1996)\citenamefont {Georges}, \citenamefont {Kotliar}, \citenamefont {Krauth},\ and\ \citenamefont {Rozenberg}}]{RevModPhys.68.13}%
  \BibitemOpen
  \bibfield  {author} {\bibinfo {author} {\bibfnamefont {A.}~\bibnamefont {Georges}}, \bibinfo {author} {\bibfnamefont {G.}~\bibnamefont {Kotliar}}, \bibinfo {author} {\bibfnamefont {W.}~\bibnamefont {Krauth}}, \ and\ \bibinfo {author} {\bibfnamefont {M.~J.}\ \bibnamefont {Rozenberg}},\ }\href {\doibase 10.1103/RevModPhys.68.13} {\bibfield  {journal} {\bibinfo  {journal} {Rev. Mod. Phys.}\ }\textbf {\bibinfo {volume} {68}},\ \bibinfo {pages} {13} (\bibinfo {year} {1996})}\BibitemShut {NoStop}%
\bibitem [{\citenamefont {Kotliar}\ \emph {et~al.}(2006)\citenamefont {Kotliar}, \citenamefont {Savrasov}, \citenamefont {Haule}, \citenamefont {Oudovenko}, \citenamefont {Parcollet},\ and\ \citenamefont {Marianetti}}]{RevModPhysKotliar}%
  \BibitemOpen
  \bibfield  {author} {\bibinfo {author} {\bibfnamefont {G.}~\bibnamefont {Kotliar}}, \bibinfo {author} {\bibfnamefont {S.~Y.}\ \bibnamefont {Savrasov}}, \bibinfo {author} {\bibfnamefont {K.}~\bibnamefont {Haule}}, \bibinfo {author} {\bibfnamefont {V.~S.}\ \bibnamefont {Oudovenko}}, \bibinfo {author} {\bibfnamefont {O.}~\bibnamefont {Parcollet}}, \ and\ \bibinfo {author} {\bibfnamefont {C.~A.}\ \bibnamefont {Marianetti}},\ }\href {\doibase 10.1103/RevModPhys.78.865} {\bibfield  {journal} {\bibinfo  {journal} {Rev. Mod. Phys.}\ }\textbf {\bibinfo {volume} {78}},\ \bibinfo {pages} {865} (\bibinfo {year} {2006})}\BibitemShut {NoStop}%
\bibitem [{\citenamefont {Fetter}\ and\ \citenamefont {Walecka}(2003)}]{fetter2003quantum}%
  \BibitemOpen
  \bibfield  {author} {\bibinfo {author} {\bibfnamefont {A.}~\bibnamefont {Fetter}}\ and\ \bibinfo {author} {\bibfnamefont {J.}~\bibnamefont {Walecka}},\ }\href {https://books.google.ch/books?id=0wekf1s83b0C} {\emph {\bibinfo {title} {Quantum Theory of Many-particle Systems}}},\ Dover Books on Physics\ (\bibinfo  {publisher} {Dover Publications},\ \bibinfo {year} {2003})\BibitemShut {NoStop}%
\bibitem [{\citenamefont {Strinati}(1988)}]{Strinati}%
  \BibitemOpen
  \bibfield  {author} {\bibinfo {author} {\bibfnamefont {G.}~\bibnamefont {Strinati}},\ }\href@noop {} {\bibfield  {journal} {\bibinfo  {journal} {La Rivista del Nuovo Cimento}\ }\textbf {\bibinfo {volume} {11}},\ \bibinfo {pages} {1} (\bibinfo {year} {1988})}\BibitemShut {NoStop}%
\bibitem [{\citenamefont {Martin}\ \emph {et~al.}(2016)\citenamefont {Martin}, \citenamefont {Reining},\ and\ \citenamefont {Ceperley}}]{luciabook}%
  \BibitemOpen
  \bibfield  {author} {\bibinfo {author} {\bibfnamefont {R.~M.}\ \bibnamefont {Martin}}, \bibinfo {author} {\bibfnamefont {L.}~\bibnamefont {Reining}}, \ and\ \bibinfo {author} {\bibfnamefont {D.~M.}\ \bibnamefont {Ceperley}},\ }\href {\doibase 10.1017/CBO9781139050807} {\emph {\bibinfo {title} {Interacting Electrons: Theory and Computational Approaches}}}\ (\bibinfo  {publisher} {Cambridge University Press},\ \bibinfo {year} {2016})\BibitemShut {NoStop}%
\bibitem [{\citenamefont {Dagotto}(2005)}]{Dagotto257}%
  \BibitemOpen
  \bibfield  {author} {\bibinfo {author} {\bibfnamefont {E.}~\bibnamefont {Dagotto}},\ }\href {\doibase 10.1126/science.1107559} {\bibfield  {journal} {\bibinfo  {journal} {Science}\ }\textbf {\bibinfo {volume} {309}},\ \bibinfo {pages} {257} (\bibinfo {year} {2005})}\BibitemShut {NoStop}%
\bibitem [{\citenamefont {Tokura}(2003)}]{Tokura2003}%
  \BibitemOpen
  \bibfield  {author} {\bibinfo {author} {\bibfnamefont {Y.}~\bibnamefont {Tokura}},\ }\href {\doibase 10.1063/1.1603080} {\bibfield  {journal} {\bibinfo  {journal} {Physics Today}\ }\textbf {\bibinfo {volume} {56}},\ \bibinfo {pages} {50} (\bibinfo {year} {2003})},\ \Eprint {http://arxiv.org/abs/https://doi.org/10.1063/1.1603080} {https://doi.org/10.1063/1.1603080} \BibitemShut {NoStop}%
\bibitem [{\citenamefont {Brahlek}\ \emph {et~al.}(2017)\citenamefont {Brahlek}, \citenamefont {Zhang}, \citenamefont {Lapano}, \citenamefont {Zhang}, \citenamefont {Engel-Herbert}, \citenamefont {Shukla}, \citenamefont {Datta}, \citenamefont {Paik},\ and\ \citenamefont {Schlom}}]{app1}%
  \BibitemOpen
  \bibfield  {author} {\bibinfo {author} {\bibfnamefont {M.}~\bibnamefont {Brahlek}}, \bibinfo {author} {\bibfnamefont {L.}~\bibnamefont {Zhang}}, \bibinfo {author} {\bibfnamefont {J.}~\bibnamefont {Lapano}}, \bibinfo {author} {\bibfnamefont {H.-T.}\ \bibnamefont {Zhang}}, \bibinfo {author} {\bibfnamefont {R.}~\bibnamefont {Engel-Herbert}}, \bibinfo {author} {\bibfnamefont {N.}~\bibnamefont {Shukla}}, \bibinfo {author} {\bibfnamefont {S.}~\bibnamefont {Datta}}, \bibinfo {author} {\bibfnamefont {H.}~\bibnamefont {Paik}}, \ and\ \bibinfo {author} {\bibfnamefont {D.~G.}\ \bibnamefont {Schlom}},\ }\href {\doibase 10.1557/mrc.2017.2} {\bibfield  {journal} {\bibinfo  {journal} {MRS Communications}\ }\textbf {\bibinfo {volume} {7}},\ \bibinfo {pages} {27} (\bibinfo {year} {2017})}\BibitemShut {NoStop}%
\bibitem [{\citenamefont {Castelli}\ \emph {et~al.}(2012)\citenamefont {Castelli}, \citenamefont {Olsen}, \citenamefont {Datta}, \citenamefont {Landis}, \citenamefont {Dahl}, \citenamefont {Thygesen},\ and\ \citenamefont {Jacobsen}}]{C1EE02717D}%
  \BibitemOpen
  \bibfield  {author} {\bibinfo {author} {\bibfnamefont {I.~E.}\ \bibnamefont {Castelli}}, \bibinfo {author} {\bibfnamefont {T.}~\bibnamefont {Olsen}}, \bibinfo {author} {\bibfnamefont {S.}~\bibnamefont {Datta}}, \bibinfo {author} {\bibfnamefont {D.~D.}\ \bibnamefont {Landis}}, \bibinfo {author} {\bibfnamefont {S.}~\bibnamefont {Dahl}}, \bibinfo {author} {\bibfnamefont {K.~S.}\ \bibnamefont {Thygesen}}, \ and\ \bibinfo {author} {\bibfnamefont {K.~W.}\ \bibnamefont {Jacobsen}},\ }\href {\doibase 10.1039/C1EE02717D} {\bibfield  {journal} {\bibinfo  {journal} {Energy Environ. Sci.}\ }\textbf {\bibinfo {volume} {5}},\ \bibinfo {pages} {5814} (\bibinfo {year} {2012})}\BibitemShut {NoStop}%
\bibitem [{\citenamefont {Anisimov}\ \emph {et~al.}(1991)\citenamefont {Anisimov}, \citenamefont {Zaanen},\ and\ \citenamefont {Andersen}}]{Anisimov_1991}%
  \BibitemOpen
  \bibfield  {author} {\bibinfo {author} {\bibfnamefont {V.~I.}\ \bibnamefont {Anisimov}}, \bibinfo {author} {\bibfnamefont {J.}~\bibnamefont {Zaanen}}, \ and\ \bibinfo {author} {\bibfnamefont {O.~K.}\ \bibnamefont {Andersen}},\ }\href {\doibase 10.1103/PhysRevB.44.943} {\bibfield  {journal} {\bibinfo  {journal} {Phys. Rev. B}\ }\textbf {\bibinfo {volume} {44}},\ \bibinfo {pages} {943} (\bibinfo {year} {1991})}\BibitemShut {NoStop}%
\bibitem [{\citenamefont {Anisimov}\ \emph {et~al.}(1993)\citenamefont {Anisimov}, \citenamefont {Solovyev}, \citenamefont {Korotin}, \citenamefont {Czy\ifmmode~\dot{z}\else \.{z}\fi{}yk},\ and\ \citenamefont {Sawatzky}}]{PhysRevB.48.16929}%
  \BibitemOpen
  \bibfield  {author} {\bibinfo {author} {\bibfnamefont {V.~I.}\ \bibnamefont {Anisimov}}, \bibinfo {author} {\bibfnamefont {I.~V.}\ \bibnamefont {Solovyev}}, \bibinfo {author} {\bibfnamefont {M.~A.}\ \bibnamefont {Korotin}}, \bibinfo {author} {\bibfnamefont {M.~T.}\ \bibnamefont {Czy\ifmmode~\dot{z}\else \.{z}\fi{}yk}}, \ and\ \bibinfo {author} {\bibfnamefont {G.~A.}\ \bibnamefont {Sawatzky}},\ }\href {\doibase 10.1103/PhysRevB.48.16929} {\bibfield  {journal} {\bibinfo  {journal} {Phys. Rev. B}\ }\textbf {\bibinfo {volume} {48}},\ \bibinfo {pages} {16929} (\bibinfo {year} {1993})}\BibitemShut {NoStop}%
\bibitem [{\citenamefont {Dudarev}\ \emph {et~al.}(1998)\citenamefont {Dudarev}, \citenamefont {Botton}, \citenamefont {Savrasov}, \citenamefont {Humphreys},\ and\ \citenamefont {Sutton}}]{PhysRevB.57.1505}%
  \BibitemOpen
  \bibfield  {author} {\bibinfo {author} {\bibfnamefont {S.~L.}\ \bibnamefont {Dudarev}}, \bibinfo {author} {\bibfnamefont {G.~A.}\ \bibnamefont {Botton}}, \bibinfo {author} {\bibfnamefont {S.~Y.}\ \bibnamefont {Savrasov}}, \bibinfo {author} {\bibfnamefont {C.~J.}\ \bibnamefont {Humphreys}}, \ and\ \bibinfo {author} {\bibfnamefont {A.~P.}\ \bibnamefont {Sutton}},\ }\href {\doibase 10.1103/PhysRevB.57.1505} {\bibfield  {journal} {\bibinfo  {journal} {Phys. Rev. B}\ }\textbf {\bibinfo {volume} {57}},\ \bibinfo {pages} {1505} (\bibinfo {year} {1998})}\BibitemShut {NoStop}%
\bibitem [{\citenamefont {Campo}\ and\ \citenamefont {Cococcioni}(2010)}]{Campo}%
  \BibitemOpen
  \bibfield  {author} {\bibinfo {author} {\bibfnamefont {V.~L.~J.}\ \bibnamefont {Campo}}\ and\ \bibinfo {author} {\bibfnamefont {M.}~\bibnamefont {Cococcioni}},\ }\href {\doibase 10.1088/0953-8984/22/5/055602} {\bibfield  {journal} {\bibinfo  {journal} {Journal of Physics: Condensed Matter}\ }\textbf {\bibinfo {volume} {22}},\ \bibinfo {pages} {055602} (\bibinfo {year} {2010})}\BibitemShut {NoStop}%
\bibitem [{\citenamefont {Gunnarsson}(1990)}]{PhysRevB.41.514}%
  \BibitemOpen
  \bibfield  {author} {\bibinfo {author} {\bibfnamefont {O.}~\bibnamefont {Gunnarsson}},\ }\href {\doibase 10.1103/PhysRevB.41.514} {\bibfield  {journal} {\bibinfo  {journal} {Phys. Rev. B}\ }\textbf {\bibinfo {volume} {41}},\ \bibinfo {pages} {514} (\bibinfo {year} {1990})}\BibitemShut {NoStop}%
\bibitem [{\citenamefont {Springer}\ and\ \citenamefont {Aryasetiawan}(1998)}]{PhysRevB.57.4364}%
  \BibitemOpen
  \bibfield  {author} {\bibinfo {author} {\bibfnamefont {M.}~\bibnamefont {Springer}}\ and\ \bibinfo {author} {\bibfnamefont {F.}~\bibnamefont {Aryasetiawan}},\ }\href {\doibase 10.1103/PhysRevB.57.4364} {\bibfield  {journal} {\bibinfo  {journal} {Phys. Rev. B}\ }\textbf {\bibinfo {volume} {57}},\ \bibinfo {pages} {4364} (\bibinfo {year} {1998})}\BibitemShut {NoStop}%
\bibitem [{\citenamefont {Anisimov}\ and\ \citenamefont {Gunnarsson}(1991)}]{PhysRevB.43.7570}%
  \BibitemOpen
  \bibfield  {author} {\bibinfo {author} {\bibfnamefont {V.~I.}\ \bibnamefont {Anisimov}}\ and\ \bibinfo {author} {\bibfnamefont {O.}~\bibnamefont {Gunnarsson}},\ }\href {\doibase 10.1103/PhysRevB.43.7570} {\bibfield  {journal} {\bibinfo  {journal} {Phys. Rev. B}\ }\textbf {\bibinfo {volume} {43}},\ \bibinfo {pages} {7570} (\bibinfo {year} {1991})}\BibitemShut {NoStop}%
\bibitem [{\citenamefont {Hybertsen}\ \emph {et~al.}(1989)\citenamefont {Hybertsen}, \citenamefont {Schl\"uter},\ and\ \citenamefont {Christensen}}]{PhysRevB.39.9028}%
  \BibitemOpen
  \bibfield  {author} {\bibinfo {author} {\bibfnamefont {M.~S.}\ \bibnamefont {Hybertsen}}, \bibinfo {author} {\bibfnamefont {M.}~\bibnamefont {Schl\"uter}}, \ and\ \bibinfo {author} {\bibfnamefont {N.~E.}\ \bibnamefont {Christensen}},\ }\href {\doibase 10.1103/PhysRevB.39.9028} {\bibfield  {journal} {\bibinfo  {journal} {Phys. Rev. B}\ }\textbf {\bibinfo {volume} {39}},\ \bibinfo {pages} {9028} (\bibinfo {year} {1989})}\BibitemShut {NoStop}%
\bibitem [{\citenamefont {Aryasetiawan}\ \emph {et~al.}(2004)\citenamefont {Aryasetiawan}, \citenamefont {Imada}, \citenamefont {Georges}, \citenamefont {Kotliar}, \citenamefont {Biermann},\ and\ \citenamefont {Lichtenstein}}]{PhysRevB.70.195104}%
  \BibitemOpen
  \bibfield  {author} {\bibinfo {author} {\bibfnamefont {F.}~\bibnamefont {Aryasetiawan}}, \bibinfo {author} {\bibfnamefont {M.}~\bibnamefont {Imada}}, \bibinfo {author} {\bibfnamefont {A.}~\bibnamefont {Georges}}, \bibinfo {author} {\bibfnamefont {G.}~\bibnamefont {Kotliar}}, \bibinfo {author} {\bibfnamefont {S.}~\bibnamefont {Biermann}}, \ and\ \bibinfo {author} {\bibfnamefont {A.~I.}\ \bibnamefont {Lichtenstein}},\ }\href {\doibase 10.1103/PhysRevB.70.195104} {\bibfield  {journal} {\bibinfo  {journal} {Phys. Rev. B}\ }\textbf {\bibinfo {volume} {70}},\ \bibinfo {pages} {195104} (\bibinfo {year} {2004})}\BibitemShut {NoStop}%
\bibitem [{\citenamefont {Cococcioni}\ and\ \citenamefont {de~Gironcoli}(2005)}]{PhysRevB.71.035105}%
  \BibitemOpen
  \bibfield  {author} {\bibinfo {author} {\bibfnamefont {M.}~\bibnamefont {Cococcioni}}\ and\ \bibinfo {author} {\bibfnamefont {S.}~\bibnamefont {de~Gironcoli}},\ }\href {\doibase 10.1103/PhysRevB.71.035105} {\bibfield  {journal} {\bibinfo  {journal} {Phys. Rev. B}\ }\textbf {\bibinfo {volume} {71}},\ \bibinfo {pages} {035105} (\bibinfo {year} {2005})}\BibitemShut {NoStop}%
\bibitem [{\citenamefont {Timrov}\ \emph {et~al.}(2018)\citenamefont {Timrov}, \citenamefont {Marzari},\ and\ \citenamefont {Cococcioni}}]{PhysRevB.98.085127}%
  \BibitemOpen
  \bibfield  {author} {\bibinfo {author} {\bibfnamefont {I.}~\bibnamefont {Timrov}}, \bibinfo {author} {\bibfnamefont {N.}~\bibnamefont {Marzari}}, \ and\ \bibinfo {author} {\bibfnamefont {M.}~\bibnamefont {Cococcioni}},\ }\href {\doibase 10.1103/PhysRevB.98.085127} {\bibfield  {journal} {\bibinfo  {journal} {Phys. Rev. B}\ }\textbf {\bibinfo {volume} {98}},\ \bibinfo {pages} {085127} (\bibinfo {year} {2018})}\BibitemShut {NoStop}%
\bibitem [{\citenamefont {Timrov}\ \emph {et~al.}(2021)\citenamefont {Timrov}, \citenamefont {Marzari},\ and\ \citenamefont {Cococcioni}}]{PhysRevB.103.045141}%
  \BibitemOpen
  \bibfield  {author} {\bibinfo {author} {\bibfnamefont {I.}~\bibnamefont {Timrov}}, \bibinfo {author} {\bibfnamefont {N.}~\bibnamefont {Marzari}}, \ and\ \bibinfo {author} {\bibfnamefont {M.}~\bibnamefont {Cococcioni}},\ }\href {\doibase 10.1103/PhysRevB.103.045141} {\bibfield  {journal} {\bibinfo  {journal} {Phys. Rev. B}\ }\textbf {\bibinfo {volume} {103}},\ \bibinfo {pages} {045141} (\bibinfo {year} {2021})}\BibitemShut {NoStop}%
\bibitem [{\citenamefont {Perdew}\ and\ \citenamefont {Zunger}(1981)}]{PhysRevB.23.5048}%
  \BibitemOpen
  \bibfield  {author} {\bibinfo {author} {\bibfnamefont {J.~P.}\ \bibnamefont {Perdew}}\ and\ \bibinfo {author} {\bibfnamefont {A.}~\bibnamefont {Zunger}},\ }\href {\doibase 10.1103/PhysRevB.23.5048} {\bibfield  {journal} {\bibinfo  {journal} {Phys. Rev. B}\ }\textbf {\bibinfo {volume} {23}},\ \bibinfo {pages} {5048} (\bibinfo {year} {1981})}\BibitemShut {NoStop}%
\bibitem [{\citenamefont {Tsuneda}\ and\ \citenamefont {Hirao}(2014)}]{doi:10.1063/1.4866996}%
  \BibitemOpen
  \bibfield  {author} {\bibinfo {author} {\bibfnamefont {T.}~\bibnamefont {Tsuneda}}\ and\ \bibinfo {author} {\bibfnamefont {K.}~\bibnamefont {Hirao}},\ }\href {\doibase 10.1063/1.4866996} {\bibfield  {journal} {\bibinfo  {journal} {The Journal of Chemical Physics}\ }\textbf {\bibinfo {volume} {140}},\ \bibinfo {pages} {18A513} (\bibinfo {year} {2014})},\ \Eprint {http://arxiv.org/abs/https://doi.org/10.1063/1.4866996} {https://doi.org/10.1063/1.4866996} \BibitemShut {NoStop}%
\bibitem [{\citenamefont {Kulik}\ \emph {et~al.}(2006)\citenamefont {Kulik}, \citenamefont {Cococcioni}, \citenamefont {Scherlis},\ and\ \citenamefont {Marzari}}]{PhysRevLett.97.103001}%
  \BibitemOpen
  \bibfield  {author} {\bibinfo {author} {\bibfnamefont {H.~J.}\ \bibnamefont {Kulik}}, \bibinfo {author} {\bibfnamefont {M.}~\bibnamefont {Cococcioni}}, \bibinfo {author} {\bibfnamefont {D.~A.}\ \bibnamefont {Scherlis}}, \ and\ \bibinfo {author} {\bibfnamefont {N.}~\bibnamefont {Marzari}},\ }\href {\doibase 10.1103/PhysRevLett.97.103001} {\bibfield  {journal} {\bibinfo  {journal} {Phys. Rev. Lett.}\ }\textbf {\bibinfo {volume} {97}},\ \bibinfo {pages} {103001} (\bibinfo {year} {2006})}\BibitemShut {NoStop}%
\bibitem [{\citenamefont {Shick}\ \emph {et~al.}(1999)\citenamefont {Shick}, \citenamefont {Liechtenstein},\ and\ \citenamefont {Pickett}}]{PhysRevB.60.10763}%
  \BibitemOpen
  \bibfield  {author} {\bibinfo {author} {\bibfnamefont {A.~B.}\ \bibnamefont {Shick}}, \bibinfo {author} {\bibfnamefont {A.~I.}\ \bibnamefont {Liechtenstein}}, \ and\ \bibinfo {author} {\bibfnamefont {W.~E.}\ \bibnamefont {Pickett}},\ }\href {\doibase 10.1103/PhysRevB.60.10763} {\bibfield  {journal} {\bibinfo  {journal} {Phys. Rev. B}\ }\textbf {\bibinfo {volume} {60}},\ \bibinfo {pages} {10763} (\bibinfo {year} {1999})}\BibitemShut {NoStop}%
\bibitem [{\citenamefont {Tran}\ \emph {et~al.}(2006)\citenamefont {Tran}, \citenamefont {Blaha}, \citenamefont {Schwarz},\ and\ \citenamefont {Nov\'ak}}]{PhysRevB.74.155108}%
  \BibitemOpen
  \bibfield  {author} {\bibinfo {author} {\bibfnamefont {F.}~\bibnamefont {Tran}}, \bibinfo {author} {\bibfnamefont {P.}~\bibnamefont {Blaha}}, \bibinfo {author} {\bibfnamefont {K.}~\bibnamefont {Schwarz}}, \ and\ \bibinfo {author} {\bibfnamefont {P.}~\bibnamefont {Nov\'ak}},\ }\href {\doibase 10.1103/PhysRevB.74.155108} {\bibfield  {journal} {\bibinfo  {journal} {Phys. Rev. B}\ }\textbf {\bibinfo {volume} {74}},\ \bibinfo {pages} {155108} (\bibinfo {year} {2006})}\BibitemShut {NoStop}%
\bibitem [{\citenamefont {Petukhov}\ \emph {et~al.}(2003)\citenamefont {Petukhov}, \citenamefont {Mazin}, \citenamefont {Chioncel},\ and\ \citenamefont {Lichtenstein}}]{PhysRevB.67.153106}%
  \BibitemOpen
  \bibfield  {author} {\bibinfo {author} {\bibfnamefont {A.~G.}\ \bibnamefont {Petukhov}}, \bibinfo {author} {\bibfnamefont {I.~I.}\ \bibnamefont {Mazin}}, \bibinfo {author} {\bibfnamefont {L.}~\bibnamefont {Chioncel}}, \ and\ \bibinfo {author} {\bibfnamefont {A.~I.}\ \bibnamefont {Lichtenstein}},\ }\href {\doibase 10.1103/PhysRevB.67.153106} {\bibfield  {journal} {\bibinfo  {journal} {Phys. Rev. B}\ }\textbf {\bibinfo {volume} {67}},\ \bibinfo {pages} {153106} (\bibinfo {year} {2003})}\BibitemShut {NoStop}%
\bibitem [{\citenamefont {Pavarini}\ \emph {et~al.}(2004)\citenamefont {Pavarini}, \citenamefont {Biermann}, \citenamefont {Poteryaev}, \citenamefont {Lichtenstein}, \citenamefont {Georges},\ and\ \citenamefont {Andersen}}]{PhysRevLett.92.176403}%
  \BibitemOpen
  \bibfield  {author} {\bibinfo {author} {\bibfnamefont {E.}~\bibnamefont {Pavarini}}, \bibinfo {author} {\bibfnamefont {S.}~\bibnamefont {Biermann}}, \bibinfo {author} {\bibfnamefont {A.}~\bibnamefont {Poteryaev}}, \bibinfo {author} {\bibfnamefont {A.~I.}\ \bibnamefont {Lichtenstein}}, \bibinfo {author} {\bibfnamefont {A.}~\bibnamefont {Georges}}, \ and\ \bibinfo {author} {\bibfnamefont {O.~K.}\ \bibnamefont {Andersen}},\ }\href {\doibase 10.1103/PhysRevLett.92.176403} {\bibfield  {journal} {\bibinfo  {journal} {Phys. Rev. Lett.}\ }\textbf {\bibinfo {volume} {92}},\ \bibinfo {pages} {176403} (\bibinfo {year} {2004})}\BibitemShut {NoStop}%
\bibitem [{\citenamefont {Sakuma}\ \emph {et~al.}(2013)\citenamefont {Sakuma}, \citenamefont {Werner},\ and\ \citenamefont {Aryasetiawan}}]{PhysRevB.88.235110}%
  \BibitemOpen
  \bibfield  {author} {\bibinfo {author} {\bibfnamefont {R.}~\bibnamefont {Sakuma}}, \bibinfo {author} {\bibfnamefont {P.}~\bibnamefont {Werner}}, \ and\ \bibinfo {author} {\bibfnamefont {F.}~\bibnamefont {Aryasetiawan}},\ }\href {\doibase 10.1103/PhysRevB.88.235110} {\bibfield  {journal} {\bibinfo  {journal} {Phys. Rev. B}\ }\textbf {\bibinfo {volume} {88}},\ \bibinfo {pages} {235110} (\bibinfo {year} {2013})}\BibitemShut {NoStop}%
\bibitem [{\citenamefont {Gatti}\ and\ \citenamefont {Guzzo}(2013)}]{GattiGuzzo}%
  \BibitemOpen
  \bibfield  {author} {\bibinfo {author} {\bibfnamefont {M.}~\bibnamefont {Gatti}}\ and\ \bibinfo {author} {\bibfnamefont {M.}~\bibnamefont {Guzzo}},\ }\href {\doibase 10.1103/PhysRevB.87.155147} {\bibfield  {journal} {\bibinfo  {journal} {Phys. Rev. B}\ }\textbf {\bibinfo {volume} {87}},\ \bibinfo {pages} {155147} (\bibinfo {year} {2013})}\BibitemShut {NoStop}%
\bibitem [{\citenamefont {Tomczak}\ \emph {et~al.}(2012)\citenamefont {Tomczak}, \citenamefont {Casula}, \citenamefont {Miyake}, \citenamefont {Aryasetiawan},\ and\ \citenamefont {Biermann}}]{Tomczak_2012}%
  \BibitemOpen
  \bibfield  {author} {\bibinfo {author} {\bibfnamefont {J.~M.}\ \bibnamefont {Tomczak}}, \bibinfo {author} {\bibfnamefont {M.}~\bibnamefont {Casula}}, \bibinfo {author} {\bibfnamefont {T.}~\bibnamefont {Miyake}}, \bibinfo {author} {\bibfnamefont {F.}~\bibnamefont {Aryasetiawan}}, \ and\ \bibinfo {author} {\bibfnamefont {S.}~\bibnamefont {Biermann}},\ }\href {\doibase 10.1209/0295-5075/100/67001} {\bibfield  {journal} {\bibinfo  {journal} {{EPL} (Europhysics Letters)}\ }\textbf {\bibinfo {volume} {100}},\ \bibinfo {pages} {67001} (\bibinfo {year} {2012})}\BibitemShut {NoStop}%
\bibitem [{\citenamefont {Tomczak}\ \emph {et~al.}(2014)\citenamefont {Tomczak}, \citenamefont {Casula}, \citenamefont {Miyake},\ and\ \citenamefont {Biermann}}]{Tomczak_review}%
  \BibitemOpen
  \bibfield  {author} {\bibinfo {author} {\bibfnamefont {J.~M.}\ \bibnamefont {Tomczak}}, \bibinfo {author} {\bibfnamefont {M.}~\bibnamefont {Casula}}, \bibinfo {author} {\bibfnamefont {T.}~\bibnamefont {Miyake}}, \ and\ \bibinfo {author} {\bibfnamefont {S.}~\bibnamefont {Biermann}},\ }\href {\doibase 10.1103/PhysRevB.90.165138} {\bibfield  {journal} {\bibinfo  {journal} {Phys. Rev. B}\ }\textbf {\bibinfo {volume} {90}},\ \bibinfo {pages} {165138} (\bibinfo {year} {2014})}\BibitemShut {NoStop}%
\bibitem [{\citenamefont {Miyake}\ \emph {et~al.}(2013)\citenamefont {Miyake}, \citenamefont {Martins}, \citenamefont {Sakuma},\ and\ \citenamefont {Aryasetiawan}}]{PhysRevB.87.115110}%
  \BibitemOpen
  \bibfield  {author} {\bibinfo {author} {\bibfnamefont {T.}~\bibnamefont {Miyake}}, \bibinfo {author} {\bibfnamefont {C.}~\bibnamefont {Martins}}, \bibinfo {author} {\bibfnamefont {R.}~\bibnamefont {Sakuma}}, \ and\ \bibinfo {author} {\bibfnamefont {F.}~\bibnamefont {Aryasetiawan}},\ }\href {\doibase 10.1103/PhysRevB.87.115110} {\bibfield  {journal} {\bibinfo  {journal} {Phys. Rev. B}\ }\textbf {\bibinfo {volume} {87}},\ \bibinfo {pages} {115110} (\bibinfo {year} {2013})}\BibitemShut {NoStop}%
\bibitem [{\citenamefont {Biermann}\ \emph {et~al.}(2003)\citenamefont {Biermann}, \citenamefont {Aryasetiawan},\ and\ \citenamefont {Georges}}]{PhysRevLett.90.086402}%
  \BibitemOpen
  \bibfield  {author} {\bibinfo {author} {\bibfnamefont {S.}~\bibnamefont {Biermann}}, \bibinfo {author} {\bibfnamefont {F.}~\bibnamefont {Aryasetiawan}}, \ and\ \bibinfo {author} {\bibfnamefont {A.}~\bibnamefont {Georges}},\ }\href {\doibase 10.1103/PhysRevLett.90.086402} {\bibfield  {journal} {\bibinfo  {journal} {Phys. Rev. Lett.}\ }\textbf {\bibinfo {volume} {90}},\ \bibinfo {pages} {086402} (\bibinfo {year} {2003})}\BibitemShut {NoStop}%
\bibitem [{\citenamefont {Fujimori}\ \emph {et~al.}(1992)\citenamefont {Fujimori}, \citenamefont {Hase}, \citenamefont {Namatame}, \citenamefont {Fujishima}, \citenamefont {Tokura}, \citenamefont {Eisaki}, \citenamefont {Uchida}, \citenamefont {Takegahara},\ and\ \citenamefont {de~Groot}}]{PhysRevLett.69.1796}%
  \BibitemOpen
  \bibfield  {author} {\bibinfo {author} {\bibfnamefont {A.}~\bibnamefont {Fujimori}}, \bibinfo {author} {\bibfnamefont {I.}~\bibnamefont {Hase}}, \bibinfo {author} {\bibfnamefont {H.}~\bibnamefont {Namatame}}, \bibinfo {author} {\bibfnamefont {Y.}~\bibnamefont {Fujishima}}, \bibinfo {author} {\bibfnamefont {Y.}~\bibnamefont {Tokura}}, \bibinfo {author} {\bibfnamefont {H.}~\bibnamefont {Eisaki}}, \bibinfo {author} {\bibfnamefont {S.}~\bibnamefont {Uchida}}, \bibinfo {author} {\bibfnamefont {K.}~\bibnamefont {Takegahara}}, \ and\ \bibinfo {author} {\bibfnamefont {F.~M.~F.}\ \bibnamefont {de~Groot}},\ }\href {\doibase 10.1103/PhysRevLett.69.1796} {\bibfield  {journal} {\bibinfo  {journal} {Phys. Rev. Lett.}\ }\textbf {\bibinfo {volume} {69}},\ \bibinfo {pages} {1796} (\bibinfo {year} {1992})}\BibitemShut {NoStop}%
\bibitem [{\citenamefont {Morikawa}\ \emph {et~al.}(1995)\citenamefont {Morikawa}, \citenamefont {Mizokawa}, \citenamefont {Kobayashi}, \citenamefont {Fujimori}, \citenamefont {Eisaki}, \citenamefont {Uchida}, \citenamefont {Iga},\ and\ \citenamefont {Nishihara}}]{PhysRevB.52.13711}%
  \BibitemOpen
  \bibfield  {author} {\bibinfo {author} {\bibfnamefont {K.}~\bibnamefont {Morikawa}}, \bibinfo {author} {\bibfnamefont {T.}~\bibnamefont {Mizokawa}}, \bibinfo {author} {\bibfnamefont {K.}~\bibnamefont {Kobayashi}}, \bibinfo {author} {\bibfnamefont {A.}~\bibnamefont {Fujimori}}, \bibinfo {author} {\bibfnamefont {H.}~\bibnamefont {Eisaki}}, \bibinfo {author} {\bibfnamefont {S.}~\bibnamefont {Uchida}}, \bibinfo {author} {\bibfnamefont {F.}~\bibnamefont {Iga}}, \ and\ \bibinfo {author} {\bibfnamefont {Y.}~\bibnamefont {Nishihara}},\ }\href {\doibase 10.1103/PhysRevB.52.13711} {\bibfield  {journal} {\bibinfo  {journal} {Phys. Rev. B}\ }\textbf {\bibinfo {volume} {52}},\ \bibinfo {pages} {13711} (\bibinfo {year} {1995})}\BibitemShut {NoStop}%
\bibitem [{\citenamefont {Rozenberg}\ \emph {et~al.}(1996)\citenamefont {Rozenberg}, \citenamefont {Inoue}, \citenamefont {Makino}, \citenamefont {Iga},\ and\ \citenamefont {Nishihara}}]{PhysRevLett.76.4781}%
  \BibitemOpen
  \bibfield  {author} {\bibinfo {author} {\bibfnamefont {M.~J.}\ \bibnamefont {Rozenberg}}, \bibinfo {author} {\bibfnamefont {I.~H.}\ \bibnamefont {Inoue}}, \bibinfo {author} {\bibfnamefont {H.}~\bibnamefont {Makino}}, \bibinfo {author} {\bibfnamefont {F.}~\bibnamefont {Iga}}, \ and\ \bibinfo {author} {\bibfnamefont {Y.}~\bibnamefont {Nishihara}},\ }\href {\doibase 10.1103/PhysRevLett.76.4781} {\bibfield  {journal} {\bibinfo  {journal} {Phys. Rev. Lett.}\ }\textbf {\bibinfo {volume} {76}},\ \bibinfo {pages} {4781} (\bibinfo {year} {1996})}\BibitemShut {NoStop}%
\bibitem [{\citenamefont {Nilsson}\ \emph {et~al.}(2017)\citenamefont {Nilsson}, \citenamefont {Boehnke}, \citenamefont {Werner},\ and\ \citenamefont {Aryasetiawan}}]{PhysRevMaterials.1.043803}%
  \BibitemOpen
  \bibfield  {author} {\bibinfo {author} {\bibfnamefont {F.}~\bibnamefont {Nilsson}}, \bibinfo {author} {\bibfnamefont {L.}~\bibnamefont {Boehnke}}, \bibinfo {author} {\bibfnamefont {P.}~\bibnamefont {Werner}}, \ and\ \bibinfo {author} {\bibfnamefont {F.}~\bibnamefont {Aryasetiawan}},\ }\href {\doibase 10.1103/PhysRevMaterials.1.043803} {\bibfield  {journal} {\bibinfo  {journal} {Phys. Rev. Materials}\ }\textbf {\bibinfo {volume} {1}},\ \bibinfo {pages} {043803} (\bibinfo {year} {2017})}\BibitemShut {NoStop}%
\bibitem [{\citenamefont {Boehnke}\ \emph {et~al.}(2016)\citenamefont {Boehnke}, \citenamefont {Nilsson}, \citenamefont {Aryasetiawan},\ and\ \citenamefont {Werner}}]{PhysRevB.94.201106}%
  \BibitemOpen
  \bibfield  {author} {\bibinfo {author} {\bibfnamefont {L.}~\bibnamefont {Boehnke}}, \bibinfo {author} {\bibfnamefont {F.}~\bibnamefont {Nilsson}}, \bibinfo {author} {\bibfnamefont {F.}~\bibnamefont {Aryasetiawan}}, \ and\ \bibinfo {author} {\bibfnamefont {P.}~\bibnamefont {Werner}},\ }\href {\doibase 10.1103/PhysRevB.94.201106} {\bibfield  {journal} {\bibinfo  {journal} {Phys. Rev. B}\ }\textbf {\bibinfo {volume} {94}},\ \bibinfo {pages} {201106} (\bibinfo {year} {2016})}\BibitemShut {NoStop}%
\bibitem [{\citenamefont {Fiorentini}\ and\ \citenamefont {Baldereschi}(1995)}]{Fiorentini}%
  \BibitemOpen
  \bibfield  {author} {\bibinfo {author} {\bibfnamefont {V.}~\bibnamefont {Fiorentini}}\ and\ \bibinfo {author} {\bibfnamefont {A.}~\bibnamefont {Baldereschi}},\ }\href {\doibase 10.1103/PhysRevB.51.17196} {\bibfield  {journal} {\bibinfo  {journal} {Phys. Rev. B}\ }\textbf {\bibinfo {volume} {51}},\ \bibinfo {pages} {17196} (\bibinfo {year} {1995})}\BibitemShut {NoStop}%
\bibitem [{\citenamefont {Gygi}\ and\ \citenamefont {Baldereschi}(1989)}]{Gygi}%
  \BibitemOpen
  \bibfield  {author} {\bibinfo {author} {\bibfnamefont {F.}~\bibnamefont {Gygi}}\ and\ \bibinfo {author} {\bibfnamefont {A.}~\bibnamefont {Baldereschi}},\ }\href {\doibase 10.1103/PhysRevLett.62.2160} {\bibfield  {journal} {\bibinfo  {journal} {Phys. Rev. Lett.}\ }\textbf {\bibinfo {volume} {62}},\ \bibinfo {pages} {2160} (\bibinfo {year} {1989})}\BibitemShut {NoStop}%
\bibitem [{\citenamefont {Massidda}\ \emph {et~al.}(1995)\citenamefont {Massidda}, \citenamefont {Continenza}, \citenamefont {Posternak},\ and\ \citenamefont {Baldereschi}}]{PhysRevLett.74.2323}%
  \BibitemOpen
  \bibfield  {author} {\bibinfo {author} {\bibfnamefont {S.}~\bibnamefont {Massidda}}, \bibinfo {author} {\bibfnamefont {A.}~\bibnamefont {Continenza}}, \bibinfo {author} {\bibfnamefont {M.}~\bibnamefont {Posternak}}, \ and\ \bibinfo {author} {\bibfnamefont {A.}~\bibnamefont {Baldereschi}},\ }\href {\doibase 10.1103/PhysRevLett.74.2323} {\bibfield  {journal} {\bibinfo  {journal} {Phys. Rev. Lett.}\ }\textbf {\bibinfo {volume} {74}},\ \bibinfo {pages} {2323} (\bibinfo {year} {1995})}\BibitemShut {NoStop}%
\bibitem [{\citenamefont {Anisimov}\ \emph {et~al.}(1997)\citenamefont {Anisimov}, \citenamefont {Aryasetiawan},\ and\ \citenamefont {Lichtenstein}}]{Anisimov_1997}%
  \BibitemOpen
  \bibfield  {author} {\bibinfo {author} {\bibfnamefont {V.~I.}\ \bibnamefont {Anisimov}}, \bibinfo {author} {\bibfnamefont {F.}~\bibnamefont {Aryasetiawan}}, \ and\ \bibinfo {author} {\bibfnamefont {A.~I.}\ \bibnamefont {Lichtenstein}},\ }\href {\doibase 10.1088/0953-8984/9/4/002} {\bibfield  {journal} {\bibinfo  {journal} {Journal of Physics: Condensed Matter}\ }\textbf {\bibinfo {volume} {9}},\ \bibinfo {pages} {767} (\bibinfo {year} {1997})}\BibitemShut {NoStop}%
\bibitem [{\citenamefont {Jiang}\ \emph {et~al.}(2010)\citenamefont {Jiang}, \citenamefont {Gomez-Abal}, \citenamefont {Rinke},\ and\ \citenamefont {Scheffler}}]{PhysRevB.82.045108}%
  \BibitemOpen
  \bibfield  {author} {\bibinfo {author} {\bibfnamefont {H.}~\bibnamefont {Jiang}}, \bibinfo {author} {\bibfnamefont {R.~I.}\ \bibnamefont {Gomez-Abal}}, \bibinfo {author} {\bibfnamefont {P.}~\bibnamefont {Rinke}}, \ and\ \bibinfo {author} {\bibfnamefont {M.}~\bibnamefont {Scheffler}},\ }\href {\doibase 10.1103/PhysRevB.82.045108} {\bibfield  {journal} {\bibinfo  {journal} {Phys. Rev. B}\ }\textbf {\bibinfo {volume} {82}},\ \bibinfo {pages} {045108} (\bibinfo {year} {2010})}\BibitemShut {NoStop}%
\bibitem [{\citenamefont {Nohara}\ \emph {et~al.}(2009)\citenamefont {Nohara}, \citenamefont {Yamamoto},\ and\ \citenamefont {Fujiwara}}]{PhysRevB.79.195110}%
  \BibitemOpen
  \bibfield  {author} {\bibinfo {author} {\bibfnamefont {Y.}~\bibnamefont {Nohara}}, \bibinfo {author} {\bibfnamefont {S.}~\bibnamefont {Yamamoto}}, \ and\ \bibinfo {author} {\bibfnamefont {T.}~\bibnamefont {Fujiwara}},\ }\href {\doibase 10.1103/PhysRevB.79.195110} {\bibfield  {journal} {\bibinfo  {journal} {Phys. Rev. B}\ }\textbf {\bibinfo {volume} {79}},\ \bibinfo {pages} {195110} (\bibinfo {year} {2009})}\BibitemShut {NoStop}%
\bibitem [{\citenamefont {Huang}\ and\ \citenamefont {Wang}(2012)}]{Huang_2012}%
  \BibitemOpen
  \bibfield  {author} {\bibinfo {author} {\bibfnamefont {L.}~\bibnamefont {Huang}}\ and\ \bibinfo {author} {\bibfnamefont {Y.}~\bibnamefont {Wang}},\ }\href {\doibase 10.1209/0295-5075/99/67003} {\bibfield  {journal} {\bibinfo  {journal} {{EPL} (Europhysics Letters)}\ }\textbf {\bibinfo {volume} {99}},\ \bibinfo {pages} {67003} (\bibinfo {year} {2012})}\BibitemShut {NoStop}%
\bibitem [{\citenamefont {Casula}\ \emph {et~al.}(2012{\natexlab{a}})\citenamefont {Casula}, \citenamefont {Rubtsov},\ and\ \citenamefont {Biermann}}]{PhysRevB.85.035115}%
  \BibitemOpen
  \bibfield  {author} {\bibinfo {author} {\bibfnamefont {M.}~\bibnamefont {Casula}}, \bibinfo {author} {\bibfnamefont {A.}~\bibnamefont {Rubtsov}}, \ and\ \bibinfo {author} {\bibfnamefont {S.}~\bibnamefont {Biermann}},\ }\href {\doibase 10.1103/PhysRevB.85.035115} {\bibfield  {journal} {\bibinfo  {journal} {Phys. Rev. B}\ }\textbf {\bibinfo {volume} {85}},\ \bibinfo {pages} {035115} (\bibinfo {year} {2012}{\natexlab{a}})}\BibitemShut {NoStop}%
\bibitem [{\citenamefont {Reining}(2018)}]{luciagw}%
  \BibitemOpen
  \bibfield  {author} {\bibinfo {author} {\bibfnamefont {L.}~\bibnamefont {Reining}},\ }\href {\doibase 10.1002/wcms.1344} {\bibfield  {journal} {\bibinfo  {journal} {WIREs Computational Molecular Science}\ }\textbf {\bibinfo {volume} {8}},\ \bibinfo {pages} {e1344} (\bibinfo {year} {2018})},\ \Eprint {http://arxiv.org/abs/https://onlinelibrary.wiley.com/doi/pdf/10.1002/wcms.1344} {https://onlinelibrary.wiley.com/doi/pdf/10.1002/wcms.1344} \BibitemShut {NoStop}%
\bibitem [{\citenamefont {Casula}\ \emph {et~al.}(2012{\natexlab{b}})\citenamefont {Casula}, \citenamefont {Werner}, \citenamefont {Vaugier}, \citenamefont {Aryasetiawan}, \citenamefont {Miyake}, \citenamefont {Millis},\ and\ \citenamefont {Biermann}}]{PhysRevLett.109.126408}%
  \BibitemOpen
  \bibfield  {author} {\bibinfo {author} {\bibfnamefont {M.}~\bibnamefont {Casula}}, \bibinfo {author} {\bibfnamefont {P.}~\bibnamefont {Werner}}, \bibinfo {author} {\bibfnamefont {L.}~\bibnamefont {Vaugier}}, \bibinfo {author} {\bibfnamefont {F.}~\bibnamefont {Aryasetiawan}}, \bibinfo {author} {\bibfnamefont {T.}~\bibnamefont {Miyake}}, \bibinfo {author} {\bibfnamefont {A.~J.}\ \bibnamefont {Millis}}, \ and\ \bibinfo {author} {\bibfnamefont {S.}~\bibnamefont {Biermann}},\ }\href {\doibase 10.1103/PhysRevLett.109.126408} {\bibfield  {journal} {\bibinfo  {journal} {Phys. Rev. Lett.}\ }\textbf {\bibinfo {volume} {109}},\ \bibinfo {pages} {126408} (\bibinfo {year} {2012}{\natexlab{b}})}\BibitemShut {NoStop}%
\bibitem [{\citenamefont {Biermann}(2014)}]{0953-8984-26-17-173202}%
  \BibitemOpen
  \bibfield  {author} {\bibinfo {author} {\bibfnamefont {S.}~\bibnamefont {Biermann}},\ }\href {http://stacks.iop.org/0953-8984/26/i=17/a=173202} {\bibfield  {journal} {\bibinfo  {journal} {Journal of Physics: Condensed Matter}\ }\textbf {\bibinfo {volume} {26}},\ \bibinfo {pages} {173202} (\bibinfo {year} {2014})}\BibitemShut {NoStop}%
\bibitem [{\citenamefont {Hedin}(1999)}]{Hedin_1999}%
  \BibitemOpen
  \bibfield  {author} {\bibinfo {author} {\bibfnamefont {L.}~\bibnamefont {Hedin}},\ }\href {\doibase 10.1088/0953-8984/11/42/201} {\bibfield  {journal} {\bibinfo  {journal} {Journal of Physics: Condensed Matter}\ }\textbf {\bibinfo {volume} {11}},\ \bibinfo {pages} {R489} (\bibinfo {year} {1999})}\BibitemShut {NoStop}%
\bibitem [{\citenamefont {Gygi}\ and\ \citenamefont {Baldereschi}(1986)}]{PhysRevB.34.4405}%
  \BibitemOpen
  \bibfield  {author} {\bibinfo {author} {\bibfnamefont {F.}~\bibnamefont {Gygi}}\ and\ \bibinfo {author} {\bibfnamefont {A.}~\bibnamefont {Baldereschi}},\ }\href {\doibase 10.1103/PhysRevB.34.4405} {\bibfield  {journal} {\bibinfo  {journal} {Phys. Rev. B}\ }\textbf {\bibinfo {volume} {34}},\ \bibinfo {pages} {4405} (\bibinfo {year} {1986})}\BibitemShut {NoStop}%
\bibitem [{\citenamefont {Marzari}\ \emph {et~al.}(2012)\citenamefont {Marzari}, \citenamefont {Mostofi}, \citenamefont {Yates}, \citenamefont {Souza},\ and\ \citenamefont {Vanderbilt}}]{Marzari_review}%
  \BibitemOpen
  \bibfield  {author} {\bibinfo {author} {\bibfnamefont {N.}~\bibnamefont {Marzari}}, \bibinfo {author} {\bibfnamefont {A.~A.}\ \bibnamefont {Mostofi}}, \bibinfo {author} {\bibfnamefont {J.~R.}\ \bibnamefont {Yates}}, \bibinfo {author} {\bibfnamefont {I.}~\bibnamefont {Souza}}, \ and\ \bibinfo {author} {\bibfnamefont {D.}~\bibnamefont {Vanderbilt}},\ }\href {\doibase 10.1103/RevModPhys.84.1419} {\bibfield  {journal} {\bibinfo  {journal} {Rev. Mod. Phys.}\ }\textbf {\bibinfo {volume} {84}},\ \bibinfo {pages} {1419} (\bibinfo {year} {2012})}\BibitemShut {NoStop}%
\bibitem [{\citenamefont {Amadon}\ \emph {et~al.}(2008)\citenamefont {Amadon}, \citenamefont {Lechermann}, \citenamefont {Georges}, \citenamefont {Jollet}, \citenamefont {Wehling},\ and\ \citenamefont {Lichtenstein}}]{Amadon_2008}%
  \BibitemOpen
  \bibfield  {author} {\bibinfo {author} {\bibfnamefont {B.}~\bibnamefont {Amadon}}, \bibinfo {author} {\bibfnamefont {F.}~\bibnamefont {Lechermann}}, \bibinfo {author} {\bibfnamefont {A.}~\bibnamefont {Georges}}, \bibinfo {author} {\bibfnamefont {F.}~\bibnamefont {Jollet}}, \bibinfo {author} {\bibfnamefont {T.~O.}\ \bibnamefont {Wehling}}, \ and\ \bibinfo {author} {\bibfnamefont {A.~I.}\ \bibnamefont {Lichtenstein}},\ }\href {\doibase 10.1103/PhysRevB.77.205112} {\bibfield  {journal} {\bibinfo  {journal} {Phys. Rev. B}\ }\textbf {\bibinfo {volume} {77}},\ \bibinfo {pages} {205112} (\bibinfo {year} {2008})}\BibitemShut {NoStop}%
\bibitem [{\citenamefont {Ehrenreich}(1966)}]{Ehrenreich}%
  \BibitemOpen
  \bibfield  {author} {\bibinfo {author} {\bibfnamefont {H.}~\bibnamefont {Ehrenreich}},\ }in\ \href@noop {} {\emph {\bibinfo {booktitle} {The Optical Properties of Solids: Proceedings of the International School of Physics ``Enrico Fermi'', held June 28 - July 10, 1965, in Varenna on Lake Como. Course XXXIV, sponsored by the Italian Physical Society. Edited by J. Tauc.}}}\ (\bibinfo  {publisher} {Academic Press, New York},\ \bibinfo {year} {1966})\BibitemShut {NoStop}%
\bibitem [{\citenamefont {Amadon}\ \emph {et~al.}(2014)\citenamefont {Amadon}, \citenamefont {Applencourt},\ and\ \citenamefont {Bruneval}}]{Amadon_2014}%
  \BibitemOpen
  \bibfield  {author} {\bibinfo {author} {\bibfnamefont {B.}~\bibnamefont {Amadon}}, \bibinfo {author} {\bibfnamefont {T.}~\bibnamefont {Applencourt}}, \ and\ \bibinfo {author} {\bibfnamefont {F.}~\bibnamefont {Bruneval}},\ }\href {\doibase 10.1103/PhysRevB.89.125110} {\bibfield  {journal} {\bibinfo  {journal} {Phys. Rev. B}\ }\textbf {\bibinfo {volume} {89}},\ \bibinfo {pages} {125110} (\bibinfo {year} {2014})}\BibitemShut {NoStop}%
\bibitem [{\citenamefont {Wang}\ \emph {et~al.}(2012)\citenamefont {Wang}, \citenamefont {Han}, \citenamefont {de' Medici}, \citenamefont {Park}, \citenamefont {Marianetti},\ and\ \citenamefont {Millis}}]{PhysRevB.86.195136}%
  \BibitemOpen
  \bibfield  {author} {\bibinfo {author} {\bibfnamefont {X.}~\bibnamefont {Wang}}, \bibinfo {author} {\bibfnamefont {M.~J.}\ \bibnamefont {Han}}, \bibinfo {author} {\bibfnamefont {L.}~\bibnamefont {de' Medici}}, \bibinfo {author} {\bibfnamefont {H.}~\bibnamefont {Park}}, \bibinfo {author} {\bibfnamefont {C.~A.}\ \bibnamefont {Marianetti}}, \ and\ \bibinfo {author} {\bibfnamefont {A.~J.}\ \bibnamefont {Millis}},\ }\href {\doibase 10.1103/PhysRevB.86.195136} {\bibfield  {journal} {\bibinfo  {journal} {Phys. Rev. B}\ }\textbf {\bibinfo {volume} {86}},\ \bibinfo {pages} {195136} (\bibinfo {year} {2012})}\BibitemShut {NoStop}%
\bibitem [{\citenamefont {Rohringer}\ \emph {et~al.}(2018)\citenamefont {Rohringer}, \citenamefont {Hafermann}, \citenamefont {Toschi}, \citenamefont {Katanin}, \citenamefont {Antipov}, \citenamefont {Katsnelson}, \citenamefont {Lichtenstein}, \citenamefont {Rubtsov},\ and\ \citenamefont {Held}}]{RevModPhys.90.025003}%
  \BibitemOpen
  \bibfield  {author} {\bibinfo {author} {\bibfnamefont {G.}~\bibnamefont {Rohringer}}, \bibinfo {author} {\bibfnamefont {H.}~\bibnamefont {Hafermann}}, \bibinfo {author} {\bibfnamefont {A.}~\bibnamefont {Toschi}}, \bibinfo {author} {\bibfnamefont {A.~A.}\ \bibnamefont {Katanin}}, \bibinfo {author} {\bibfnamefont {A.~E.}\ \bibnamefont {Antipov}}, \bibinfo {author} {\bibfnamefont {M.~I.}\ \bibnamefont {Katsnelson}}, \bibinfo {author} {\bibfnamefont {A.~I.}\ \bibnamefont {Lichtenstein}}, \bibinfo {author} {\bibfnamefont {A.~N.}\ \bibnamefont {Rubtsov}}, \ and\ \bibinfo {author} {\bibfnamefont {K.}~\bibnamefont {Held}},\ }\href {\doibase 10.1103/RevModPhys.90.025003} {\bibfield  {journal} {\bibinfo  {journal} {Rev. Mod. Phys.}\ }\textbf {\bibinfo {volume} {90}},\ \bibinfo {pages} {025003} (\bibinfo {year} {2018})}\BibitemShut {NoStop}%
\bibitem [{\citenamefont {Belozerov}\ \emph {et~al.}(2012)\citenamefont {Belozerov}, \citenamefont {Korotin}, \citenamefont {Anisimov},\ and\ \citenamefont {Poteryaev}}]{PhysRevB.85.045109}%
  \BibitemOpen
  \bibfield  {author} {\bibinfo {author} {\bibfnamefont {A.~S.}\ \bibnamefont {Belozerov}}, \bibinfo {author} {\bibfnamefont {M.~A.}\ \bibnamefont {Korotin}}, \bibinfo {author} {\bibfnamefont {V.~I.}\ \bibnamefont {Anisimov}}, \ and\ \bibinfo {author} {\bibfnamefont {A.~I.}\ \bibnamefont {Poteryaev}},\ }\href {\doibase 10.1103/PhysRevB.85.045109} {\bibfield  {journal} {\bibinfo  {journal} {Phys. Rev. B}\ }\textbf {\bibinfo {volume} {85}},\ \bibinfo {pages} {045109} (\bibinfo {year} {2012})}\BibitemShut {NoStop}%
\bibitem [{\citenamefont {van Roekeghem}\ \emph {et~al.}(2014)\citenamefont {van Roekeghem}, \citenamefont {Ayral}, \citenamefont {Tomczak}, \citenamefont {Casula}, \citenamefont {Xu}, \citenamefont {Ding}, \citenamefont {Ferrero}, \citenamefont {Parcollet}, \citenamefont {Jiang},\ and\ \citenamefont {Biermann}}]{PhysRevLett.113.266403}%
  \BibitemOpen
  \bibfield  {author} {\bibinfo {author} {\bibfnamefont {A.}~\bibnamefont {van Roekeghem}}, \bibinfo {author} {\bibfnamefont {T.}~\bibnamefont {Ayral}}, \bibinfo {author} {\bibfnamefont {J.~M.}\ \bibnamefont {Tomczak}}, \bibinfo {author} {\bibfnamefont {M.}~\bibnamefont {Casula}}, \bibinfo {author} {\bibfnamefont {N.}~\bibnamefont {Xu}}, \bibinfo {author} {\bibfnamefont {H.}~\bibnamefont {Ding}}, \bibinfo {author} {\bibfnamefont {M.}~\bibnamefont {Ferrero}}, \bibinfo {author} {\bibfnamefont {O.}~\bibnamefont {Parcollet}}, \bibinfo {author} {\bibfnamefont {H.}~\bibnamefont {Jiang}}, \ and\ \bibinfo {author} {\bibfnamefont {S.}~\bibnamefont {Biermann}},\ }\href {\doibase 10.1103/PhysRevLett.113.266403} {\bibfield  {journal} {\bibinfo  {journal} {Phys. Rev. Lett.}\ }\textbf {\bibinfo {volume} {113}},\ \bibinfo {pages} {266403} (\bibinfo {year} {2014})}\BibitemShut {NoStop}%
\bibitem [{\citenamefont {van Roekeghem}\ and\ \citenamefont {Biermann}(2014)}]{van_Roekeghem_2014}%
  \BibitemOpen
  \bibfield  {author} {\bibinfo {author} {\bibfnamefont {A.}~\bibnamefont {van Roekeghem}}\ and\ \bibinfo {author} {\bibfnamefont {S.}~\bibnamefont {Biermann}},\ }\href {\doibase 10.1209/0295-5075/108/57003} {\bibfield  {journal} {\bibinfo  {journal} {{EPL} (Europhysics Letters)}\ }\textbf {\bibinfo {volume} {108}},\ \bibinfo {pages} {57003} (\bibinfo {year} {2014})}\BibitemShut {NoStop}%
\bibitem [{\citenamefont {Colonna}\ \emph {et~al.}(2019)\citenamefont {Colonna}, \citenamefont {Nguyen}, \citenamefont {Ferretti},\ and\ \citenamefont {Marzari}}]{NicolaGW100}%
  \BibitemOpen
  \bibfield  {author} {\bibinfo {author} {\bibfnamefont {N.}~\bibnamefont {Colonna}}, \bibinfo {author} {\bibfnamefont {N.~L.}\ \bibnamefont {Nguyen}}, \bibinfo {author} {\bibfnamefont {A.}~\bibnamefont {Ferretti}}, \ and\ \bibinfo {author} {\bibfnamefont {N.}~\bibnamefont {Marzari}},\ }\href {\doibase 10.1021/acs.jctc.8b00976} {\bibfield  {journal} {\bibinfo  {journal} {J. Chem. Theory Comput.}\ }\textbf {\bibinfo {volume} {15}},\ \bibinfo {pages} {1905} (\bibinfo {year} {2019})}\BibitemShut {NoStop}%
\bibitem [{\citenamefont {Farid}(2002)}]{Farid}%
  \BibitemOpen
  \bibfield  {author} {\bibinfo {author} {\bibfnamefont {B.}~\bibnamefont {Farid}},\ }\href {\doibase 10.1080/13642810208222682} {\bibfield  {journal} {\bibinfo  {journal} {Philosophical Magazine B}\ }\textbf {\bibinfo {volume} {82}},\ \bibinfo {pages} {1413} (\bibinfo {year} {2002})},\ \Eprint {http://arxiv.org/abs/https://doi.org/10.1080/13642810208222682} {https://doi.org/10.1080/13642810208222682} \BibitemShut {NoStop}%
\bibitem [{\citenamefont {Karolak}\ \emph {et~al.}(2010)\citenamefont {Karolak}, \citenamefont {Ulm}, \citenamefont {Wehling}, \citenamefont {Mazurenko}, \citenamefont {Poteryaev},\ and\ \citenamefont {Lichtenstein}}]{KAROLAK201011}%
  \BibitemOpen
  \bibfield  {author} {\bibinfo {author} {\bibfnamefont {M.}~\bibnamefont {Karolak}}, \bibinfo {author} {\bibfnamefont {G.}~\bibnamefont {Ulm}}, \bibinfo {author} {\bibfnamefont {T.}~\bibnamefont {Wehling}}, \bibinfo {author} {\bibfnamefont {V.}~\bibnamefont {Mazurenko}}, \bibinfo {author} {\bibfnamefont {A.}~\bibnamefont {Poteryaev}}, \ and\ \bibinfo {author} {\bibfnamefont {A.}~\bibnamefont {Lichtenstein}},\ }\href {\doibase https://doi.org/10.1016/j.elspec.2010.05.021} {\bibfield  {journal} {\bibinfo  {journal} {Journal of Electron Spectroscopy and Related Phenomena}\ }\textbf {\bibinfo {volume} {181}},\ \bibinfo {pages} {11 } (\bibinfo {year} {2010})},\ \bibinfo {note} {proceedings of International Workshop on Strong Correlations and Angle-Resolved Photoemission Spectroscopy 2009}\BibitemShut {NoStop}%
\bibitem [{\citenamefont {Ryee}\ and\ \citenamefont {Han}(2018)}]{Ryee}%
  \BibitemOpen
  \bibfield  {author} {\bibinfo {author} {\bibfnamefont {S.}~\bibnamefont {Ryee}}\ and\ \bibinfo {author} {\bibfnamefont {M.~J.}\ \bibnamefont {Han}},\ }\href {\doibase 10.1038/s41598-018-27731-4} {\bibfield  {journal} {\bibinfo  {journal} {Scientific Reports}\ }\textbf {\bibinfo {volume} {8}},\ \bibinfo {pages} {9559} (\bibinfo {year} {2018})}\BibitemShut {NoStop}%
\bibitem [{\citenamefont {Haule}(2015)}]{PhysRevLett.115.196403}%
  \BibitemOpen
  \bibfield  {author} {\bibinfo {author} {\bibfnamefont {K.}~\bibnamefont {Haule}},\ }\href {\doibase 10.1103/PhysRevLett.115.196403} {\bibfield  {journal} {\bibinfo  {journal} {Phys. Rev. Lett.}\ }\textbf {\bibinfo {volume} {115}},\ \bibinfo {pages} {196403} (\bibinfo {year} {2015})}\BibitemShut {NoStop}%
\bibitem [{\citenamefont {Deng}\ \emph {et~al.}(2012)\citenamefont {Deng}, \citenamefont {Ferrero}, \citenamefont {Mravlje}, \citenamefont {Aichhorn},\ and\ \citenamefont {Georges}}]{Deng2012}%
  \BibitemOpen
  \bibfield  {author} {\bibinfo {author} {\bibfnamefont {X.}~\bibnamefont {Deng}}, \bibinfo {author} {\bibfnamefont {M.}~\bibnamefont {Ferrero}}, \bibinfo {author} {\bibfnamefont {J.}~\bibnamefont {Mravlje}}, \bibinfo {author} {\bibfnamefont {M.}~\bibnamefont {Aichhorn}}, \ and\ \bibinfo {author} {\bibfnamefont {A.}~\bibnamefont {Georges}},\ }\href {\doibase 10.1103/PhysRevB.85.125137} {\bibfield  {journal} {\bibinfo  {journal} {Phys. Rev. B}\ }\textbf {\bibinfo {volume} {85}},\ \bibinfo {pages} {125137} (\bibinfo {year} {2012})}\BibitemShut {NoStop}%
\bibitem [{\citenamefont {Czy\ifmmode~\dot{z}\else \.{z}\fi{}yk}\ and\ \citenamefont {Sawatzky}(1994)}]{AMF}%
  \BibitemOpen
  \bibfield  {author} {\bibinfo {author} {\bibfnamefont {M.~T.}\ \bibnamefont {Czy\ifmmode~\dot{z}\else \.{z}\fi{}yk}}\ and\ \bibinfo {author} {\bibfnamefont {G.~A.}\ \bibnamefont {Sawatzky}},\ }\href {\doibase 10.1103/PhysRevB.49.14211} {\bibfield  {journal} {\bibinfo  {journal} {Phys. Rev. B}\ }\textbf {\bibinfo {volume} {49}},\ \bibinfo {pages} {14211} (\bibinfo {year} {1994})}\BibitemShut {NoStop}%
\bibitem [{\citenamefont {Luttinger}\ and\ \citenamefont {Ward}(1960)}]{PhysRev.118.1417}%
  \BibitemOpen
  \bibfield  {author} {\bibinfo {author} {\bibfnamefont {J.~M.}\ \bibnamefont {Luttinger}}\ and\ \bibinfo {author} {\bibfnamefont {J.~C.}\ \bibnamefont {Ward}},\ }\href {\doibase 10.1103/PhysRev.118.1417} {\bibfield  {journal} {\bibinfo  {journal} {Phys. Rev.}\ }\textbf {\bibinfo {volume} {118}},\ \bibinfo {pages} {1417} (\bibinfo {year} {1960})}\BibitemShut {NoStop}%
\bibitem [{\citenamefont {Takizawa}\ \emph {et~al.}(2009)\citenamefont {Takizawa}, \citenamefont {Minohara}, \citenamefont {Kumigashira}, \citenamefont {Toyota}, \citenamefont {Oshima}, \citenamefont {Wadati}, \citenamefont {Yoshida}, \citenamefont {Fujimori}, \citenamefont {Lippmaa}, \citenamefont {Kawasaki}, \citenamefont {Koinuma}, \citenamefont {Sordi},\ and\ \citenamefont {Rozenberg}}]{PhysRevB.80.235104}%
  \BibitemOpen
  \bibfield  {author} {\bibinfo {author} {\bibfnamefont {M.}~\bibnamefont {Takizawa}}, \bibinfo {author} {\bibfnamefont {M.}~\bibnamefont {Minohara}}, \bibinfo {author} {\bibfnamefont {H.}~\bibnamefont {Kumigashira}}, \bibinfo {author} {\bibfnamefont {D.}~\bibnamefont {Toyota}}, \bibinfo {author} {\bibfnamefont {M.}~\bibnamefont {Oshima}}, \bibinfo {author} {\bibfnamefont {H.}~\bibnamefont {Wadati}}, \bibinfo {author} {\bibfnamefont {T.}~\bibnamefont {Yoshida}}, \bibinfo {author} {\bibfnamefont {A.}~\bibnamefont {Fujimori}}, \bibinfo {author} {\bibfnamefont {M.}~\bibnamefont {Lippmaa}}, \bibinfo {author} {\bibfnamefont {M.}~\bibnamefont {Kawasaki}}, \bibinfo {author} {\bibfnamefont {H.}~\bibnamefont {Koinuma}}, \bibinfo {author} {\bibfnamefont {G.}~\bibnamefont {Sordi}}, \ and\ \bibinfo {author} {\bibfnamefont {M.}~\bibnamefont {Rozenberg}},\ }\href {\doibase 10.1103/PhysRevB.80.235104} {\bibfield  {journal} {\bibinfo  {journal} {Phys. Rev. B}\ }\textbf {\bibinfo {volume} {80}},\ \bibinfo {pages} {235104}
  (\bibinfo {year} {2009})}\BibitemShut {NoStop}%
\bibitem [{\citenamefont {Yoshida}\ \emph {et~al.}(2005)\citenamefont {Yoshida}, \citenamefont {Tanaka}, \citenamefont {Yagi}, \citenamefont {Ino}, \citenamefont {Eisaki}, \citenamefont {Fujimori},\ and\ \citenamefont {Shen}}]{PhysRevLett.95.146404}%
  \BibitemOpen
  \bibfield  {author} {\bibinfo {author} {\bibfnamefont {T.}~\bibnamefont {Yoshida}}, \bibinfo {author} {\bibfnamefont {K.}~\bibnamefont {Tanaka}}, \bibinfo {author} {\bibfnamefont {H.}~\bibnamefont {Yagi}}, \bibinfo {author} {\bibfnamefont {A.}~\bibnamefont {Ino}}, \bibinfo {author} {\bibfnamefont {H.}~\bibnamefont {Eisaki}}, \bibinfo {author} {\bibfnamefont {A.}~\bibnamefont {Fujimori}}, \ and\ \bibinfo {author} {\bibfnamefont {Z.-X.}\ \bibnamefont {Shen}},\ }\href {\doibase 10.1103/PhysRevLett.95.146404} {\bibfield  {journal} {\bibinfo  {journal} {Phys. Rev. Lett.}\ }\textbf {\bibinfo {volume} {95}},\ \bibinfo {pages} {146404} (\bibinfo {year} {2005})}\BibitemShut {NoStop}%
\bibitem [{\citenamefont {Backes}\ \emph {et~al.}(2016)\citenamefont {Backes}, \citenamefont {R\"odel}, \citenamefont {Fortuna}, \citenamefont {Frantzeskakis}, \citenamefont {Le~F\`evre}, \citenamefont {Bertran}, \citenamefont {Kobayashi}, \citenamefont {Yukawa}, \citenamefont {Mitsuhashi}, \citenamefont {Kitamura}, \citenamefont {Horiba}, \citenamefont {Kumigashira}, \citenamefont {Saint-Martin}, \citenamefont {Fouchet}, \citenamefont {Berini}, \citenamefont {Dumont}, \citenamefont {Kim}, \citenamefont {Lechermann}, \citenamefont {Jeschke}, \citenamefont {Rozenberg}, \citenamefont {Valent\'{\i}},\ and\ \citenamefont {Santander-Syro}}]{PhysRevB.94.241110}%
  \BibitemOpen
  \bibfield  {author} {\bibinfo {author} {\bibfnamefont {S.}~\bibnamefont {Backes}}, \bibinfo {author} {\bibfnamefont {T.~C.}\ \bibnamefont {R\"odel}}, \bibinfo {author} {\bibfnamefont {F.}~\bibnamefont {Fortuna}}, \bibinfo {author} {\bibfnamefont {E.}~\bibnamefont {Frantzeskakis}}, \bibinfo {author} {\bibfnamefont {P.}~\bibnamefont {Le~F\`evre}}, \bibinfo {author} {\bibfnamefont {F.}~\bibnamefont {Bertran}}, \bibinfo {author} {\bibfnamefont {M.}~\bibnamefont {Kobayashi}}, \bibinfo {author} {\bibfnamefont {R.}~\bibnamefont {Yukawa}}, \bibinfo {author} {\bibfnamefont {T.}~\bibnamefont {Mitsuhashi}}, \bibinfo {author} {\bibfnamefont {M.}~\bibnamefont {Kitamura}}, \bibinfo {author} {\bibfnamefont {K.}~\bibnamefont {Horiba}}, \bibinfo {author} {\bibfnamefont {H.}~\bibnamefont {Kumigashira}}, \bibinfo {author} {\bibfnamefont {R.}~\bibnamefont {Saint-Martin}}, \bibinfo {author} {\bibfnamefont {A.}~\bibnamefont {Fouchet}}, \bibinfo {author} {\bibfnamefont {B.}~\bibnamefont {Berini}}, \bibinfo {author} {\bibfnamefont
  {Y.}~\bibnamefont {Dumont}}, \bibinfo {author} {\bibfnamefont {A.~J.}\ \bibnamefont {Kim}}, \bibinfo {author} {\bibfnamefont {F.}~\bibnamefont {Lechermann}}, \bibinfo {author} {\bibfnamefont {H.~O.}\ \bibnamefont {Jeschke}}, \bibinfo {author} {\bibfnamefont {M.~J.}\ \bibnamefont {Rozenberg}}, \bibinfo {author} {\bibfnamefont {R.}~\bibnamefont {Valent\'{\i}}}, \ and\ \bibinfo {author} {\bibfnamefont {A.~F.}\ \bibnamefont {Santander-Syro}},\ }\href {\doibase 10.1103/PhysRevB.94.241110} {\bibfield  {journal} {\bibinfo  {journal} {Phys. Rev. B}\ }\textbf {\bibinfo {volume} {94}},\ \bibinfo {pages} {241110} (\bibinfo {year} {2016})}\BibitemShut {NoStop}%
\bibitem [{\citenamefont {Nowadnick}\ \emph {et~al.}(2015)\citenamefont {Nowadnick}, \citenamefont {Ruf}, \citenamefont {Park}, \citenamefont {King}, \citenamefont {Schlom}, \citenamefont {Shen},\ and\ \citenamefont {Millis}}]{PhysRevB.92.245109}%
  \BibitemOpen
  \bibfield  {author} {\bibinfo {author} {\bibfnamefont {E.~A.}\ \bibnamefont {Nowadnick}}, \bibinfo {author} {\bibfnamefont {J.~P.}\ \bibnamefont {Ruf}}, \bibinfo {author} {\bibfnamefont {H.}~\bibnamefont {Park}}, \bibinfo {author} {\bibfnamefont {P.~D.~C.}\ \bibnamefont {King}}, \bibinfo {author} {\bibfnamefont {D.~G.}\ \bibnamefont {Schlom}}, \bibinfo {author} {\bibfnamefont {K.~M.}\ \bibnamefont {Shen}}, \ and\ \bibinfo {author} {\bibfnamefont {A.~J.}\ \bibnamefont {Millis}},\ }\href {\doibase 10.1103/PhysRevB.92.245109} {\bibfield  {journal} {\bibinfo  {journal} {Phys. Rev. B}\ }\textbf {\bibinfo {volume} {92}},\ \bibinfo {pages} {245109} (\bibinfo {year} {2015})}\BibitemShut {NoStop}%
\bibitem [{\citenamefont {Maiti}\ \emph {et~al.}(1998)\citenamefont {Maiti}, \citenamefont {Mahadevan},\ and\ \citenamefont {Sarma}}]{PhysRevLett.80.2885}%
  \BibitemOpen
  \bibfield  {author} {\bibinfo {author} {\bibfnamefont {K.}~\bibnamefont {Maiti}}, \bibinfo {author} {\bibfnamefont {P.}~\bibnamefont {Mahadevan}}, \ and\ \bibinfo {author} {\bibfnamefont {D.~D.}\ \bibnamefont {Sarma}},\ }\href {\doibase 10.1103/PhysRevLett.80.2885} {\bibfield  {journal} {\bibinfo  {journal} {Phys. Rev. Lett.}\ }\textbf {\bibinfo {volume} {80}},\ \bibinfo {pages} {2885} (\bibinfo {year} {1998})}\BibitemShut {NoStop}%
\bibitem [{\citenamefont {Sekiyama}\ \emph {et~al.}(2004)\citenamefont {Sekiyama}, \citenamefont {Fujiwara}, \citenamefont {Imada}, \citenamefont {Suga}, \citenamefont {Eisaki}, \citenamefont {Uchida}, \citenamefont {Takegahara}, \citenamefont {Harima}, \citenamefont {Saitoh}, \citenamefont {Nekrasov}, \citenamefont {Keller}, \citenamefont {Kondakov}, \citenamefont {Kozhevnikov}, \citenamefont {Pruschke}, \citenamefont {Held}, \citenamefont {Vollhardt},\ and\ \citenamefont {Anisimov}}]{PhysRevLett.93.156402}%
  \BibitemOpen
  \bibfield  {author} {\bibinfo {author} {\bibfnamefont {A.}~\bibnamefont {Sekiyama}}, \bibinfo {author} {\bibfnamefont {H.}~\bibnamefont {Fujiwara}}, \bibinfo {author} {\bibfnamefont {S.}~\bibnamefont {Imada}}, \bibinfo {author} {\bibfnamefont {S.}~\bibnamefont {Suga}}, \bibinfo {author} {\bibfnamefont {H.}~\bibnamefont {Eisaki}}, \bibinfo {author} {\bibfnamefont {S.~I.}\ \bibnamefont {Uchida}}, \bibinfo {author} {\bibfnamefont {K.}~\bibnamefont {Takegahara}}, \bibinfo {author} {\bibfnamefont {H.}~\bibnamefont {Harima}}, \bibinfo {author} {\bibfnamefont {Y.}~\bibnamefont {Saitoh}}, \bibinfo {author} {\bibfnamefont {I.~A.}\ \bibnamefont {Nekrasov}}, \bibinfo {author} {\bibfnamefont {G.}~\bibnamefont {Keller}}, \bibinfo {author} {\bibfnamefont {D.~E.}\ \bibnamefont {Kondakov}}, \bibinfo {author} {\bibfnamefont {A.~V.}\ \bibnamefont {Kozhevnikov}}, \bibinfo {author} {\bibfnamefont {T.}~\bibnamefont {Pruschke}}, \bibinfo {author} {\bibfnamefont {K.}~\bibnamefont {Held}}, \bibinfo {author} {\bibfnamefont
  {D.}~\bibnamefont {Vollhardt}}, \ and\ \bibinfo {author} {\bibfnamefont {V.~I.}\ \bibnamefont {Anisimov}},\ }\href {\doibase 10.1103/PhysRevLett.93.156402} {\bibfield  {journal} {\bibinfo  {journal} {Phys. Rev. Lett.}\ }\textbf {\bibinfo {volume} {93}},\ \bibinfo {pages} {156402} (\bibinfo {year} {2004})}\BibitemShut {NoStop}%
\bibitem [{\citenamefont {Torrance}\ \emph {et~al.}(1992)\citenamefont {Torrance}, \citenamefont {Lacorre}, \citenamefont {Nazzal}, \citenamefont {Ansaldo},\ and\ \citenamefont {Niedermayer}}]{PhysRevB.45.8209}%
  \BibitemOpen
  \bibfield  {author} {\bibinfo {author} {\bibfnamefont {J.~B.}\ \bibnamefont {Torrance}}, \bibinfo {author} {\bibfnamefont {P.}~\bibnamefont {Lacorre}}, \bibinfo {author} {\bibfnamefont {A.~I.}\ \bibnamefont {Nazzal}}, \bibinfo {author} {\bibfnamefont {E.~J.}\ \bibnamefont {Ansaldo}}, \ and\ \bibinfo {author} {\bibfnamefont {C.}~\bibnamefont {Niedermayer}},\ }\href {\doibase 10.1103/PhysRevB.45.8209} {\bibfield  {journal} {\bibinfo  {journal} {Phys. Rev. B}\ }\textbf {\bibinfo {volume} {45}},\ \bibinfo {pages} {8209} (\bibinfo {year} {1992})}\BibitemShut {NoStop}%
\bibitem [{\citenamefont {Gou}\ \emph {et~al.}(2011)\citenamefont {Gou}, \citenamefont {Grinberg}, \citenamefont {Rappe},\ and\ \citenamefont {Rondinelli}}]{Gou2011}%
  \BibitemOpen
  \bibfield  {author} {\bibinfo {author} {\bibfnamefont {G.}~\bibnamefont {Gou}}, \bibinfo {author} {\bibfnamefont {I.}~\bibnamefont {Grinberg}}, \bibinfo {author} {\bibfnamefont {A.~M.}\ \bibnamefont {Rappe}}, \ and\ \bibinfo {author} {\bibfnamefont {J.~M.}\ \bibnamefont {Rondinelli}},\ }\href {\doibase 10.1103/PhysRevB.84.144101} {\bibfield  {journal} {\bibinfo  {journal} {Phys. Rev. B}\ }\textbf {\bibinfo {volume} {84}},\ \bibinfo {pages} {144101} (\bibinfo {year} {2011})}\BibitemShut {NoStop}%
\bibitem [{\citenamefont {Eguchi}\ \emph {et~al.}(2009)\citenamefont {Eguchi}, \citenamefont {Chainani}, \citenamefont {Taguchi}, \citenamefont {Matsunami}, \citenamefont {Ishida}, \citenamefont {Horiba}, \citenamefont {Senba}, \citenamefont {Ohashi},\ and\ \citenamefont {Shin}}]{Eguchi2009}%
  \BibitemOpen
  \bibfield  {author} {\bibinfo {author} {\bibfnamefont {R.}~\bibnamefont {Eguchi}}, \bibinfo {author} {\bibfnamefont {A.}~\bibnamefont {Chainani}}, \bibinfo {author} {\bibfnamefont {M.}~\bibnamefont {Taguchi}}, \bibinfo {author} {\bibfnamefont {M.}~\bibnamefont {Matsunami}}, \bibinfo {author} {\bibfnamefont {Y.}~\bibnamefont {Ishida}}, \bibinfo {author} {\bibfnamefont {K.}~\bibnamefont {Horiba}}, \bibinfo {author} {\bibfnamefont {Y.}~\bibnamefont {Senba}}, \bibinfo {author} {\bibfnamefont {H.}~\bibnamefont {Ohashi}}, \ and\ \bibinfo {author} {\bibfnamefont {S.}~\bibnamefont {Shin}},\ }\href {\doibase 10.1103/PhysRevB.79.115122} {\bibfield  {journal} {\bibinfo  {journal} {Phys. Rev. B}\ }\textbf {\bibinfo {volume} {79}},\ \bibinfo {pages} {115122} (\bibinfo {year} {2009})}\BibitemShut {NoStop}%
\bibitem [{\citenamefont {Horiba}\ \emph {et~al.}(2007)\citenamefont {Horiba}, \citenamefont {Eguchi}, \citenamefont {Taguchi}, \citenamefont {Chainani}, \citenamefont {Kikkawa}, \citenamefont {Senba}, \citenamefont {Ohashi},\ and\ \citenamefont {Shin}}]{Horiba2007}%
  \BibitemOpen
  \bibfield  {author} {\bibinfo {author} {\bibfnamefont {K.}~\bibnamefont {Horiba}}, \bibinfo {author} {\bibfnamefont {R.}~\bibnamefont {Eguchi}}, \bibinfo {author} {\bibfnamefont {M.}~\bibnamefont {Taguchi}}, \bibinfo {author} {\bibfnamefont {A.}~\bibnamefont {Chainani}}, \bibinfo {author} {\bibfnamefont {A.}~\bibnamefont {Kikkawa}}, \bibinfo {author} {\bibfnamefont {Y.}~\bibnamefont {Senba}}, \bibinfo {author} {\bibfnamefont {H.}~\bibnamefont {Ohashi}}, \ and\ \bibinfo {author} {\bibfnamefont {S.}~\bibnamefont {Shin}},\ }\href {\doibase 10.1103/PhysRevB.76.155104} {\bibfield  {journal} {\bibinfo  {journal} {Phys. Rev. B}\ }\textbf {\bibinfo {volume} {76}},\ \bibinfo {pages} {155104} (\bibinfo {year} {2007})}\BibitemShut {NoStop}%
\bibitem [{\citenamefont {King}\ \emph {et~al.}(2014)\citenamefont {King}, \citenamefont {Wei}, \citenamefont {Nie}, \citenamefont {Uchida}, \citenamefont {Adamo}, \citenamefont {Zhu}, \citenamefont {He}, \citenamefont {Bozovic}, \citenamefont {Schlom},\ and\ \citenamefont {Shen}}]{KingNature}%
  \BibitemOpen
  \bibfield  {author} {\bibinfo {author} {\bibfnamefont {P.~D.~C.}\ \bibnamefont {King}}, \bibinfo {author} {\bibfnamefont {H.~I.}\ \bibnamefont {Wei}}, \bibinfo {author} {\bibfnamefont {Y.~F.}\ \bibnamefont {Nie}}, \bibinfo {author} {\bibfnamefont {M.}~\bibnamefont {Uchida}}, \bibinfo {author} {\bibfnamefont {C.}~\bibnamefont {Adamo}}, \bibinfo {author} {\bibfnamefont {S.}~\bibnamefont {Zhu}}, \bibinfo {author} {\bibfnamefont {X.}~\bibnamefont {He}}, \bibinfo {author} {\bibfnamefont {I.}~\bibnamefont {Bozovic}}, \bibinfo {author} {\bibfnamefont {D.~G.}\ \bibnamefont {Schlom}}, \ and\ \bibinfo {author} {\bibfnamefont {K.~M.}\ \bibnamefont {Shen}},\ }\href {\doibase 10.1038/nnano.2014.59} {\bibfield  {journal} {\bibinfo  {journal} {Nature Nanotechnology}\ }\textbf {\bibinfo {volume} {9}},\ \bibinfo {pages} {443} (\bibinfo {year} {2014})}\BibitemShut {NoStop}%
\bibitem [{\citenamefont {Perdew}\ \emph {et~al.}(1996)\citenamefont {Perdew}, \citenamefont {Burke},\ and\ \citenamefont {Ernzerhof}}]{PhysRevLett.77.3865}%
  \BibitemOpen
  \bibfield  {author} {\bibinfo {author} {\bibfnamefont {J.~P.}\ \bibnamefont {Perdew}}, \bibinfo {author} {\bibfnamefont {K.}~\bibnamefont {Burke}}, \ and\ \bibinfo {author} {\bibfnamefont {M.}~\bibnamefont {Ernzerhof}},\ }\href {\doibase 10.1103/PhysRevLett.77.3865} {\bibfield  {journal} {\bibinfo  {journal} {Phys. Rev. Lett.}\ }\textbf {\bibinfo {volume} {77}},\ \bibinfo {pages} {3865} (\bibinfo {year} {1996})}\BibitemShut {NoStop}%
\bibitem [{\citenamefont {Guan}\ \emph {et~al.}(2010)\citenamefont {Guan}, \citenamefont {Liu}, \citenamefont {Jin}, \citenamefont {Guo}, \citenamefont {Zhao}, \citenamefont {Wang},\ and\ \citenamefont {Fu}}]{GUAN20102011}%
  \BibitemOpen
  \bibfield  {author} {\bibinfo {author} {\bibfnamefont {L.}~\bibnamefont {Guan}}, \bibinfo {author} {\bibfnamefont {B.}~\bibnamefont {Liu}}, \bibinfo {author} {\bibfnamefont {L.}~\bibnamefont {Jin}}, \bibinfo {author} {\bibfnamefont {J.}~\bibnamefont {Guo}}, \bibinfo {author} {\bibfnamefont {Q.}~\bibnamefont {Zhao}}, \bibinfo {author} {\bibfnamefont {Y.}~\bibnamefont {Wang}}, \ and\ \bibinfo {author} {\bibfnamefont {G.}~\bibnamefont {Fu}},\ }\href {\doibase https://doi.org/10.1016/j.ssc.2010.08.016} {\bibfield  {journal} {\bibinfo  {journal} {Solid State Communications}\ }\textbf {\bibinfo {volume} {150}},\ \bibinfo {pages} {2011 } (\bibinfo {year} {2010})}\BibitemShut {NoStop}%
\bibitem [{\citenamefont {Alves}\ \emph {et~al.}(2019)\citenamefont {Alves}, \citenamefont {Martins}, \citenamefont {Domenech},\ and\ \citenamefont {Abbate}}]{ALVES20192952}%
  \BibitemOpen
  \bibfield  {author} {\bibinfo {author} {\bibfnamefont {E.}~\bibnamefont {Alves}}, \bibinfo {author} {\bibfnamefont {H.}~\bibnamefont {Martins}}, \bibinfo {author} {\bibfnamefont {S.}~\bibnamefont {Domenech}}, \ and\ \bibinfo {author} {\bibfnamefont {M.}~\bibnamefont {Abbate}},\ }\href {\doibase https://doi.org/10.1016/j.physleta.2019.06.012} {\bibfield  {journal} {\bibinfo  {journal} {Physics Letters A}\ }\textbf {\bibinfo {volume} {383}},\ \bibinfo {pages} {2952 } (\bibinfo {year} {2019})}\BibitemShut {NoStop}%
\bibitem [{\citenamefont {Sarma}\ \emph {et~al.}(1995)\citenamefont {Sarma}, \citenamefont {Shanthi}, \citenamefont {Barman}, \citenamefont {Hamada}, \citenamefont {Sawada},\ and\ \citenamefont {Terakura}}]{Sarma1995}%
  \BibitemOpen
  \bibfield  {author} {\bibinfo {author} {\bibfnamefont {D.~D.}\ \bibnamefont {Sarma}}, \bibinfo {author} {\bibfnamefont {N.}~\bibnamefont {Shanthi}}, \bibinfo {author} {\bibfnamefont {S.~R.}\ \bibnamefont {Barman}}, \bibinfo {author} {\bibfnamefont {N.}~\bibnamefont {Hamada}}, \bibinfo {author} {\bibfnamefont {H.}~\bibnamefont {Sawada}}, \ and\ \bibinfo {author} {\bibfnamefont {K.}~\bibnamefont {Terakura}},\ }\href {\doibase 10.1103/PhysRevLett.75.1126} {\bibfield  {journal} {\bibinfo  {journal} {Phys. Rev. Lett.}\ }\textbf {\bibinfo {volume} {75}},\ \bibinfo {pages} {1126} (\bibinfo {year} {1995})}\BibitemShut {NoStop}%
\bibitem [{\citenamefont {Solovyev}\ \emph {et~al.}(1996)\citenamefont {Solovyev}, \citenamefont {Hamada},\ and\ \citenamefont {Terakura}}]{PhysRevB.53.7158}%
  \BibitemOpen
  \bibfield  {author} {\bibinfo {author} {\bibfnamefont {I.}~\bibnamefont {Solovyev}}, \bibinfo {author} {\bibfnamefont {N.}~\bibnamefont {Hamada}}, \ and\ \bibinfo {author} {\bibfnamefont {K.}~\bibnamefont {Terakura}},\ }\href {\doibase 10.1103/PhysRevB.53.7158} {\bibfield  {journal} {\bibinfo  {journal} {Phys. Rev. B}\ }\textbf {\bibinfo {volume} {53}},\ \bibinfo {pages} {7158} (\bibinfo {year} {1996})}\BibitemShut {NoStop}%
\bibitem [{\citenamefont {Petocchi}\ \emph {et~al.}(2020)\citenamefont {Petocchi}, \citenamefont {Nilsson}, \citenamefont {Aryasetiawan},\ and\ \citenamefont {~}}]{PhysRevResearch.2.013191}%
  \BibitemOpen
  \bibfield  {author} {\bibinfo {author} {\bibfnamefont {F.}~\bibnamefont {Petocchi}}, \bibinfo {author} {\bibfnamefont {F.}~\bibnamefont {Nilsson}}, \bibinfo {author} {\bibfnamefont {F.}~\bibnamefont {Aryasetiawan}}, \ and\ \bibinfo {author} {\bibfnamefont {P.}~\bibnamefont {~}},\ }\href {\doibase 10.1103/PhysRevResearch.2.013191} {\bibfield  {journal} {\bibinfo  {journal} {Phys. Rev. Research}\ }\textbf {\bibinfo {volume} {2}},\ \bibinfo {pages} {013191} (\bibinfo {year} {2020})}\BibitemShut {NoStop}%
\bibitem [{\citenamefont {Janson}\ \emph {et~al.}(2018)\citenamefont {Janson}, \citenamefont {Zhong}, \citenamefont {Sangiovanni},\ and\ \citenamefont {Held}}]{Janson2018}%
  \BibitemOpen
  \bibfield  {author} {\bibinfo {author} {\bibfnamefont {O.}~\bibnamefont {Janson}}, \bibinfo {author} {\bibfnamefont {Z.}~\bibnamefont {Zhong}}, \bibinfo {author} {\bibfnamefont {G.}~\bibnamefont {Sangiovanni}}, \ and\ \bibinfo {author} {\bibfnamefont {K.}~\bibnamefont {Held}},\ }\enquote {\bibinfo {title} {Dynamical mean field theory for oxide heterostructures},}\ in\ \href {\doibase 10.1007/978-3-319-74989-1_9} {\emph {\bibinfo {booktitle} {Spectroscopy of Complex Oxide Interfaces: Photoemission and Related Spectroscopies}}},\ \bibinfo {editor} {edited by\ \bibinfo {editor} {\bibfnamefont {C.}~\bibnamefont {Cancellieri}}\ and\ \bibinfo {editor} {\bibfnamefont {V.~N.}\ \bibnamefont {Strocov}}}\ (\bibinfo  {publisher} {Springer International Publishing},\ \bibinfo {address} {Cham},\ \bibinfo {year} {2018})\ pp.\ \bibinfo {pages} {215--243}\BibitemShut {NoStop}%
\bibitem [{\citenamefont {He}\ and\ \citenamefont {Franchini}(2012)}]{PhysRevB.86.235117}%
  \BibitemOpen
  \bibfield  {author} {\bibinfo {author} {\bibfnamefont {J.}~\bibnamefont {He}}\ and\ \bibinfo {author} {\bibfnamefont {C.}~\bibnamefont {Franchini}},\ }\href {\doibase 10.1103/PhysRevB.86.235117} {\bibfield  {journal} {\bibinfo  {journal} {Phys. Rev. B}\ }\textbf {\bibinfo {volume} {86}},\ \bibinfo {pages} {235117} (\bibinfo {year} {2012})}\BibitemShut {NoStop}%
\bibitem [{\citenamefont {Karolak}\ \emph {et~al.}(2011)\citenamefont {Karolak}, \citenamefont {Wehling}, \citenamefont {Lechermann},\ and\ \citenamefont {Lichtenstein}}]{Karolak_2011}%
  \BibitemOpen
  \bibfield  {author} {\bibinfo {author} {\bibfnamefont {M.}~\bibnamefont {Karolak}}, \bibinfo {author} {\bibfnamefont {T.~O.}\ \bibnamefont {Wehling}}, \bibinfo {author} {\bibfnamefont {F.}~\bibnamefont {Lechermann}}, \ and\ \bibinfo {author} {\bibfnamefont {A.~I.}\ \bibnamefont {Lichtenstein}},\ }\href {\doibase 10.1088/0953-8984/23/8/085601} {\bibfield  {journal} {\bibinfo  {journal} {Journal of Physics: Condensed Matter}\ }\textbf {\bibinfo {volume} {23}},\ \bibinfo {pages} {085601} (\bibinfo {year} {2011})}\BibitemShut {NoStop}%
\bibitem [{\citenamefont {Nekrasov}\ \emph {et~al.}(2006)\citenamefont {Nekrasov}, \citenamefont {Held}, \citenamefont {Keller}, \citenamefont {Kondakov}, \citenamefont {Pruschke}, \citenamefont {Kollar}, \citenamefont {Andersen}, \citenamefont {Anisimov},\ and\ \citenamefont {Vollhardt}}]{PhysRevB.73.155112}%
  \BibitemOpen
  \bibfield  {author} {\bibinfo {author} {\bibfnamefont {I.~A.}\ \bibnamefont {Nekrasov}}, \bibinfo {author} {\bibfnamefont {K.}~\bibnamefont {Held}}, \bibinfo {author} {\bibfnamefont {G.}~\bibnamefont {Keller}}, \bibinfo {author} {\bibfnamefont {D.~E.}\ \bibnamefont {Kondakov}}, \bibinfo {author} {\bibfnamefont {T.}~\bibnamefont {Pruschke}}, \bibinfo {author} {\bibfnamefont {M.}~\bibnamefont {Kollar}}, \bibinfo {author} {\bibfnamefont {O.~K.}\ \bibnamefont {Andersen}}, \bibinfo {author} {\bibfnamefont {V.~I.}\ \bibnamefont {Anisimov}}, \ and\ \bibinfo {author} {\bibfnamefont {D.}~\bibnamefont {Vollhardt}},\ }\href {\doibase 10.1103/PhysRevB.73.155112} {\bibfield  {journal} {\bibinfo  {journal} {Phys. Rev. B}\ }\textbf {\bibinfo {volume} {73}},\ \bibinfo {pages} {155112} (\bibinfo {year} {2006})}\BibitemShut {NoStop}%
\bibitem [{\citenamefont {Kohiki}\ \emph {et~al.}(2000)\citenamefont {Kohiki}, \citenamefont {Arai}, \citenamefont {Yoshikawa}, \citenamefont {Fukushima}, \citenamefont {Oku},\ and\ \citenamefont {Waseda}}]{PhysRevB.62.7964}%
  \BibitemOpen
  \bibfield  {author} {\bibinfo {author} {\bibfnamefont {S.}~\bibnamefont {Kohiki}}, \bibinfo {author} {\bibfnamefont {M.}~\bibnamefont {Arai}}, \bibinfo {author} {\bibfnamefont {H.}~\bibnamefont {Yoshikawa}}, \bibinfo {author} {\bibfnamefont {S.}~\bibnamefont {Fukushima}}, \bibinfo {author} {\bibfnamefont {M.}~\bibnamefont {Oku}}, \ and\ \bibinfo {author} {\bibfnamefont {Y.}~\bibnamefont {Waseda}},\ }\href {\doibase 10.1103/PhysRevB.62.7964} {\bibfield  {journal} {\bibinfo  {journal} {Phys. Rev. B}\ }\textbf {\bibinfo {volume} {62}},\ \bibinfo {pages} {7964} (\bibinfo {year} {2000})}\BibitemShut {NoStop}%
\bibitem [{\citenamefont {Han}\ \emph {et~al.}(2014)\citenamefont {Han}, \citenamefont {Kino},\ and\ \citenamefont {Kotani}}]{PhysRevB.90.035127}%
  \BibitemOpen
  \bibfield  {author} {\bibinfo {author} {\bibfnamefont {M.~J.}\ \bibnamefont {Han}}, \bibinfo {author} {\bibfnamefont {H.}~\bibnamefont {Kino}}, \ and\ \bibinfo {author} {\bibfnamefont {T.}~\bibnamefont {Kotani}},\ }\href {\doibase 10.1103/PhysRevB.90.035127} {\bibfield  {journal} {\bibinfo  {journal} {Phys. Rev. B}\ }\textbf {\bibinfo {volume} {90}},\ \bibinfo {pages} {035127} (\bibinfo {year} {2014})}\BibitemShut {NoStop}%
\bibitem [{\citenamefont {Miyake}\ and\ \citenamefont {Aryasetiawan}(2008)}]{PhysRevB.77.085122}%
  \BibitemOpen
  \bibfield  {author} {\bibinfo {author} {\bibfnamefont {T.}~\bibnamefont {Miyake}}\ and\ \bibinfo {author} {\bibfnamefont {F.}~\bibnamefont {Aryasetiawan}},\ }\href {\doibase 10.1103/PhysRevB.77.085122} {\bibfield  {journal} {\bibinfo  {journal} {Phys. Rev. B}\ }\textbf {\bibinfo {volume} {77}},\ \bibinfo {pages} {085122} (\bibinfo {year} {2008})}\BibitemShut {NoStop}%
\bibitem [{\citenamefont {Yoshida}\ \emph {et~al.}(2010)\citenamefont {Yoshida}, \citenamefont {Hashimoto}, \citenamefont {Takizawa}, \citenamefont {Fujimori}, \citenamefont {Kubota}, \citenamefont {Ono},\ and\ \citenamefont {Eisaki}}]{PhysRevB.82.085119}%
  \BibitemOpen
  \bibfield  {author} {\bibinfo {author} {\bibfnamefont {T.}~\bibnamefont {Yoshida}}, \bibinfo {author} {\bibfnamefont {M.}~\bibnamefont {Hashimoto}}, \bibinfo {author} {\bibfnamefont {T.}~\bibnamefont {Takizawa}}, \bibinfo {author} {\bibfnamefont {A.}~\bibnamefont {Fujimori}}, \bibinfo {author} {\bibfnamefont {M.}~\bibnamefont {Kubota}}, \bibinfo {author} {\bibfnamefont {K.}~\bibnamefont {Ono}}, \ and\ \bibinfo {author} {\bibfnamefont {H.}~\bibnamefont {Eisaki}},\ }\href {\doibase 10.1103/PhysRevB.82.085119} {\bibfield  {journal} {\bibinfo  {journal} {Phys. Rev. B}\ }\textbf {\bibinfo {volume} {82}},\ \bibinfo {pages} {085119} (\bibinfo {year} {2010})}\BibitemShut {NoStop}%
\bibitem [{\citenamefont {Mossanek}\ \emph {et~al.}(2009)\citenamefont {Mossanek}, \citenamefont {Abbate}, \citenamefont {Yoshida}, \citenamefont {Fujimori}, \citenamefont {Yoshida}, \citenamefont {Shirakawa}, \citenamefont {Eisaki}, \citenamefont {Kohno}, \citenamefont {Fonseca},\ and\ \citenamefont {Vicentin}}]{PhysRevB.79.033104}%
  \BibitemOpen
  \bibfield  {author} {\bibinfo {author} {\bibfnamefont {R.~J.~O.}\ \bibnamefont {Mossanek}}, \bibinfo {author} {\bibfnamefont {M.}~\bibnamefont {Abbate}}, \bibinfo {author} {\bibfnamefont {T.}~\bibnamefont {Yoshida}}, \bibinfo {author} {\bibfnamefont {A.}~\bibnamefont {Fujimori}}, \bibinfo {author} {\bibfnamefont {Y.}~\bibnamefont {Yoshida}}, \bibinfo {author} {\bibfnamefont {N.}~\bibnamefont {Shirakawa}}, \bibinfo {author} {\bibfnamefont {H.}~\bibnamefont {Eisaki}}, \bibinfo {author} {\bibfnamefont {S.}~\bibnamefont {Kohno}}, \bibinfo {author} {\bibfnamefont {P.~T.}\ \bibnamefont {Fonseca}}, \ and\ \bibinfo {author} {\bibfnamefont {F.~C.}\ \bibnamefont {Vicentin}},\ }\href {\doibase 10.1103/PhysRevB.79.033104} {\bibfield  {journal} {\bibinfo  {journal} {Phys. Rev. B}\ }\textbf {\bibinfo {volume} {79}},\ \bibinfo {pages} {033104} (\bibinfo {year} {2009})}\BibitemShut {NoStop}%
\bibitem [{\citenamefont {Giannozzi}\ \emph {et~al.}(2009)\citenamefont {Giannozzi}, \citenamefont {Baroni}, \citenamefont {Bonini}, \citenamefont {Calandra}, \citenamefont {Car}, \citenamefont {Cavazzoni}, \citenamefont {Ceresoli}, \citenamefont {Chiarotti}, \citenamefont {Cococcioni}, \citenamefont {Dabo}, \citenamefont {{Dal Corso}}, \citenamefont {de~Gironcoli}, \citenamefont {Fabris}, \citenamefont {Fratesi}, \citenamefont {Gebauer}, \citenamefont {Gerstmann}, \citenamefont {Gougoussis}, \citenamefont {Kokalj}, \citenamefont {Lazzeri}, \citenamefont {Martin-Samos}, \citenamefont {Marzari}, \citenamefont {Mauri}, \citenamefont {Mazzarello}, \citenamefont {Paolini}, \citenamefont {Pasquarello}, \citenamefont {Paulatto}, \citenamefont {Sbraccia}, \citenamefont {Scandolo}, \citenamefont {Sclauzero}, \citenamefont {Seitsonen}, \citenamefont {Smogunov}, \citenamefont {Umari},\ and\ \citenamefont {Wentzcovitch}}]{QE-2009}%
  \BibitemOpen
  \bibfield  {author} {\bibinfo {author} {\bibfnamefont {P.}~\bibnamefont {Giannozzi}}, \bibinfo {author} {\bibfnamefont {S.}~\bibnamefont {Baroni}}, \bibinfo {author} {\bibfnamefont {N.}~\bibnamefont {Bonini}}, \bibinfo {author} {\bibfnamefont {M.}~\bibnamefont {Calandra}}, \bibinfo {author} {\bibfnamefont {R.}~\bibnamefont {Car}}, \bibinfo {author} {\bibfnamefont {C.}~\bibnamefont {Cavazzoni}}, \bibinfo {author} {\bibfnamefont {D.}~\bibnamefont {Ceresoli}}, \bibinfo {author} {\bibfnamefont {G.~L.}\ \bibnamefont {Chiarotti}}, \bibinfo {author} {\bibfnamefont {M.}~\bibnamefont {Cococcioni}}, \bibinfo {author} {\bibfnamefont {I.}~\bibnamefont {Dabo}}, \bibinfo {author} {\bibfnamefont {A.}~\bibnamefont {{Dal Corso}}}, \bibinfo {author} {\bibfnamefont {S.}~\bibnamefont {de~Gironcoli}}, \bibinfo {author} {\bibfnamefont {S.}~\bibnamefont {Fabris}}, \bibinfo {author} {\bibfnamefont {G.}~\bibnamefont {Fratesi}}, \bibinfo {author} {\bibfnamefont {R.}~\bibnamefont {Gebauer}}, \bibinfo {author} {\bibfnamefont
  {U.}~\bibnamefont {Gerstmann}}, \bibinfo {author} {\bibfnamefont {C.}~\bibnamefont {Gougoussis}}, \bibinfo {author} {\bibfnamefont {A.}~\bibnamefont {Kokalj}}, \bibinfo {author} {\bibfnamefont {M.}~\bibnamefont {Lazzeri}}, \bibinfo {author} {\bibfnamefont {L.}~\bibnamefont {Martin-Samos}}, \bibinfo {author} {\bibfnamefont {N.}~\bibnamefont {Marzari}}, \bibinfo {author} {\bibfnamefont {F.}~\bibnamefont {Mauri}}, \bibinfo {author} {\bibfnamefont {R.}~\bibnamefont {Mazzarello}}, \bibinfo {author} {\bibfnamefont {S.}~\bibnamefont {Paolini}}, \bibinfo {author} {\bibfnamefont {A.}~\bibnamefont {Pasquarello}}, \bibinfo {author} {\bibfnamefont {L.}~\bibnamefont {Paulatto}}, \bibinfo {author} {\bibfnamefont {C.}~\bibnamefont {Sbraccia}}, \bibinfo {author} {\bibfnamefont {S.}~\bibnamefont {Scandolo}}, \bibinfo {author} {\bibfnamefont {G.}~\bibnamefont {Sclauzero}}, \bibinfo {author} {\bibfnamefont {A.~P.}\ \bibnamefont {Seitsonen}}, \bibinfo {author} {\bibfnamefont {A.}~\bibnamefont {Smogunov}}, \bibinfo {author}
  {\bibfnamefont {P.}~\bibnamefont {Umari}}, \ and\ \bibinfo {author} {\bibfnamefont {R.~M.}\ \bibnamefont {Wentzcovitch}},\ }\href {http://www.quantum-espresso.org} {\bibfield  {journal} {\bibinfo  {journal} {Journal of Physics: Condensed Matter}\ }\textbf {\bibinfo {volume} {21}},\ \bibinfo {pages} {395502 (19pp)} (\bibinfo {year} {2009})}\BibitemShut {NoStop}%
\bibitem [{\citenamefont {Giannozzi}\ \emph {et~al.}(2017)\citenamefont {Giannozzi}, \citenamefont {Andreussi}, \citenamefont {Brumme}, \citenamefont {Bunau}, \citenamefont {Nardelli}, \citenamefont {Calandra}, \citenamefont {Car}, \citenamefont {Cavazzoni}, \citenamefont {Ceresoli}, \citenamefont {Cococcioni}, \citenamefont {Colonna}, \citenamefont {Carnimeo}, \citenamefont {Corso}, \citenamefont {de~Gironcoli}, \citenamefont {Delugas}, \citenamefont {Jr}, \citenamefont {Ferretti}, \citenamefont {Floris}, \citenamefont {Fratesi}, \citenamefont {Fugallo}, \citenamefont {Gebauer}, \citenamefont {Gerstmann}, \citenamefont {Giustino}, \citenamefont {Gorni}, \citenamefont {Jia}, \citenamefont {Kawamura}, \citenamefont {Ko}, \citenamefont {Kokalj}, \citenamefont {KÃ¼Ã§Ã¼kbenli}, \citenamefont {Lazzeri}, \citenamefont {Marsili}, \citenamefont {Marzari}, \citenamefont {Mauri}, \citenamefont {Nguyen}, \citenamefont {Nguyen}, \citenamefont {de-la Roza}, \citenamefont {Paulatto}, \citenamefont {PoncÃ©}, \citenamefont
  {Rocca}, \citenamefont {Sabatini}, \citenamefont {Santra}, \citenamefont {Schlipf}, \citenamefont {Seitsonen}, \citenamefont {Smogunov}, \citenamefont {Timrov}, \citenamefont {Thonhauser}, \citenamefont {Umari}, \citenamefont {Vast}, \citenamefont {Wu},\ and\ \citenamefont {Baroni}}]{QE-2017}%
  \BibitemOpen
  \bibfield  {author} {\bibinfo {author} {\bibfnamefont {P.}~\bibnamefont {Giannozzi}}, \bibinfo {author} {\bibfnamefont {O.}~\bibnamefont {Andreussi}}, \bibinfo {author} {\bibfnamefont {T.}~\bibnamefont {Brumme}}, \bibinfo {author} {\bibfnamefont {O.}~\bibnamefont {Bunau}}, \bibinfo {author} {\bibfnamefont {M.~B.}\ \bibnamefont {Nardelli}}, \bibinfo {author} {\bibfnamefont {M.}~\bibnamefont {Calandra}}, \bibinfo {author} {\bibfnamefont {R.}~\bibnamefont {Car}}, \bibinfo {author} {\bibfnamefont {C.}~\bibnamefont {Cavazzoni}}, \bibinfo {author} {\bibfnamefont {D.}~\bibnamefont {Ceresoli}}, \bibinfo {author} {\bibfnamefont {M.}~\bibnamefont {Cococcioni}}, \bibinfo {author} {\bibfnamefont {N.}~\bibnamefont {Colonna}}, \bibinfo {author} {\bibfnamefont {I.}~\bibnamefont {Carnimeo}}, \bibinfo {author} {\bibfnamefont {A.~D.}\ \bibnamefont {Corso}}, \bibinfo {author} {\bibfnamefont {S.}~\bibnamefont {de~Gironcoli}}, \bibinfo {author} {\bibfnamefont {P.}~\bibnamefont {Delugas}}, \bibinfo {author} {\bibfnamefont
  {R.~A.~D.}\ \bibnamefont {Jr}}, \bibinfo {author} {\bibfnamefont {A.}~\bibnamefont {Ferretti}}, \bibinfo {author} {\bibfnamefont {A.}~\bibnamefont {Floris}}, \bibinfo {author} {\bibfnamefont {G.}~\bibnamefont {Fratesi}}, \bibinfo {author} {\bibfnamefont {G.}~\bibnamefont {Fugallo}}, \bibinfo {author} {\bibfnamefont {R.}~\bibnamefont {Gebauer}}, \bibinfo {author} {\bibfnamefont {U.}~\bibnamefont {Gerstmann}}, \bibinfo {author} {\bibfnamefont {F.}~\bibnamefont {Giustino}}, \bibinfo {author} {\bibfnamefont {T.}~\bibnamefont {Gorni}}, \bibinfo {author} {\bibfnamefont {J.}~\bibnamefont {Jia}}, \bibinfo {author} {\bibfnamefont {M.}~\bibnamefont {Kawamura}}, \bibinfo {author} {\bibfnamefont {H.-Y.}\ \bibnamefont {Ko}}, \bibinfo {author} {\bibfnamefont {A.}~\bibnamefont {Kokalj}}, \bibinfo {author} {\bibfnamefont {E.}~\bibnamefont {KÃ¼Ã§Ã¼kbenli}}, \bibinfo {author} {\bibfnamefont {M.}~\bibnamefont {Lazzeri}}, \bibinfo {author} {\bibfnamefont {M.}~\bibnamefont {Marsili}}, \bibinfo {author} {\bibfnamefont
  {N.}~\bibnamefont {Marzari}}, \bibinfo {author} {\bibfnamefont {F.}~\bibnamefont {Mauri}}, \bibinfo {author} {\bibfnamefont {N.~L.}\ \bibnamefont {Nguyen}}, \bibinfo {author} {\bibfnamefont {H.-V.}\ \bibnamefont {Nguyen}}, \bibinfo {author} {\bibfnamefont {A.~O.}\ \bibnamefont {de-la Roza}}, \bibinfo {author} {\bibfnamefont {L.}~\bibnamefont {Paulatto}}, \bibinfo {author} {\bibfnamefont {S.}~\bibnamefont {PoncÃ©}}, \bibinfo {author} {\bibfnamefont {D.}~\bibnamefont {Rocca}}, \bibinfo {author} {\bibfnamefont {R.}~\bibnamefont {Sabatini}}, \bibinfo {author} {\bibfnamefont {B.}~\bibnamefont {Santra}}, \bibinfo {author} {\bibfnamefont {M.}~\bibnamefont {Schlipf}}, \bibinfo {author} {\bibfnamefont {A.~P.}\ \bibnamefont {Seitsonen}}, \bibinfo {author} {\bibfnamefont {A.}~\bibnamefont {Smogunov}}, \bibinfo {author} {\bibfnamefont {I.}~\bibnamefont {Timrov}}, \bibinfo {author} {\bibfnamefont {T.}~\bibnamefont {Thonhauser}}, \bibinfo {author} {\bibfnamefont {P.}~\bibnamefont {Umari}}, \bibinfo {author}
  {\bibfnamefont {N.}~\bibnamefont {Vast}}, \bibinfo {author} {\bibfnamefont {X.}~\bibnamefont {Wu}}, \ and\ \bibinfo {author} {\bibfnamefont {S.}~\bibnamefont {Baroni}},\ }\href {http://stacks.iop.org/0953-8984/29/i=46/a=465901} {\bibfield  {journal} {\bibinfo  {journal} {Journal of Physics: Condensed Matter}\ }\textbf {\bibinfo {volume} {29}},\ \bibinfo {pages} {465901} (\bibinfo {year} {2017})}\BibitemShut {NoStop}%
\bibitem [{\citenamefont {Pizzi}\ \emph {et~al.}(2020)\citenamefont {Pizzi}, \citenamefont {Vitale}, \citenamefont {Arita}, \citenamefont {BlÃ¼gel}, \citenamefont {Freimuth}, \citenamefont {G{\'{e}}ranton}, \citenamefont {Gibertini}, \citenamefont {Gresch}, \citenamefont {Johnson}, \citenamefont {Koretsune}, \citenamefont {Iba{\~{n}}ez-Azpiroz}, \citenamefont {Lee}, \citenamefont {Lihm}, \citenamefont {Marchand}, \citenamefont {Marrazzo}, \citenamefont {Mokrousov}, \citenamefont {Mustafa}, \citenamefont {Nohara}, \citenamefont {Nomura}, \citenamefont {Paulatto}, \citenamefont {Ponc{\'{e}}}, \citenamefont {Ponweiser}, \citenamefont {Qiao}, \citenamefont {ThÃ¶le}, \citenamefont {Tsirkin}, \citenamefont {Wierzbowska}, \citenamefont {Marzari}, \citenamefont {Vanderbilt}, \citenamefont {Souza}, \citenamefont {Mostofi},\ and\ \citenamefont {Yates}}]{Pizzi_2020}%
  \BibitemOpen
  \bibfield  {author} {\bibinfo {author} {\bibfnamefont {G.}~\bibnamefont {Pizzi}}, \bibinfo {author} {\bibfnamefont {V.}~\bibnamefont {Vitale}}, \bibinfo {author} {\bibfnamefont {R.}~\bibnamefont {Arita}}, \bibinfo {author} {\bibfnamefont {S.}~\bibnamefont {BlÃ¼gel}}, \bibinfo {author} {\bibfnamefont {F.}~\bibnamefont {Freimuth}}, \bibinfo {author} {\bibfnamefont {G.}~\bibnamefont {G{\'{e}}ranton}}, \bibinfo {author} {\bibfnamefont {M.}~\bibnamefont {Gibertini}}, \bibinfo {author} {\bibfnamefont {D.}~\bibnamefont {Gresch}}, \bibinfo {author} {\bibfnamefont {C.}~\bibnamefont {Johnson}}, \bibinfo {author} {\bibfnamefont {T.}~\bibnamefont {Koretsune}}, \bibinfo {author} {\bibfnamefont {J.}~\bibnamefont {Iba{\~{n}}ez-Azpiroz}}, \bibinfo {author} {\bibfnamefont {H.}~\bibnamefont {Lee}}, \bibinfo {author} {\bibfnamefont {J.-M.}\ \bibnamefont {Lihm}}, \bibinfo {author} {\bibfnamefont {D.}~\bibnamefont {Marchand}}, \bibinfo {author} {\bibfnamefont {A.}~\bibnamefont {Marrazzo}}, \bibinfo {author} {\bibfnamefont
  {Y.}~\bibnamefont {Mokrousov}}, \bibinfo {author} {\bibfnamefont {J.~I.}\ \bibnamefont {Mustafa}}, \bibinfo {author} {\bibfnamefont {Y.}~\bibnamefont {Nohara}}, \bibinfo {author} {\bibfnamefont {Y.}~\bibnamefont {Nomura}}, \bibinfo {author} {\bibfnamefont {L.}~\bibnamefont {Paulatto}}, \bibinfo {author} {\bibfnamefont {S.}~\bibnamefont {Ponc{\'{e}}}}, \bibinfo {author} {\bibfnamefont {T.}~\bibnamefont {Ponweiser}}, \bibinfo {author} {\bibfnamefont {J.}~\bibnamefont {Qiao}}, \bibinfo {author} {\bibfnamefont {F.}~\bibnamefont {ThÃ¶le}}, \bibinfo {author} {\bibfnamefont {S.~S.}\ \bibnamefont {Tsirkin}}, \bibinfo {author} {\bibfnamefont {M.}~\bibnamefont {Wierzbowska}}, \bibinfo {author} {\bibfnamefont {N.}~\bibnamefont {Marzari}}, \bibinfo {author} {\bibfnamefont {D.}~\bibnamefont {Vanderbilt}}, \bibinfo {author} {\bibfnamefont {I.}~\bibnamefont {Souza}}, \bibinfo {author} {\bibfnamefont {A.~A.}\ \bibnamefont {Mostofi}}, \ and\ \bibinfo {author} {\bibfnamefont {J.~R.}\ \bibnamefont {Yates}},\ }\href {\doibase
  10.1088/1361-648x/ab51ff} {\bibfield  {journal} {\bibinfo  {journal} {Journal of Physics: Condensed Matter}\ }\textbf {\bibinfo {volume} {32}},\ \bibinfo {pages} {165902} (\bibinfo {year} {2020})}\BibitemShut {NoStop}%
\bibitem [{\citenamefont {Prandini}\ \emph {et~al.}(2018)\citenamefont {Prandini}, \citenamefont {Marrazzo}, \citenamefont {Castelli}, \citenamefont {Mounet},\ and\ \citenamefont {Marzari}}]{PSP}%
  \BibitemOpen
  \bibfield  {author} {\bibinfo {author} {\bibfnamefont {G.}~\bibnamefont {Prandini}}, \bibinfo {author} {\bibfnamefont {A.}~\bibnamefont {Marrazzo}}, \bibinfo {author} {\bibfnamefont {I.~E.}\ \bibnamefont {Castelli}}, \bibinfo {author} {\bibfnamefont {N.}~\bibnamefont {Mounet}}, \ and\ \bibinfo {author} {\bibfnamefont {N.}~\bibnamefont {Marzari}},\ }\href {https://doi.org/10.1038/s41524-018-0127-2} {\bibfield  {journal} {\bibinfo  {journal} {npj Computational Materials}\ }\textbf {\bibinfo {volume} {4}},\ \bibinfo {pages} {72} (\bibinfo {year} {2018})}\BibitemShut {NoStop}%
\bibitem [{\citenamefont {Gonze}\ \emph {et~al.}(2020)\citenamefont {Gonze}, \citenamefont {Amadon}, \citenamefont {Antonius}, \citenamefont {Arnardi}, \citenamefont {Baguet}, \citenamefont {Beuken}, \citenamefont {Bieder}, \citenamefont {Bottin}, \citenamefont {Bouchet}, \citenamefont {Bousquet}, \citenamefont {Brouwer}, \citenamefont {Bruneval}, \citenamefont {Brunin}, \citenamefont {Cavignac}, \citenamefont {Charraud}, \citenamefont {Chen}, \citenamefont {CÃ´tÃ©}, \citenamefont {Cottenier}, \citenamefont {Denier}, \citenamefont {Geneste}, \citenamefont {Ghosez}, \citenamefont {Giantomassi}, \citenamefont {Gillet}, \citenamefont {Gingras}, \citenamefont {Hamann}, \citenamefont {Hautier}, \citenamefont {He}, \citenamefont {Helbig}, \citenamefont {Holzwarth}, \citenamefont {Jia}, \citenamefont {Jollet}, \citenamefont {Lafargue-Dit-Hauret}, \citenamefont {Lejaeghere}, \citenamefont {Marques}, \citenamefont {Martin}, \citenamefont {Martins}, \citenamefont {Miranda}, \citenamefont {Naccarato}, \citenamefont
  {Persson}, \citenamefont {Petretto}, \citenamefont {Planes}, \citenamefont {Pouillon}, \citenamefont {Prokhorenko}, \citenamefont {Ricci}, \citenamefont {Rignanese}, \citenamefont {Romero}, \citenamefont {Schmitt}, \citenamefont {Torrent}, \citenamefont {van Setten}, \citenamefont {Troeye}, \citenamefont {Verstraete}, \citenamefont {ZÃ©rah},\ and\ \citenamefont {Zwanziger}}]{Gonze2020}%
  \BibitemOpen
  \bibfield  {author} {\bibinfo {author} {\bibfnamefont {X.}~\bibnamefont {Gonze}}, \bibinfo {author} {\bibfnamefont {B.}~\bibnamefont {Amadon}}, \bibinfo {author} {\bibfnamefont {G.}~\bibnamefont {Antonius}}, \bibinfo {author} {\bibfnamefont {F.}~\bibnamefont {Arnardi}}, \bibinfo {author} {\bibfnamefont {L.}~\bibnamefont {Baguet}}, \bibinfo {author} {\bibfnamefont {J.-M.}\ \bibnamefont {Beuken}}, \bibinfo {author} {\bibfnamefont {J.}~\bibnamefont {Bieder}}, \bibinfo {author} {\bibfnamefont {F.}~\bibnamefont {Bottin}}, \bibinfo {author} {\bibfnamefont {J.}~\bibnamefont {Bouchet}}, \bibinfo {author} {\bibfnamefont {E.}~\bibnamefont {Bousquet}}, \bibinfo {author} {\bibfnamefont {N.}~\bibnamefont {Brouwer}}, \bibinfo {author} {\bibfnamefont {F.}~\bibnamefont {Bruneval}}, \bibinfo {author} {\bibfnamefont {G.}~\bibnamefont {Brunin}}, \bibinfo {author} {\bibfnamefont {T.}~\bibnamefont {Cavignac}}, \bibinfo {author} {\bibfnamefont {J.-B.}\ \bibnamefont {Charraud}}, \bibinfo {author} {\bibfnamefont {W.}~\bibnamefont
  {Chen}}, \bibinfo {author} {\bibfnamefont {M.}~\bibnamefont {CÃ´tÃ©}}, \bibinfo {author} {\bibfnamefont {S.}~\bibnamefont {Cottenier}}, \bibinfo {author} {\bibfnamefont {J.}~\bibnamefont {Denier}}, \bibinfo {author} {\bibfnamefont {G.}~\bibnamefont {Geneste}}, \bibinfo {author} {\bibfnamefont {P.}~\bibnamefont {Ghosez}}, \bibinfo {author} {\bibfnamefont {M.}~\bibnamefont {Giantomassi}}, \bibinfo {author} {\bibfnamefont {Y.}~\bibnamefont {Gillet}}, \bibinfo {author} {\bibfnamefont {O.}~\bibnamefont {Gingras}}, \bibinfo {author} {\bibfnamefont {D.~R.}\ \bibnamefont {Hamann}}, \bibinfo {author} {\bibfnamefont {G.}~\bibnamefont {Hautier}}, \bibinfo {author} {\bibfnamefont {X.}~\bibnamefont {He}}, \bibinfo {author} {\bibfnamefont {N.}~\bibnamefont {Helbig}}, \bibinfo {author} {\bibfnamefont {N.}~\bibnamefont {Holzwarth}}, \bibinfo {author} {\bibfnamefont {Y.}~\bibnamefont {Jia}}, \bibinfo {author} {\bibfnamefont {F.}~\bibnamefont {Jollet}}, \bibinfo {author} {\bibfnamefont {W.}~\bibnamefont
  {Lafargue-Dit-Hauret}}, \bibinfo {author} {\bibfnamefont {K.}~\bibnamefont {Lejaeghere}}, \bibinfo {author} {\bibfnamefont {M.~A.~L.}\ \bibnamefont {Marques}}, \bibinfo {author} {\bibfnamefont {A.}~\bibnamefont {Martin}}, \bibinfo {author} {\bibfnamefont {C.}~\bibnamefont {Martins}}, \bibinfo {author} {\bibfnamefont {H.~P.~C.}\ \bibnamefont {Miranda}}, \bibinfo {author} {\bibfnamefont {F.}~\bibnamefont {Naccarato}}, \bibinfo {author} {\bibfnamefont {K.}~\bibnamefont {Persson}}, \bibinfo {author} {\bibfnamefont {G.}~\bibnamefont {Petretto}}, \bibinfo {author} {\bibfnamefont {V.}~\bibnamefont {Planes}}, \bibinfo {author} {\bibfnamefont {Y.}~\bibnamefont {Pouillon}}, \bibinfo {author} {\bibfnamefont {S.}~\bibnamefont {Prokhorenko}}, \bibinfo {author} {\bibfnamefont {F.}~\bibnamefont {Ricci}}, \bibinfo {author} {\bibfnamefont {G.-M.}\ \bibnamefont {Rignanese}}, \bibinfo {author} {\bibfnamefont {A.~H.}\ \bibnamefont {Romero}}, \bibinfo {author} {\bibfnamefont {M.~M.}\ \bibnamefont {Schmitt}}, \bibinfo {author}
  {\bibfnamefont {M.}~\bibnamefont {Torrent}}, \bibinfo {author} {\bibfnamefont {M.~J.}\ \bibnamefont {van Setten}}, \bibinfo {author} {\bibfnamefont {B.~V.}\ \bibnamefont {Troeye}}, \bibinfo {author} {\bibfnamefont {M.~J.}\ \bibnamefont {Verstraete}}, \bibinfo {author} {\bibfnamefont {G.}~\bibnamefont {ZÃ©rah}}, \ and\ \bibinfo {author} {\bibfnamefont {J.~W.}\ \bibnamefont {Zwanziger}},\ }\href {https://doi.org/10.1016/j.cpc.2019.107042} {\bibfield  {journal} {\bibinfo  {journal} {Comput. Phys. Commun.}\ }\textbf {\bibinfo {volume} {248}},\ \bibinfo {pages} {107042} (\bibinfo {year} {2020})}\BibitemShut {NoStop}%
\bibitem [{\citenamefont {Torrent}\ \emph {et~al.}(2008)\citenamefont {Torrent}, \citenamefont {Jollet}, \citenamefont {Bottin}, \citenamefont {ZÃ©rah},\ and\ \citenamefont {Gonze}}]{Torrent2008}%
  \BibitemOpen
  \bibfield  {author} {\bibinfo {author} {\bibfnamefont {M.}~\bibnamefont {Torrent}}, \bibinfo {author} {\bibfnamefont {F.}~\bibnamefont {Jollet}}, \bibinfo {author} {\bibfnamefont {F.}~\bibnamefont {Bottin}}, \bibinfo {author} {\bibfnamefont {G.}~\bibnamefont {ZÃ©rah}}, \ and\ \bibinfo {author} {\bibfnamefont {X.}~\bibnamefont {Gonze}},\ }\href {\doibase 10.1016/j.commatsci.2007.07.020} {\bibfield  {journal} {\bibinfo  {journal} {Computational Materials Science}\ }\textbf {\bibinfo {volume} {42}},\ \bibinfo {pages} {337} (\bibinfo {year} {2008})}\BibitemShut {NoStop}%
\bibitem [{\citenamefont {Jollet}\ \emph {et~al.}(2014)\citenamefont {Jollet}, \citenamefont {Torrent},\ and\ \citenamefont {Holzwarth}}]{JOLLET20141246}%
  \BibitemOpen
  \bibfield  {author} {\bibinfo {author} {\bibfnamefont {F.}~\bibnamefont {Jollet}}, \bibinfo {author} {\bibfnamefont {M.}~\bibnamefont {Torrent}}, \ and\ \bibinfo {author} {\bibfnamefont {N.}~\bibnamefont {Holzwarth}},\ }\href {\doibase https://doi.org/10.1016/j.cpc.2013.12.023} {\bibfield  {journal} {\bibinfo  {journal} {Computer Physics Communications}\ }\textbf {\bibinfo {volume} {185}},\ \bibinfo {pages} {1246 } (\bibinfo {year} {2014})}\BibitemShut {NoStop}%
\bibitem [{\citenamefont {Itoh}(1993)}]{Itoh}%
  \BibitemOpen
  \bibfield  {author} {\bibinfo {author} {\bibfnamefont {S.}~\bibnamefont {Itoh}},\ }\href {\doibase 10.1016/0038-1098(93)90042-L} {\bibfield  {journal} {\bibinfo  {journal} {Solid State Communications}\ }\textbf {\bibinfo {volume} {88}},\ \bibinfo {pages} {525} (\bibinfo {year} {1993})}\BibitemShut {NoStop}%
\bibitem [{\citenamefont {Chamberland}\ and\ \citenamefont {Danielson}(1971)}]{CHAMBERLAND1971243}%
  \BibitemOpen
  \bibfield  {author} {\bibinfo {author} {\bibfnamefont {B.}~\bibnamefont {Chamberland}}\ and\ \bibinfo {author} {\bibfnamefont {P.}~\bibnamefont {Danielson}},\ }\href {\doibase https://doi.org/10.1016/0022-4596(71)90035-1} {\bibfield  {journal} {\bibinfo  {journal} {Journal of Solid State Chemistry}\ }\textbf {\bibinfo {volume} {3}},\ \bibinfo {pages} {243 } (\bibinfo {year} {1971})}\BibitemShut {NoStop}%
\bibitem [{\citenamefont {Monkhorst}\ and\ \citenamefont {Pack}(1976)}]{PhysRevB.13.5188}%
  \BibitemOpen
  \bibfield  {author} {\bibinfo {author} {\bibfnamefont {H.~J.}\ \bibnamefont {Monkhorst}}\ and\ \bibinfo {author} {\bibfnamefont {J.~D.}\ \bibnamefont {Pack}},\ }\href {\doibase 10.1103/PhysRevB.13.5188} {\bibfield  {journal} {\bibinfo  {journal} {Phys. Rev. B}\ }\textbf {\bibinfo {volume} {13}},\ \bibinfo {pages} {5188} (\bibinfo {year} {1976})}\BibitemShut {NoStop}%
\bibitem [{\citenamefont {Garc\'{\i}a-Mu\~noz}\ \emph {et~al.}(1992)\citenamefont {Garc\'{\i}a-Mu\~noz}, \citenamefont {Rodr\'{\i}guez-Carvajal}, \citenamefont {Lacorre},\ and\ \citenamefont {Torrance}}]{PhysRevB.46.4414}%
  \BibitemOpen
  \bibfield  {author} {\bibinfo {author} {\bibfnamefont {J.~L.}\ \bibnamefont {Garc\'{\i}a-Mu\~noz}}, \bibinfo {author} {\bibfnamefont {J.}~\bibnamefont {Rodr\'{\i}guez-Carvajal}}, \bibinfo {author} {\bibfnamefont {P.}~\bibnamefont {Lacorre}}, \ and\ \bibinfo {author} {\bibfnamefont {J.~B.}\ \bibnamefont {Torrance}},\ }\href {\doibase 10.1103/PhysRevB.46.4414} {\bibfield  {journal} {\bibinfo  {journal} {Phys. Rev. B}\ }\textbf {\bibinfo {volume} {46}},\ \bibinfo {pages} {4414} (\bibinfo {year} {1992})}\BibitemShut {NoStop}%
\bibitem [{\citenamefont {Masys}\ and\ \citenamefont {Jonauskas}(2015)}]{MASYS2015153}%
  \BibitemOpen
  \bibfield  {author} {\bibinfo {author} {\bibfnamefont {Å.}~\bibnamefont {Masys}}\ and\ \bibinfo {author} {\bibfnamefont {V.}~\bibnamefont {Jonauskas}},\ }\href {\doibase https://doi.org/10.1016/j.commatsci.2015.06.034} {\bibfield  {journal} {\bibinfo  {journal} {Computational Materials Science}\ }\textbf {\bibinfo {volume} {108}},\ \bibinfo {pages} {153 } (\bibinfo {year} {2015})}\BibitemShut {NoStop}%
\bibitem [{\citenamefont {Hamada}(1993)}]{HAMADA19931157}%
  \BibitemOpen
  \bibfield  {author} {\bibinfo {author} {\bibfnamefont {N.}~\bibnamefont {Hamada}},\ }\href {\doibase https://doi.org/10.1016/0022-3697(93)90159-O} {\bibfield  {journal} {\bibinfo  {journal} {Journal of Physics and Chemistry of Solids}\ }\textbf {\bibinfo {volume} {54}},\ \bibinfo {pages} {1157 } (\bibinfo {year} {1993})},\ \bibinfo {note} {special Issue Spectroscopies in Novel Superconductors}\BibitemShut {NoStop}%
\bibitem [{\citenamefont {Kulik}\ and\ \citenamefont {Marzari}(2011)}]{Kulik}%
  \BibitemOpen
  \bibfield  {author} {\bibinfo {author} {\bibfnamefont {H.~J.}\ \bibnamefont {Kulik}}\ and\ \bibinfo {author} {\bibfnamefont {N.}~\bibnamefont {Marzari}},\ }\href {\doibase 10.1063/1.3559452} {\bibfield  {journal} {\bibinfo  {journal} {The Journal of Chemical Physics}\ }\textbf {\bibinfo {volume} {134}},\ \bibinfo {pages} {094103} (\bibinfo {year} {2011})},\ \Eprint {http://arxiv.org/abs/https://doi.org/10.1063/1.3559452} {https://doi.org/10.1063/1.3559452} \BibitemShut {NoStop}%
\end{thebibliography}
